\documentclass[ALICE,manyauthors]{cernphprep}
\usepackage[comma,square,numbers,sort&compress]{natbib}
\usepackage{hyperref}
\usepackage{lineno}
\usepackage{multirow}
\usepackage{xspace}
\usepackage{float}
\usepackage[T1]{fontenc}
\usepackage{orcidlink}

\begin{document}
%

\newcommand{\pp}           {pp\xspace}
\newcommand{\ppbar}        {\mbox{$\mathrm {p\overline{p}}$}\xspace}
\newcommand{\XeXe}         {\mbox{Xe--Xe}\xspace}
\newcommand{\PbPb}         {\mbox{Pb--Pb}\xspace}
\newcommand{\pA}           {\mbox{pA}\xspace}
\newcommand{\pPb}          {\mbox{p--Pb}\xspace}
\newcommand{\AuAu}         {\mbox{Au--Au}\xspace}
\newcommand{\dAu}          {\mbox{d--Au}\xspace}

\newcommand{\s}            {\ensuremath{\sqrt{s}}\xspace}
\newcommand{\snn}          {\ensuremath{\sqrt{s_{\mathrm{NN}}}}\xspace}
\newcommand{\pt}           {\ensuremath{p_{\rm T}}\xspace}
\newcommand{\meanpt}       {$\langle p_{\mathrm{T}}\rangle$\xspace}
\newcommand{\ycms}         {\ensuremath{y_{\rm CMS}}\xspace}
\newcommand{\ylab}         {\ensuremath{y_{\rm lab}}\xspace}
\newcommand{\etarange}[1]  {\mbox{$\left | \eta \right |~<~#1$}}
\newcommand{\yrange}[1]    {\mbox{$\left | y \right |~<~#1$}}
\newcommand{\dndy}         {\ensuremath{\mathrm{d}N_\mathrm{ch}/\mathrm{d}y}\xspace}
\newcommand{\dndeta}       {\ensuremath{\mathrm{d}N_\mathrm{ch}/\mathrm{d}\eta}\xspace}
\newcommand{\avdndeta}     {\ensuremath{\langle\dndeta\rangle}\xspace}
\newcommand{\dNdy}         {\ensuremath{\mathrm{d}N_\mathrm{ch}/\mathrm{d}y}\xspace}
\newcommand{\Npart}        {\ensuremath{N_\mathrm{part}}\xspace}
\newcommand{\Ncoll}        {\ensuremath{N_\mathrm{coll}}\xspace}
\newcommand{\dEdx}         {\ensuremath{\textrm{d}E/\textrm{d}x}\xspace}
\newcommand{\RpPb}         {\ensuremath{R_{\rm pPb}}\xspace}

\newcommand{\nineH}        {$\sqrt{s}~=~0.9$~Te\kern-.1emV\xspace}
\newcommand{\seven}        {$\sqrt{s}~=~7$~Te\kern-.1emV\xspace}
\newcommand{\twoH}         {$\sqrt{s}~=~0.2$~Te\kern-.1emV\xspace}
\newcommand{\twosevensix}  {$\sqrt{s}~=~2.76$~Te\kern-.1emV\xspace}
\newcommand{\five}         {$\sqrt{s}~=~5.02$~Te\kern-.1emV\xspace}
\newcommand{\twosevensixnn}{$\sqrt{s_{\mathrm{NN}}}~=~2.76$~Te\kern-.1emV\xspace}
\newcommand{\fivenn}       {$\sqrt{s_{\mathrm{NN}}}~=~5.02$~Te\kern-.1emV\xspace}
\newcommand{\LT}           {L{\'e}vy-Tsallis\xspace}
\newcommand{\GeVc}         {Ge\kern-.1emV/$c$\xspace}
\newcommand{\MeVc}         {Me\kern-.1emV/$c$\xspace}
\newcommand{\TeV}          {Te\kern-.1emV\xspace}
\newcommand{\GeV}          {Ge\kern-.1emV\xspace}
\newcommand{\MeV}          {Me\kern-.1emV\xspace}
\newcommand{\GeVmass}      {Ge\kern-.2emV/$c^2$\xspace}
\newcommand{\MeVmass}      {Me\kern-.2emV/$c^2$\xspace}
\newcommand{\lumi}         {\ensuremath{\mathcal{L}}\xspace}

\newcommand{\ITS}          {\rm{ITS}\xspace}
\newcommand{\TOF}          {\rm{TOF}\xspace}
\newcommand{\ZDC}          {\rm{ZDC}\xspace}
\newcommand{\ZDCs}         {\rm{ZDCs}\xspace}
\newcommand{\ZNA}          {\rm{ZNA}\xspace}
\newcommand{\ZNC}          {\rm{ZNC}\xspace}
\newcommand{\SPD}          {\rm{SPD}\xspace}
\newcommand{\SDD}          {\rm{SDD}\xspace}
\newcommand{\SSD}          {\rm{SSD}\xspace}
\newcommand{\TPC}          {\rm{TPC}\xspace}
\newcommand{\TRD}          {\rm{TRD}\xspace}
\newcommand{\VZERO}        {\rm{V0}\xspace}
\newcommand{\VZEROA}       {\rm{V0A}\xspace}
\newcommand{\VZEROC}       {\rm{V0C}\xspace}
\newcommand{\Vdecay} 	   {\ensuremath{V^{0}}\xspace}

\newcommand{\ee}           {\ensuremath{e^{+}e^{-}}} 
\newcommand{\pip}          {\ensuremath{\pi^{+}}\xspace}
\newcommand{\pim}          {\ensuremath{\pi^{-}}\xspace}
\newcommand{\kap}          {\ensuremath{\rm{K}^{+}}\xspace}
\newcommand{\kam}          {\ensuremath{\rm{K}^{-}}\xspace}
\newcommand{\pbar}         {\ensuremath{\rm\overline{p}}\xspace}
\newcommand{\kzero}        {\ensuremath{{\rm K}^{0}_{\rm{S}}}\xspace}
\newcommand{\lmb}          {\ensuremath{\Lambda}\xspace}
\newcommand{\almb}         {\ensuremath{\overline{\Lambda}}\xspace}
\newcommand{\Om}           {\ensuremath{\Omega^-}\xspace}
\newcommand{\Mo}           {\ensuremath{\overline{\Omega}^+}\xspace}
\newcommand{\X}            {\ensuremath{\Xi^-}\xspace}
\newcommand{\Ix}           {\ensuremath{\overline{\Xi}^+}\xspace}
\newcommand{\Xis}          {\ensuremath{\Xi^{\pm}}\xspace}
\newcommand{\Oms}          {\ensuremath{\Omega^{\pm}}\xspace}
\newcommand{\degree}       {\ensuremath{^{\rm o}}\xspace}

\newcommand{\percent}{\% }
\newcommand{\h}{\langle}
\newcommand{\ia}{\rangle }
\newcommand{\towards}{$\mid\Delta\varphi\mid<60^{\circ},$}

\newcommand{\transverse}{$60^{\circ}<\mid\Delta\varphi\mid<120^{\circ},$ }
\newcommand{\away}{$\mid\Delta\varphi\mid>120^{\circ}.$ }
\newcommand{\nch}{$N_{ch}(\ p_{T,LT} )\ \Delta\eta\Delta\varphi N_{ev}(\ p_{T,LT} ) \ $ }

\newcommand{\etta}{$\mid\eta\mid< 0.8$ }
\newcommand{\PT}{$ p_{\rm T}$}
\newcommand{\PTmin}{$ p_{\rm {(T,min)}} $ }
\newcommand{\PTleading}{$ p_{\rm T}^{\rm leading}$ }
\newcommand{\ptt}{\ensuremath{p_{\mathrm{T}}^{\rm trig}}\xspace}
\newcommand{\py}{PYTHIA~8\xspace}
\newcommand{\ep}{EPOS~LHC\xspace}
\newcommand{\AD}        {\rm{AD}\xspace}
\newcommand{\ADA}        {\rm{ADA}\xspace}
\newcommand{\ADC}        {\rm{ADC}\xspace}

\begin{titlepage}
\PHyear{2022}       
\PHnumber{041}      
\PHdate{11 March}  

\title{Underlying-event properties in \pp and \pPb collisions at $\sqrt{\mathbf{\it s_{\rm NN}}}$ = 5.02\,\TeV}
\ShortTitle{Underlying-event properties in \pp and \pPb collisions at $\sqrt{s_{\rm NN}}$ = 5.02\,\TeV}   

\Collaboration{ALICE Collaboration\thanks{See Appendix~\ref{app:collab} for the list of collaboration members}}
\ShortAuthor{ALICE Collaboration} 

\begin{abstract}

We report about the properties of the underlying event measured with ALICE at the LHC in \pp and \pPb collisions at $\sqrt{s_{\rm NN}}=5.02$ \TeV. The event activity, quantified by charged-particle number and summed-$p_{\rm T}$ densities, is measured as a function of the leading-particle transverse momentum ($p_{\rm T}^{\rm trig}$). These quantities are studied in three azimuthal-angle regions relative to the leading particle in the event: toward, away, and transverse. Results are presented for three different $p_{\rm T}$ thresholds (0.15, 0.5 and 1 \,GeV/$c$) at mid-pseudorapidity ($|\eta|<0.8$). The event activity in the transverse region, which is the most sensitive to the underlying event, exhibits similar behaviour in both \pp and \pPb collisions, namely, a steep increase with $p_{\rm T}^{\rm trig}$ for low \ptt, followed by a saturation at $p_{\rm T}^{\rm trig} \approx 5$\,\GeVc. The results from \pp collisions are compared with existing measurements at other centre-of-mass energies. The quantities in the toward and away regions are also analyzed after the subtraction of the contribution measured in the transverse region. The remaining jet-like particle densities are consistent in \pp and \pPb collisions for $p_{\rm T}^{\rm trig}>10$\,\GeVc, whereas for lower  $p_{\rm T}^{\rm trig}$ values the event activity is slightly higher in \pPb than in \pp collisions. The measurements are compared with predictions from the \py and \ep Monte Carlo event generators.

\end{abstract}
\end{titlepage}

\setcounter{page}{2} 


\section{Introduction} 

\begin{figure}[ht]
\centering
\includegraphics[width=0.4\textwidth]{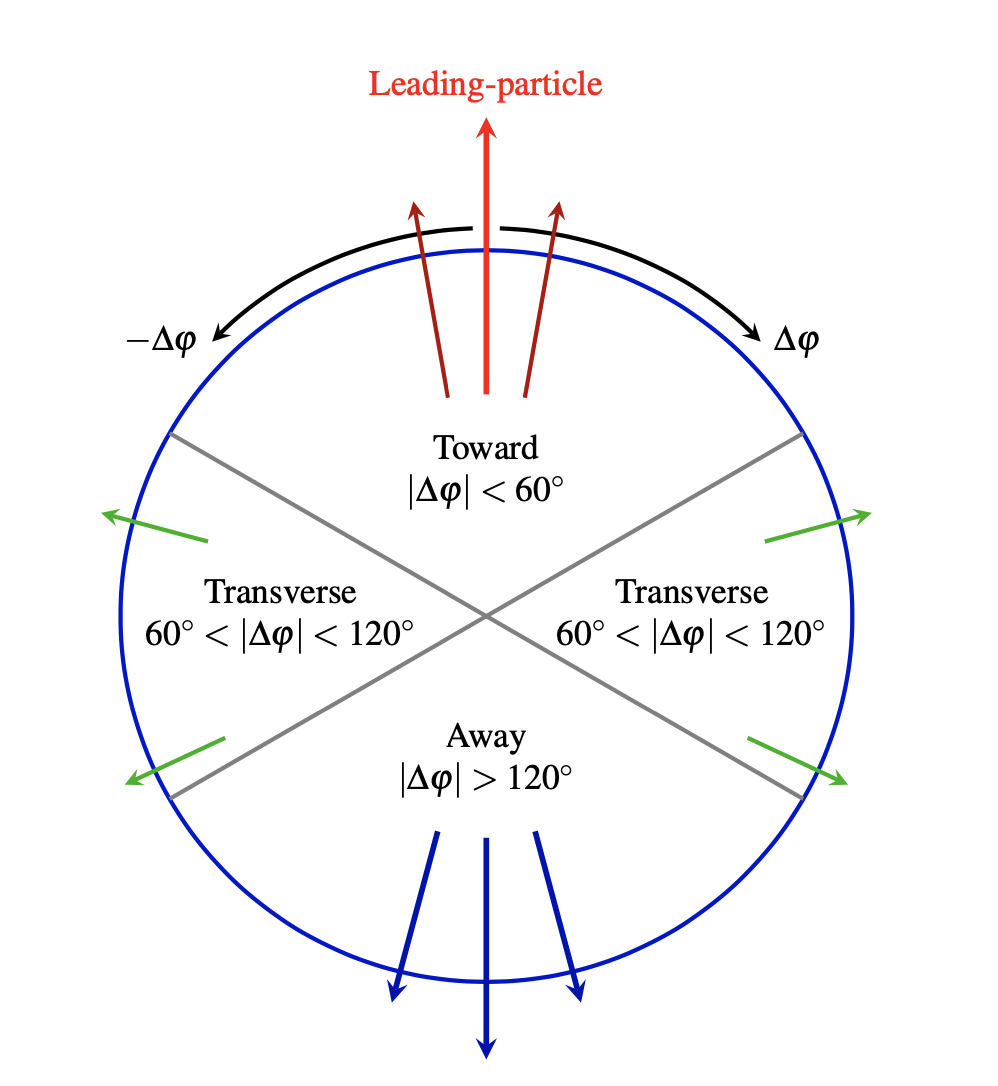}
\caption{Illustration of the toward, transverse, and away regions in the azimuthal plane with respect to the
leading particle direction. The figure has been taken from Ref.~\cite{ALICE:2019mmy}.}
\label{fig:f51001}
\end{figure}

In non-diffractive proton--proton (pp) collisions at high energies, the underlying event (UE) consists of the set of particles that arise from the proton break-up (beam remnants), and from other semi-hard scatterings, in a scenario of multiparton interactions (MPI)~\cite{Sjostrand:1987su}. The UE activity accompanies high transverse momentum (\pt) particles produced by the main partonic scattering (jets). Experimental studies aimed at probing the UE component are commonly performed in azimuthal-angle regions where the contribution from the hard scattering is expected to be minimal. The present study follows the strategy originally introduced by the CDF collaboration~\cite{Affolder:2001xt}. Firstly, the leading particle (or trigger particle) in the event is found, i.e., the charged particle with the highest transverse momentum in the collision (\ptt). Secondly, the associated particles for three different thresholds of the transverse momentum, $\pt>0.15$, 0.5, and 1\,GeV/$c$, are grouped in three classes depending on their relative azimuthal angle with respect to the leading particle, $|\Delta\varphi| =|\varphi^{\mathrm{assoc}}-\varphi^{\mathrm{trig}}|$:

\begin{itemize}
\item toward: \towards
\item transverse: \transverse and
\item away: \away 
\end{itemize}
The three topological regions corresponding to the azimuthal-angle intervals defined above are illustrated in Fig.~\ref{fig:f51001}. The toward region contains the primary jet of the
collision, while the away region contains the recoiled jet~\cite{Adam:2019xpp}. In contrast, the transverse region is mostly dominated by the UE dynamics, but it also includes contributions from initial- and final-state radiation (ISR and FSR)~\cite{Buttar:2005gdq}. This strategy~\cite{Affolder:2001xt} has been used by several experiments at RHIC~\cite{STAR:2019cie}, the Tevatron~\cite{Affolder:2001xt,CDF:2004jod,Field:2011iq,CDF:2010pdo}, and the LHC~\cite{CMS:2010rux,ATLAS:2010kmf,ALICE:2011ac,ATLAS:2014iez,ATLAS:2017blj,ALICE:2019mmy}. The studies include measurements in events with Drell-Yan~\cite{CMS:2012oqb} and Z-boson~\cite{ATLAS:2014yqy,CMS:2017ngy,ATLAS:2019ocl} production.

Experimental results have shown that the event activity, quantified by charged-particle number or summed-\pt densities, in the transverse region rises steeply with increasing \ptt at low \ptt ($<5$\,\GeVc), and then it roughly saturates (plateau) for larger \ptt~\cite{ALICE:2019mmy}. This saturation is expected in models that include the concept of impact parameter such that the requirement of the presence of a high-\pt particle in a \pp collision biases the selection of collisions towards those with a small impact parameter~\cite{Strikman:2011cx}. Based on UE observables measured at LHC centre-of-mass energies, $\sqrt{s}=0.9$, 7, and 13 \TeV, the event activity in the plateau region has been found to increase faster with increasing \s than in minimum-bias \pp collisions~\cite{ALICE:2011ac,ALICE:2019mmy}. An analogous study for \pPb collisions has never  been performed, although an attempt to determine the correlation between the impact parameter of the collision and the charged-particle multiplicity has been reported~\cite{ALICE:2014xsp}.

The measurements performed at RHIC and LHC in \pp, p--A, and d--A collisions have shown for high-particle multiplicities similar phenomena as were originally observed only in A--A collisions and have been attributed there to collective effects~\cite{Nagle:2018nvi}.
Thus, investigating \pp and \pPb collisions has become ever more pertinent in order to understand the origin of these effects~\cite{Nagle:2018nvi,ALICE:2017jyt,Adam:2016dau,Acharya:2018orn, Khachatryan:2010gv, Khachatryan:2016txc}. In QCD (quantum chromodynamics)-inspired Monte Carlo (MC) generators like \textsc{PYTHIA}~8~\cite{Sjostrand:2014zea}, outgoing partons originating from MPI are allowed to interact with those from the main partonic scattering. This mechanism, known as colour reconnection, produces effects resembling collective behaviour in pp collisions~\cite{OrtizVelasquez:2013ofg}. Given the dynamics encoded in the transverse region, the colour reconnection effects are expected to be more relevant in such a topological region~\cite{Ortiz:2018vgc}. Therefore, beyond the importance of UE measurements for MC tuning~\cite{Skands:2014pea,Gieseke:2012ft}, the study of the event activity in the transverse region is important to contribute to the understanding of the new effects observed in high-multiplicity \pp and \pPb collisions~\cite{Martin:2016igp}. 

In this paper, measurements of the event activity as a function of \ptt in \pp and \pPb collisions at the same centre-of-mass energy per nucleon pair ($\sqrt{s_{\rm NN}}=5.02$\,\TeV) are reported. The event activity for each topological region in \pPb collisions is compared with that in \pp collisions at the same \ptt. In order to search for a possible system size dependence of the jet-like particle densities, the jet-like signal (in the toward and away regions) is further isolated by subtracting the UE contribution estimated from the transverse region. The results from \pp collisions are compared to predictions from the EPOS tune LHC~\cite{Pierog:2013ria} and PYTHIA 8.244 (Monash 2013 tune~\cite{Skands:2014pea}) Monte Carlo event generators, hereinafter referred to as \ep and \py/Monash, respectively. For \pPb collisions, data are compared to EPOS~LHC and PYTHIA~8/Angantyr~\cite{Bierlich:2018xfw}. 

For \pp collisions, the modelling of UE in \py/Monash considers an impact-parameter dependent MPI activity.  The partonic configuration is hadronised using string fragmentation as described by the Lund string model~\cite{ANDERSSON198331}, followed by the decays of unstable particles. In collisions with several MPI, individual long strings connected to the remnants are replaced by shorter additional strings connecting partons from different semi-hard scatterings (colour reconnection). The Monash 2013 tune used minimum-bias, Drell-Yan, and UE data from the LHC to constrain the initial-state radiation and multiparton interactions, combined with data from the SPS and the Tevatron to constrain the scaling with the collision energy. The simulation of \pPb collisions was performed with the recent model named Angantyr~\cite{Bierlich:2018xfw}, which is based on an extrapolation of \pp dynamics with a minimum number of free parameters. The model does not assume the formation of a hot thermalised medium, instead, the generalisation to collisions involving nuclei is inspired by the Fritiof model~\cite{PI1992173} and the notion of ``wounded'' or ``participating'' nucleons. The number of wounded nucleons is calculated from the Glauber model in impact parameter space. With these assumptions, the model is able to give a good description of general final-state properties such as multiplicity and transverse momentum distributions of particles produced in interactions involving heavy nuclei. 

In \ep, the description of multiple partonic scatterings is based on a combination of Gribov-Regge theory and pQCD~\cite{Drescher:2000ha}. An elementary scattering corresponds to a parton ladder, containing a hard scattering which is calculated based on pQCD, including initial- and final-state radiation. Parton ladders that are formed in parallel to each other share the total collision energy leading to consistent treatment of energy conservation in hadronic collisions. String hadronisation in EPOS is based on the local density of the string segments per unit volume with respect to a critical-density parameter. Event-by-event, string segments in low-density regions hadronise normally and independently, creating the so-called corona, while string segments in high-density regions are used to create a core with collective expansion resulting in radial and longitudinal flow effects. The \ep tune considered here~\cite{Pierog:2013ria} is based on a dedicated parameter set used to describe data from different centre-of-mass energies and collision systems at the LHC.

The paper is organised as follows. The ALICE detectors used in the analysis are described in Section~\ref{sec:4}. Section~\ref{sec:5} is dedicated to illustrate the analysis technique, the data correction procedures, and the evaluation of the systematic uncertainties. The results are presented and discussed in Section~\ref{sec:6}, and the conclusions are summarised in Section~\ref{sec:7}.

\section{Experimental setup}\label{sec:4}

The main ALICE detectors used in the present work are the Inner Tracking System (ITS), the Time Projection Chamber (TPC), and the \VZERO detector. The ITS and TPC detectors are both used for primary vertex and track reconstruction. The \VZERO detector is used for triggering and for beam background rejection. More details concerning the ALICE detector system and its performance can be found in Refs.~\cite{ALICE:2008ngc,ALICE:2014sbx}. 

The ITS and TPC detectors are the main tracking devices covering the pseudorapidity region $|\eta|< 0.8$ for full-length tracks. They are located inside a solenoidal magnet providing a 0.5\,T magnetic field, allowing the tracking of particles with $p_{\rm T} \gtrsim 0.15$\,\GeVc. The ITS is composed of six cylindrical layers of high-resolution silicon tracking detectors. The innermost layers consist of two arrays of hybrid Silicon Pixel Detectors (SPD) located at an average radial distance of 3.9\,cm and 7.6\,cm from the beam axis and covering $|\eta|< 2$ and $|\eta|< 1.4$, respectively. The TPC has an active radial range from about 85 to 250\,cm, and an overall length along the beam direction of 500\,cm.  The TPC readout chambers have 159 tangential pad rows and thus a charged particle can, ideally, produce 159 clusters within the TPC volume. The readout chambers are mounted into 18 trapezoidal sectors at each end plate~\cite{ALICE:2014sbx}. The \VZERO detector consists of two sub-detectors placed on each side of the interaction point covering the full azimuthal acceptance and the pseudorapidity intervals of $2.8<\eta<5.1$ (\VZEROA) and $-3.7<\eta<-1.7$ (\VZEROC).

This analysis is based on the data recorded by the ALICE apparatus during the \pp run at $\sqrt{s}=5.02$\,\TeV in 2015, and the \pPb run at \snn = 5.02\,\TeV in 2016. The data were collected using a minimum-bias trigger, which required a signal in both \VZEROA and \VZEROC detectors. Only events with a reconstructed vertex within $\pm10$\,cm from the nominal interaction point along the beam direction are used. Runs with a low number of interactions per bunch crossing ($\mu$) were  selected resulting in average $\mu$ values of 0.020 and 0.005 for \pp and \pPb collisions,  respectively. Therefore, events with multiple collisions (pile-up) constitute a small fraction of the triggered events. They are identified and rejected based on the presence of multiple interaction vertices reconstructed using the SPD information. The remaining undetected pile-up is negligible for this analysis, while  the fraction of wrongly-tagged events due to SPD vertex splitting from a single interaction is $<10^{-4}$  for both collision systems. The offline event selection is optimised to reject beam-induced background by exploiting the timing signals in the two \VZERO sub-detectors. The event selection also requires at least one track with a minimum transverse momentum ($p_{\rm T} = 0.15$, $0.5$, and $1.0$\,\GeVc) in the acceptance range $|\eta| < 0.8$. The results presented in this article were obtained from the analysis of about 180 and 332 million minimum-bias \pp and \pPb collisions, respectively.

\section{Analysis details}\label{sec:5}

\subsection{Track reconstruction and selection}
The event properties are studied from the number and the momenta of the primary charged particles in the pseudorapidity interval $|\eta|<0.8$. Primary particles are defined as particles with a mean proper lifetime larger than 1\,cm/$c$, which are either produced directly in the interaction or from decays of particles with a mean proper lifetime smaller than 1\,cm/$c$~\cite{ALICE-PUBLIC-2017-005}. Charged particles are reconstructed with the ITS and TPC detectors, providing a measurement of the track transverse momentum \pt and azimuthal angle $\varphi$, which are used in the analysis. Tracks are required to have at least two hits in the ITS detector, of which at least one in either of the two innermost layers. The ratio of crossed TPC pad rows to the number of findable TPC clusters is required to be larger than 0.8, and the fraction of TPC clusters shared with another track should be less than 0.4. In addition, tracks are required to have a number of crossed TPC pad rows larger than $0.85 \times L$, where $L$ (in cm) is the geometrical track length calculated in the TPC readout plane, excluding the information from the pads at the sector boundaries ($\approx 3$\,cm from the sector edges). The number of TPC clusters associated to the track is required to be larger than $0.7 \times L$. The fit quality for the ITS and TPC track points must satisfy $\chi^{2}_{\mathrm{ITS}}/N_{\mathrm{hits}}<36$ and $\chi^{2}_{\mathrm{TPC}}/N_{\mathrm {clusters}}<4$, respectively, where $N_{\mathrm{hits}}$ and $N_{\mathrm{clusters}}$ are the number of hits in the ITS and the number of clusters in the TPC, respectively.
To select primary particles, tracks having a large distance of closest approach (DCA) to the reconstructed vertex in the longitudinal ($d_{\rm z}>2$\,cm) and radial  ($d_{\rm{xy}}>0.018$\,cm $+0.035$\,cm$\times$(\GeVc)$\times p_{\rm T}^{-1}$) directions are rejected. To further reduce the contamination from secondary particles, only tracks with $\chi^{2}_{{\rm TPC}-{\rm ITS}}<36$ are included in the analysis, where $\chi^{2}_{{\rm TPC}-{\rm ITS}}$  is calculated by comparing the track parameters from the combined ITS and TPC track reconstruction to those derived only from the TPC and constrained by the interaction point~\cite{Abelev:2012hxa}.  For this track selection~\cite{Acharya:2018qsh}, the momentum resolution is approximately 3-4\% at \pt$ = 0.15$\,GeV/$c$, it has a minimum of 1.0\% at \pt$ = 1.0$\,GeV/c, and increases linearly for larger \pt, approaching 3-10\% at 50\,GeV/$c$, depending on collision energy, and collision system. The measurements presented in this work are not corrected for momentum resolution but the effects are included by the systematic uncertainties. 

\subsection{Underlying-event observables}
The transverse momentum spectra (\pt) as a function of \ptt are corrected for all \ptt intervals and are extracted for each topological region. Then, both the primary charged-particle number and the summed transverse-momentum densities are calculated from the \pt spectra. The event activity in each topological region is measured as a function of \ptt. It is quantified with the primary charged-particle number density:
\begin{equation}
  \Big \langle \frac{ {\rm d}^{2} N_{\rm ch}}{{\rm d}\eta {\rm d}\varphi} \Big \rangle (p_{\rm T}^{\rm trig}) = \frac{1}{ \Delta\eta\Delta\varphi}  \frac{1}{N_{\rm ev} (p_{\rm T}^{\rm trig}  )\ }N_{\rm ch} (p_{\rm T}^{\rm trig}),
  \label{eq1}
\end{equation}
and the summed transverse-momentum density: 
\begin{equation}
   \Big \langle \frac{ {\rm d}^{2} \sum p_{\rm T}}{{\rm d}\eta {\rm d}\varphi} \Big \rangle (p_{\rm T}^{\rm trig}) =  \frac{1}{ \Delta\eta\Delta\varphi}  \frac{1}{N_{\rm ev} (p_{\rm T}^{\rm trig})\ } \sum p_{\rm T}(p_{\rm T}^{\rm trig}),
   \label{eq2}
\end{equation}
where $N_{\rm ev} (p_{\rm T}^{\rm trig})$ is the total number of events with the leading particle in a given \ptt interval; $N_{\rm ch}$(\ptt) and $\sum p_{\rm T}$(\ptt) stand for multiplicity and sum of the \pt of all reconstructed tracks within a given topological region, respectively.  Finally, $\Delta\eta$ is the pseudorapidity interval used in the analysis. 

This paper also reports the charged-particle number and the summed-\pt densities in the toward and away regions after the subtraction of the event activity in the transverse region. All these quantities are measured as a function of \ptt.

The charged-particle number density in the jet-like signal is derived from the difference between the number density in the toward (or away) region and that in the transverse region:
\begin{equation}
   \Big \langle \frac{ {\rm d}^{2} N_{\rm ch}}{{\rm d}\eta {\rm d}\varphi} \Big \rangle^{\rm jet \, toward(away)} (\ptt) = \Big[ \Big \langle \frac{ {\rm d}^{2} N_{\rm ch}}{{\rm d}\eta {\rm d}\varphi} \Big \rangle^{\rm toward(away)} -  \Big \langle \frac{ {\rm d}^{2} N_{\rm ch}}{{\rm d}\eta {\rm d}\varphi} \Big \rangle^{\rm transverse}  \Big] (p_{\rm T}^{\rm trig}).
   \label{eq3}
\end{equation}

In the same way, the summed-\pt density in the jet-like signal is obtained as follows: 

\begin{equation}
   \Big \langle \frac{ {\rm d}^{2} \sum p_{\rm T}}{{\rm d}\eta {\rm d}\varphi} \Big \rangle^{\rm jet \, toward(away)}  (\ptt)= \Big[ \Big \langle \frac{ {\rm d}^{2} \sum p_{\rm T}}{{\rm d}\eta {\rm d}\varphi} \Big \rangle^{\rm toward(away)} -  \Big \langle \frac{ {\rm d}^{2} \sum p_{\rm T}}{{\rm d}\eta {\rm d}\varphi} \Big \rangle^{\rm transverse}  \Big] (p_{\rm T}^{\rm trig}).
    \label{eq4}
\end{equation}

The ratio between these two quantities gives the average transverse momentum in the jet-like signal: 

\begin{equation}
    \langle p_{\rm T} \rangle^{\rm jet \, toward(away)} = \Big \langle \frac{ {\rm d}^{2} \sum p_{\rm T}}{{\rm d}\eta {\rm d}\varphi} \Big \rangle^{\rm jet \, toward(away)}  \Big /  \Big \langle \frac{ {\rm d}^{2} N_{\rm ch}}{{\rm d}\eta {\rm d}\varphi} \Big \rangle^{\rm jet \, toward(away)}.
    \label{eq5}
\end{equation}

\subsection{Corrections}

The correction of \pt spectra of charged particles follows the standard procedure of the ALICE collaboration~\cite{Acharya:2019mzb,Acharya:2018qsh}. The raw yields are corrected for efficiency and contamination from secondary particles. The efficiency correction is calculated from Monte Carlo simulations including the propagation of particles through the detector using GEANT~3~\cite{Brun:1082634}. For \pp and \pPb collisions the \py and \ep Monte Carlo event generators are used for this purpose, respectively. As the relative abundances of different charged particle species are different in the data and in the simulations, the efficiency obtained from the simulations is re-weighted considering the primary charged particle composition measured by ALICE~\cite{ALICE:2019hno}, as described in Ref.~\cite{Acharya:2018qsh}. At \pt$=0.15$\,GeV/$c$ the efficiency correction amounts to $\approx35$\%, which is related to the strong track curvature caused by the magnetic field and to the energy loss in the detector material. It is followed by a maximum value of 78\% at  \pt$\approx0.4$\,GeV/$c$, and a minimum ($\approx53$\%) at around \pt of 1\,GeV/$c$ which is caused primarily by the track length requirement. At higher \pt the efficiency correction reaches an asymptotic value of 70\% which reflects the acceptance limitations (detector boundaries and active channels) of the measurement~\cite{Acharya:2018qsh}. The residual contamination from secondary particles in the sample of selected tracks is estimated via a fit to the measured $d_{\rm{xy}}$ distributions by a combination of the $d_{\rm{xy}}$ distributions (templates) of primary and secondary particles obtained from the simulations~\cite{Acharya:2018qsh}.

Due to the finite acceptance and the efficiency of the detection apparatus, the leading particle may not be detected, and a track with lower \pt could be considered as the trigger particle. If the misidentified leading particle has a different \pt but roughly the same direction as the true leading particle, this leads to a small effect on the UE observables~\cite{ALICE:2011ac}. On the other hand, if the misidentified leading particle has a significantly different direction than the true one, this will cause a rotation of the event topology and a bias on the UE observables. Therefore, the particle densities are corrected for these effects using a data-driven procedure described in detail in Ref.~\cite{ALICE:2011ac}. A minor correction due to  the finite vertex reconstruction efficiency is also applied to the UE observables. The statistical uncertainty from the MC simulations was properly propagated to the final statistical uncertainties on data, but the analysis was performed in such a way that the correction factors are not strongly affected by the MC statistical uncertainties.

\subsection{Systematic uncertainties}

\begin{table}[b]
\begin{center}
\caption{Main sources and values of the relative systematic uncertainties of the charged-particle number and summed-\pt densities in \pp and \pPb collisions at $\sqrt{s_{\rm NN}}=5.02$\,\TeV. The average values of the uncertainties for the \ptt intervals 0.5--2\,\GeVc and 5--40\,\GeVc are displayed in the left and right columns, respectively. These uncertainties correspond to the transverse momentum threshold $\pt>0.5$\,GeV/$c$. When more than
one number is quoted, the values refer to the uncertainty in toward, transverse, and away regions, respectively;
they are independent of the azimuthal region in all other cases.}
\label{tab:3-1}
\begin{tabular}{l|c|c|c|c}
    \hline
     \multirow{2}{3em}{{\bf pp collisions}}& \multicolumn{2}{c|}{ Number density} & \multicolumn{2}{c}{ Summed-\pt\,density} \\
     & \ptt $<$ 2\,\GeVc & \ptt $>$ 5\,\GeVc & \ptt $<$ 2\,\GeVc &\ptt $>$ 5\,\GeVc \\
    \hline
    Sec. contamination     & \small 0.4\% & \small negligible & \small 0.4\% & \small negligible\\
    Correction method      & \small 3.7\%, 3.5\%, 3.3\%  & \small 1.7\%, 0.1\%, 0.6\% &\small 1.8\%, 1.5\%, 1.6\% & \small 2.0\%, 0.2\%, 0.3\%  \\    
    Track cuts             &  \small2.1\% & \small2.8\% &\small 2.1\% &\small 2.9\%  \\ 
    ITS-TPC track matching & \small 0.9\% & \small 0.8\% & \small 0.9\% &\small 0.8\%  \\
    Misidentification bias & \small 2.3\% &\small negligible &\small 2.8\% & \small 0.2\%  \\    
    Event selection & \small negligible &\small negligible &\small negligible & \small negligible  \\ 
   \hline
    Total uncertainty      & \small 4.9\%, 4.8\%, 4.6\% &\small 3.4\%, 2.9\%, 3.0\% & \small 4.1\%, 3.9\%, 4.0\% & \small 3.6\%, 3.0\%, 3.0\%  \\
    \hline
\end{tabular}
\\
\begin{tabular}{l|c|c|c|c}
    \hline
     \multirow{2}{3em}{{\bf p--Pb collisions}}& \multicolumn{2}{c|}{ Number density} & \multicolumn{2}{c}{ Summed-\pt\,density} \\
     & \ptt $<$ 2\,\GeVc & \ptt $>$ 5\,\GeVc & \ptt $<$ 2\,\GeVc &\ptt $>$ 5\,\GeVc \\
     \hline
    Sec. contamination     & \small 0.5\% & \small negligible  & \small 0.4\% &\small negligible  \\
    Correction method  & \small 2.1\%, 2.0\%, 2.0\%  & \small 0.8\%, 0.6\%, 0.7\% & \small 0.8\%, 0.6\%, 0.7\% &\small 0.9\%, negl., 0.1\%  \\    
    Track cuts             &  \small 1.5\% & \small3.1\% &\small 1.4\% &\small 3.2\%  \\ 
        ITS-TPC track matching & \small 1.8\% & \small 2.0\% & \small 1.9\% &\small 2.0\%  \\
    Misidentification bias & \small 3.9\% & \small 0.1\% & \small 1.8\% &\small 0.2\%  \\        
    Event selection & \small negligible &\small negligible &\small negligible &\small negligible  \\
    \hline
    Total uncertainty      & \small 5.0\%, 5.0\%, 5.0\% & \small 3.8\%, 3.7\%, 3.8\% & \small 3.1\%, 3.1\%, 3.1\% & \small 3.9\%, 3.8\%, 3.8\%  \\
    \hline
\end{tabular}

\end{center}
\end{table}

The relative systematic uncertainties on the quantities presented in Eqs.~\ref{eq1},~\ref{eq2},~\ref{eq3},~\ref{eq4}, and~\ref{eq5} are summarised in Table~\ref{tab:3-1} for \pp and \pPb collisions. The details about the different conditions varied in the analysis to estimate the systematic uncertainties are described below. 

\begin{itemize}

\item {Secondary contamination:} the fits to the $d_{\rm xy}$ distributions with the templates from the simulations were repeated using different fit intervals, namely $-1<d_{\rm xy}<1$ cm and $-2<d_{\rm xy}<2$ cm, instead of the default interval $-3<d_{\rm xy}<3$ cm. The maximum deviation with respect to the result obtained with the default fit range was assigned as systematic uncertainty. 

\item {Correction method:} a possible bias introduced by imperfections in the correction procedure was estimated by performing the analysis on a Monte Carlo sample of pp collisions simulated with a given event generator. The generated particles were propagated through the detector and the reconstructed quantities were corrected with the same procedure applied to the real data utilising the correction factors extracted from simulations performed with the same event generator. The reconstructed quantities considers all tracks which satisfies the selection criteria, they include tracks from primary and secondary particles, as well as fake tracks. Momentum resolution effects are also considered in the reconstructed quantities. With this approach, one expects to reproduce the generated yields, i.e. the yields obtained from the event generator without any detector effect, within statistical uncertainty. This consideration holds only if each correction is evaluated with respect to all the variables to which the given correction is sensitive. Any statistically significant difference between input and corrected distributions is added in quadrature to the total systematic uncertainty. This uncertainty is the only one which is different for each topological region.

\item {Track selection:} the systematic uncertainty related to the track selection criteria was determined by varying the track quality cuts~\cite{Acharya:2018qsh,Acharya:2019mzb}. In particular, the upper limits of the track fit quality parameters in the ITS ($\chi^2_{\rm {ITS}}/N_{\rm {hits}}$) and the TPC ($\chi^2_{\rm {TPC}}/N_{\rm{clusters}}$) were varied in the ranges of 25--49 and 3--5, respectively. The minimum ratio of crossed TPC pad rows to the number of findable TPC clusters was varied within (0.7--0.9). The maximum fraction of shared TPC clusters was varied between 0.2 and 1, and the maximum $d_{\rm z}$ was varied within 1--5\,cm. The impact on the results due to the hit requirement in the SPD was also evaluated by removing that requirement from the track selection. The maximum deviation of the results obtained varying the selections (only a single track cut at a time) with respect to the result obtained using the default track selection criteria was assigned as systematic uncertainty on each individual track-quality variable. The total uncertainty is then calculated as the sum in quadrature of each contribution.  

\item {ITS-TPC track matching efficiency:} a systematic uncertainty on the track reconstruction efficiency originates from possible differences in the probability to match the TPC tracks to the ITS hits in data and in simulations. 
It was estimated by  comparing the matching efficiency in data and simulations and propagating their difference to the underlying-event observables used in the analysis.

\item {Leading-particle misidentification bias:} the uncertainty on the leading-track misidentification correction is estimated from the discrepancy between the data-driven correction used in the analysis and the correction obtained from simulated data where the true leading particle is known.

\item {Event-selection bias:} the systematic uncertainty due to event selection is obtained by varying from 5 to 15\,cm the cut on the absolute value of the $z$ component of the vertex position. The maximum deviation of the results obtained varying the vertex position with respect to the result obtained using the default cut (10\,cm) is assigned as systematic uncertainty. This contribution is found to be negligible.

\end{itemize}

\section{Results and discussion}\label{sec:6}

\begin{figure}[tb]
    \begin{center}
    \includegraphics[width=7.90cm, height=6.90cm]{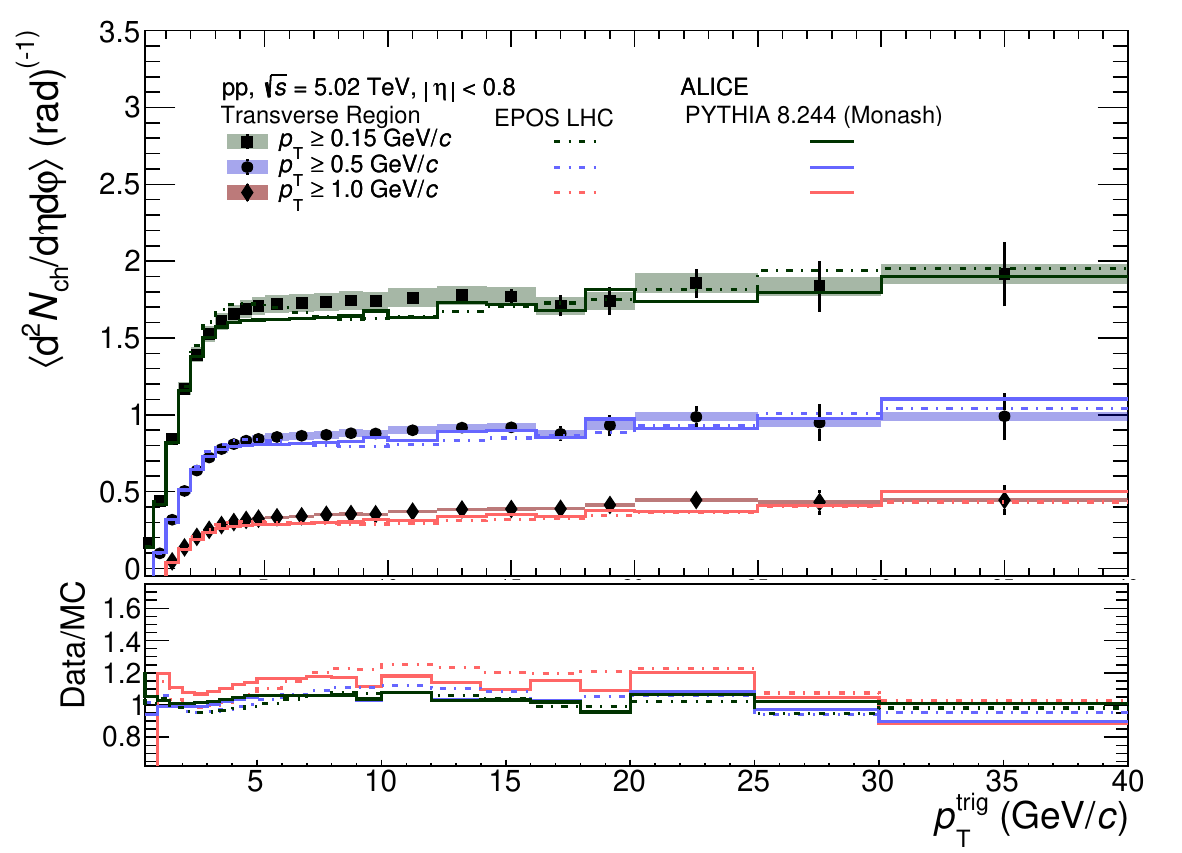}
    \includegraphics[width=7.90cm, height=6.90cm]{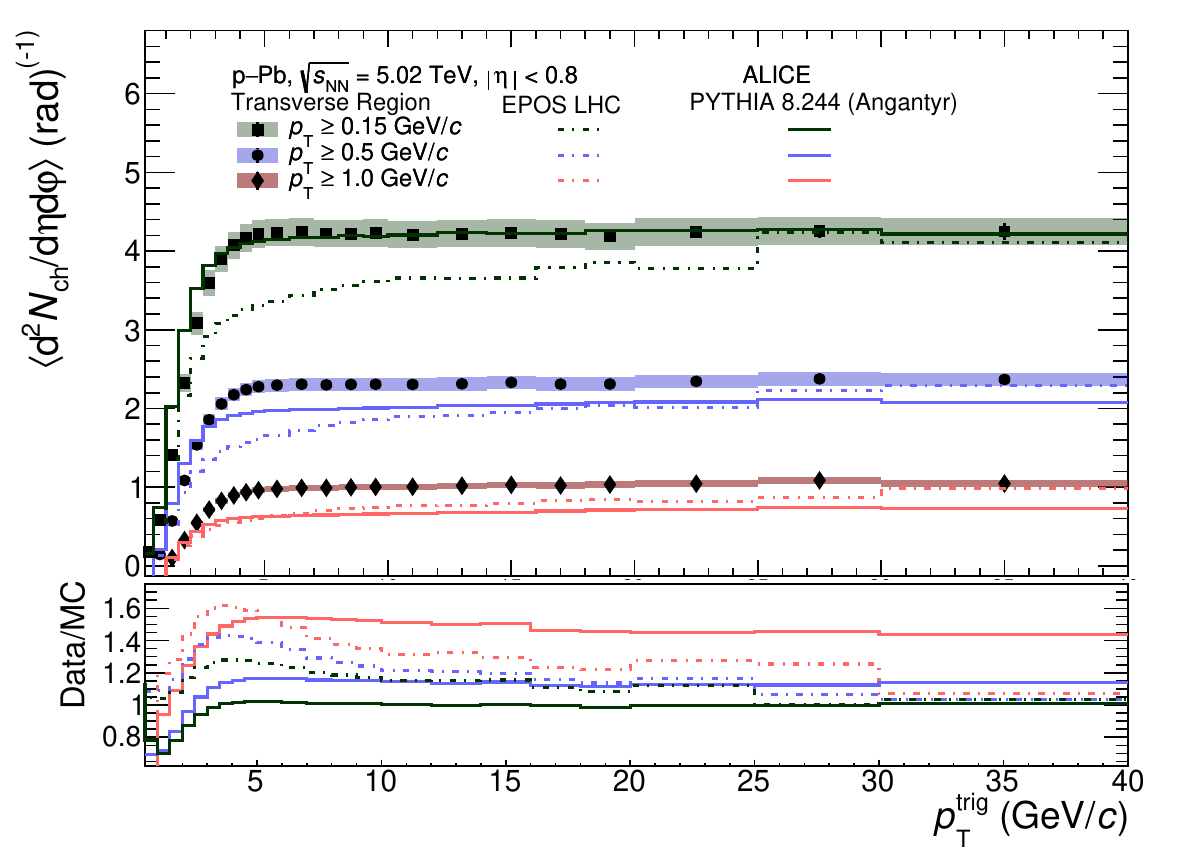}
    \end{center}
    \caption{Charged-particle number density as a function of \ptt measured in the transverse region for \pp (left) and \pPb collisions (right) at $\sqrt{s_{\rm NN}}=5.02$\,\TeV. Measurements were performed considering three \pt thresholds for associated charged particles: $\pt >0.15$\,\GeVc, $\pt >0.5$\,\GeVc, and $\pt >1$\,\GeVc. Data are compared with \py and \ep predictions. The coloured boxes and the error bars represent the systematic and statistical uncertainties, respectively.}
    \label{fig:10012}
\end{figure}

\subsection{Underlying-event observables:  \pp compared to \pPb collisions}

Figure~\ref{fig:10012} compares the charged-particle number density as a function of \ptt in the transverse region for the three  $p_{\rm T}$ thresholds: 0.15\,\GeV/$c$, 0.5\,\GeV/$c$, and 1\,\GeV/$c$. Figure~\ref{fig:10012} also shows the predictions of the event generators. For all the figures presented in this paper the statistical uncertainties in Monte Carlo predictions are not shown to facilitate the visualization of the data. In general the statistical error in MC is up to 2\% and 7\% for \ptt values of 10 and 20\,GeV/$c$, respectively.  The results from \pp and \pPb collisions exhibit similar behaviour: the number density steeply rises for low \ptt, and it flattens at $\ptt \approx 5$\,\GeVc (plateau region). In the plateau region, the event activity in \pPb collisions is $\approx 2$ times larger than the one measured in \pp collisions. This increase is smaller than that (about a factor of 3) observed for the charged-particle multiplicity densities ${\rm d}N_{\rm ch}/{\rm d}\eta$ in non-single-diffractive \pPb collisions compared to \pp collisions at the same nucleon--nucleon centre-of-mass energy~\cite{CMS:2017shj}.  It should also be noted that for both collision systems increasing the \pt threshold from 0.15\,\GeVc to 1.0\,\GeVc reduces the charged-particle number density by about a factor of 4. For \pp collisions, the charged-particle number density shows a slightly increasing trend with increasing \ptt in the plateau region ($\ptt>5$\,GeV/$c$). This increase is more pronounced for larger values of the \pt threshold for associated tracks, indicating an increased contribution of correlated hard processes (initial- and final-state radiation) to the transverse region. For example, for the \pt threshold $\pt>1$\,\GeVc, the charged-particle number density increases from 0.3 to 0.45 (i.e., by about 50\%) when \ptt is increased from 5 to 40\,\GeVc. Whereas for the \pt threshold $\pt>0.15$\,\GeVc, the increase is less than 10\%. In contrast, for \pPb collisions the charged-particle number density in the plateau region is flat for all the \pt thresholds. This behaviour also suggests that the contamination from the main partonic scattering in the transverse region is smaller in \pPb than in \pp collisions. Figure~\ref{fig:10012} also shows the predictions of the event generators. \ep better describes the \ptt dependence of the charged-particle multiplicity density for \pp collisions relative to \pPb collisions. For \pPb collisions, \ep significantly underestimates the charged-particle number density, and it does not reproduce the trend with \ptt and the value at the plateau observed in data. In contrast, \py/Angantyr qualitatively reproduces the measured trends with \ptt in \pPb collisions, providing a good quantitative description of the data for the lowest \pt threshold ($\pt>0.15$\,GeV/$c$), while it underestimates the measured densities in the plateau region for higher \pt thresholds.   In the following, measurements with the \pt threshold requirement of $0.5$\,\GeVc for associated particles are reported and discussed.  Results for other \pt thresholds ($p_{\rm T} >0.15$\,\GeVc and $>1.0$\,\GeVc) are presented in Appendix~\ref{app:101}.

\begin{figure}[tb]
    \begin{center}
    \includegraphics[width=7.90cm, height=6.90cm]{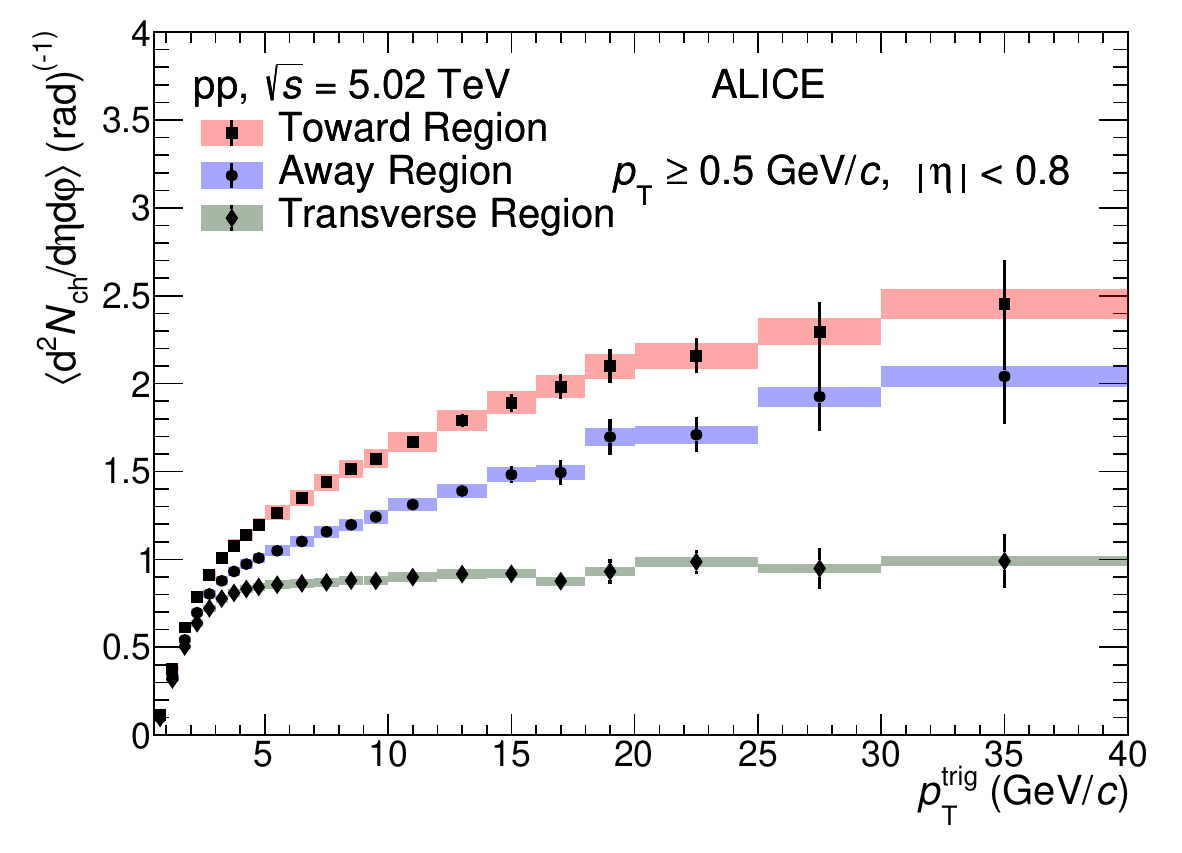}
    \includegraphics[width=7.90cm, height=6.90cm]{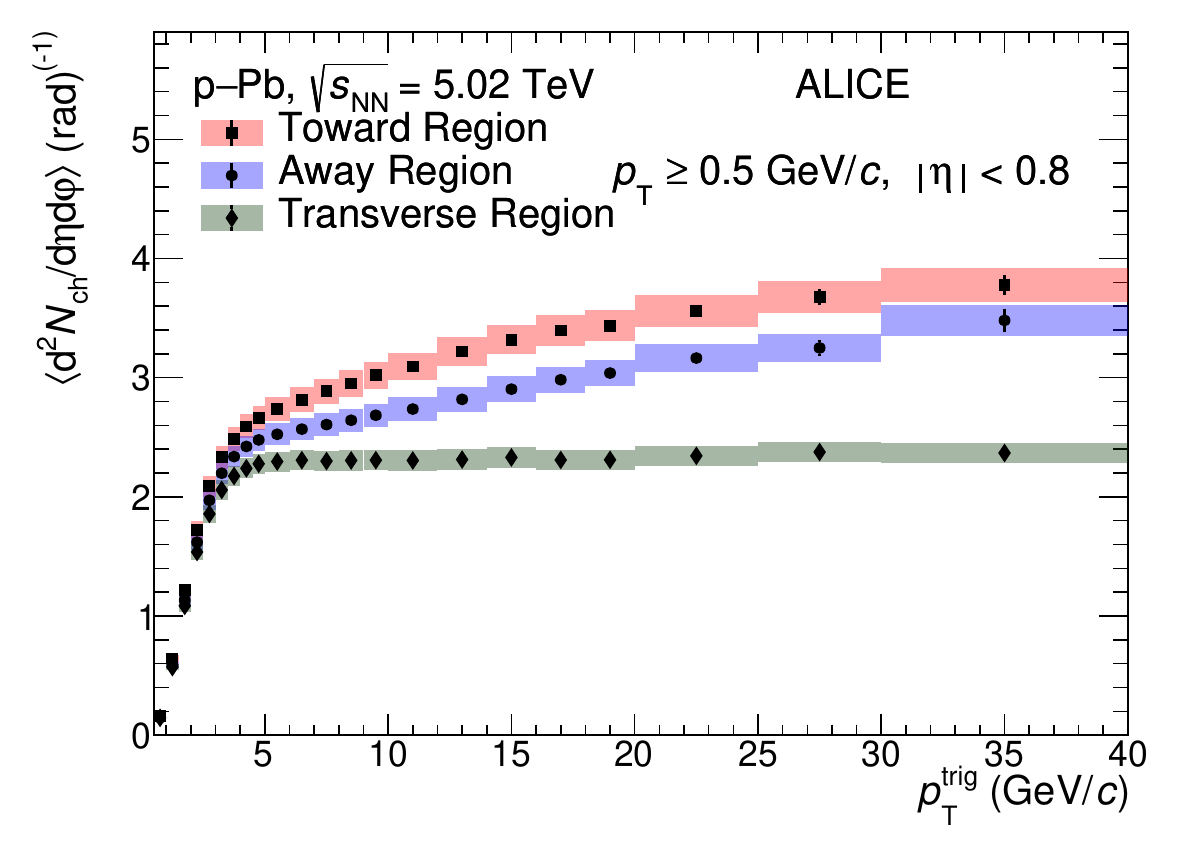}
    \end{center}
    \caption{Charged-particle number density as a function of \ptt measured in \pp (left) and \pPb collisions (right) at $\sqrt{s_{\rm NN}}=5.02$\,\TeV. Measurements were performed considering associated charged particles with $\pt>0.5$\,\GeVc. Results for the toward, transverse, and away regions are displayed. The coloured boxes and the error bars represent the systematic and statistical uncertainties, respectively.}
    \label{fig:10011}
\end{figure}

\begin{figure}[!hp]
\begin{center}
  \includegraphics[width=7.90cm, height=6.80cm]{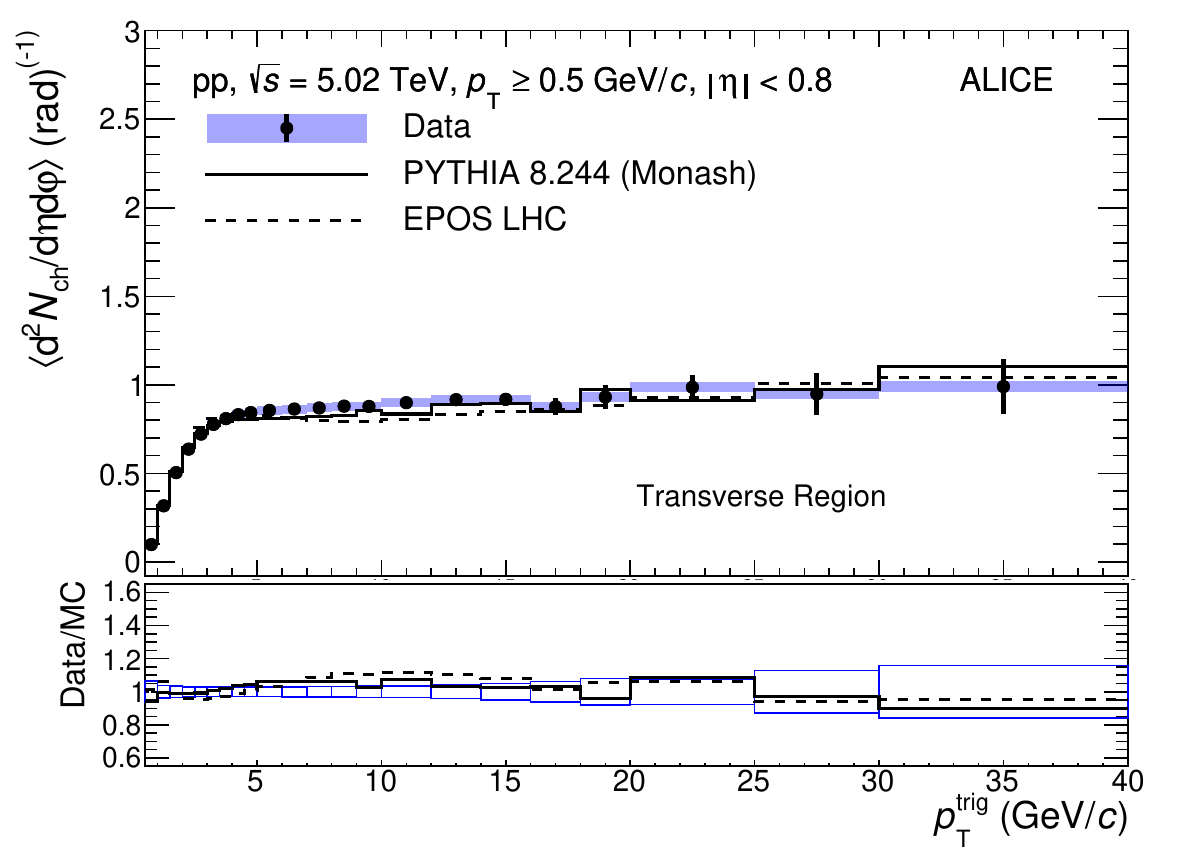}
  \includegraphics[width=7.90cm, height=6.80cm]{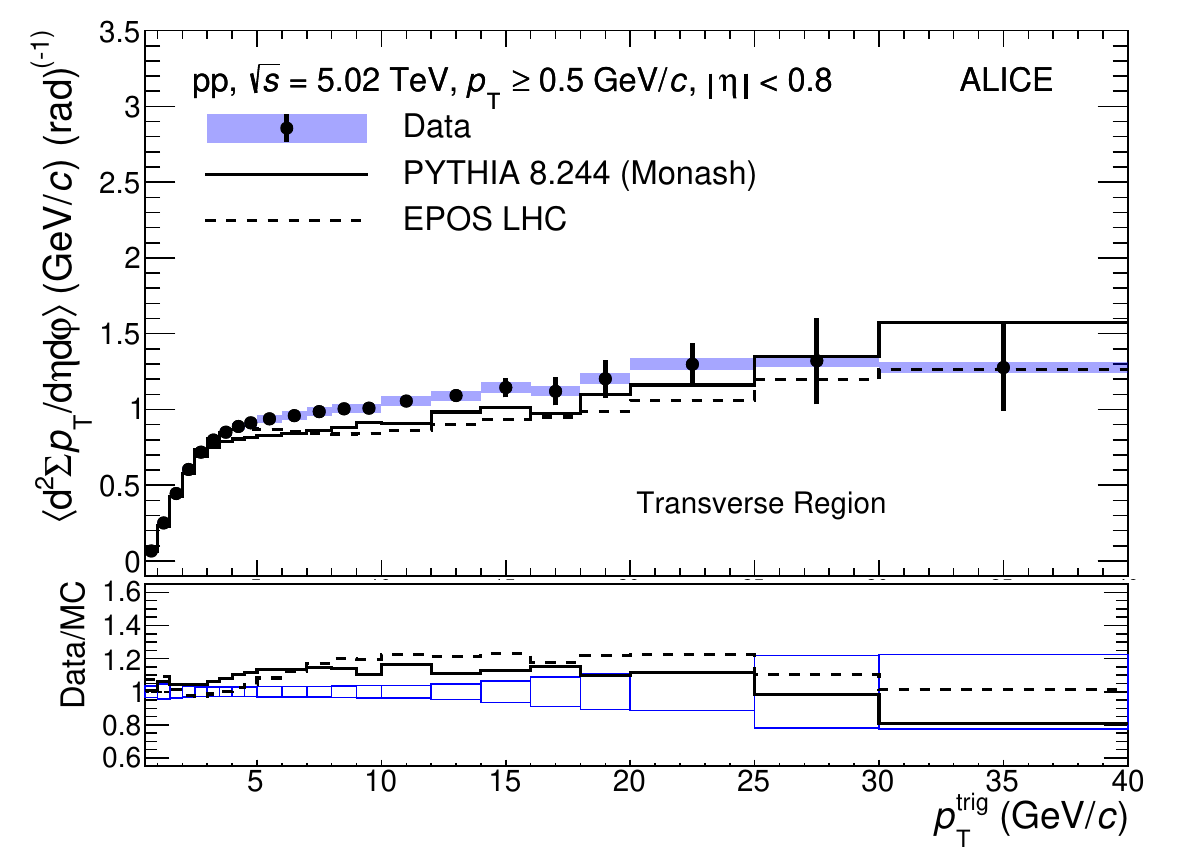}
  \includegraphics[width=7.90cm, height=6.80cm]{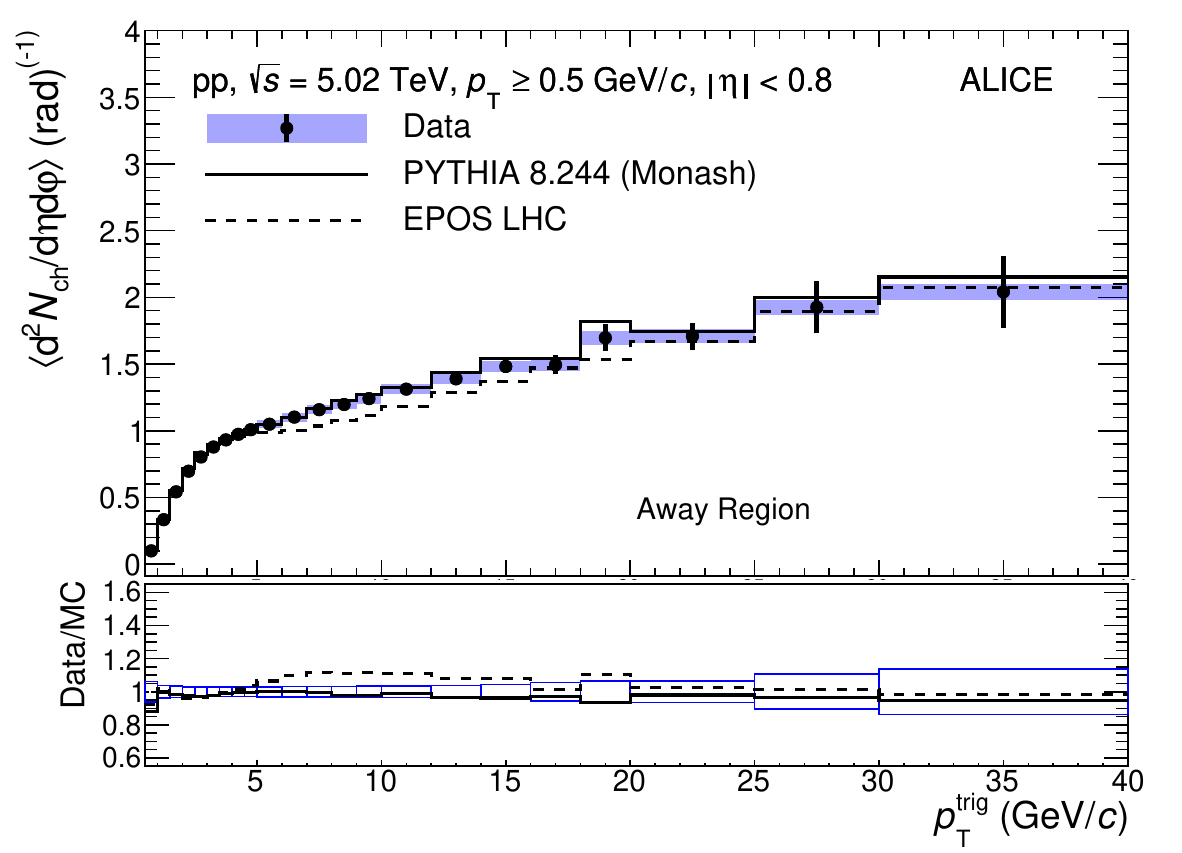}
  \includegraphics[width=7.90cm, height=6.80cm]{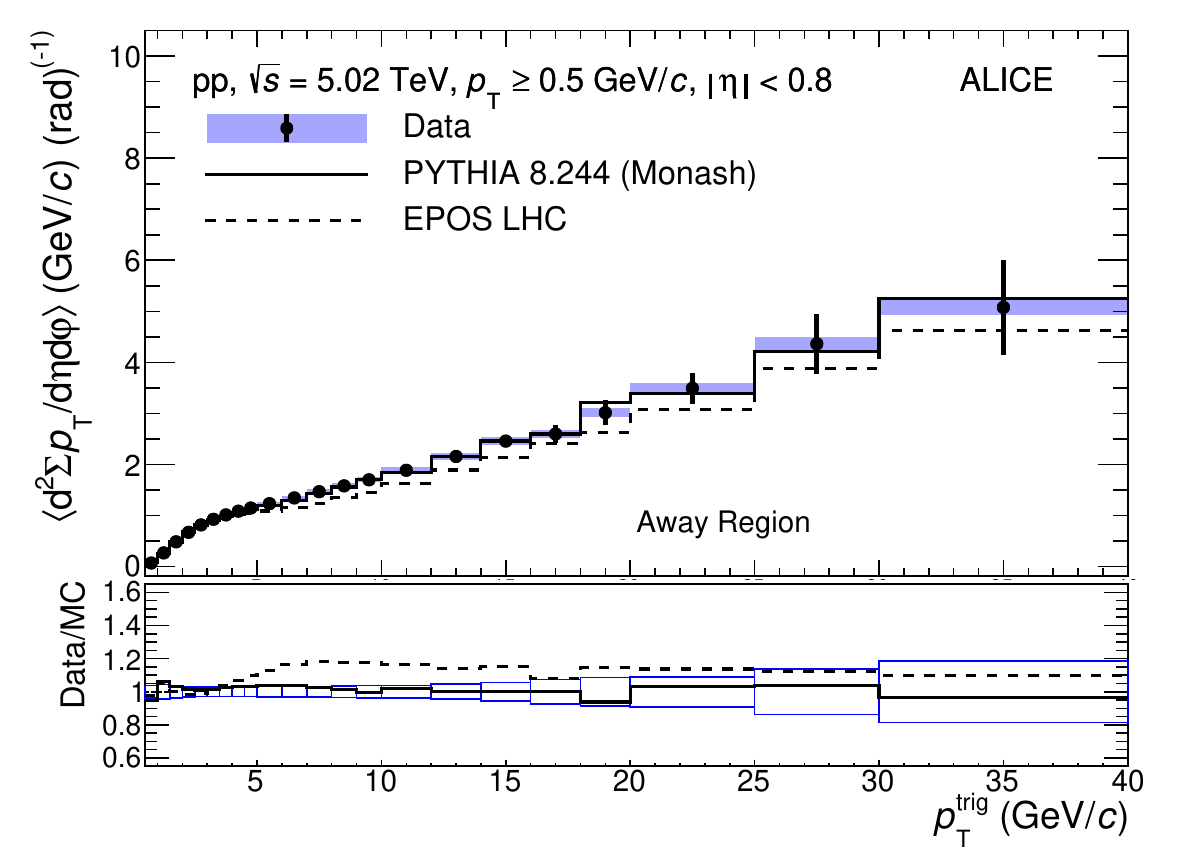}
  \includegraphics[width=7.90cm, height=6.80cm]{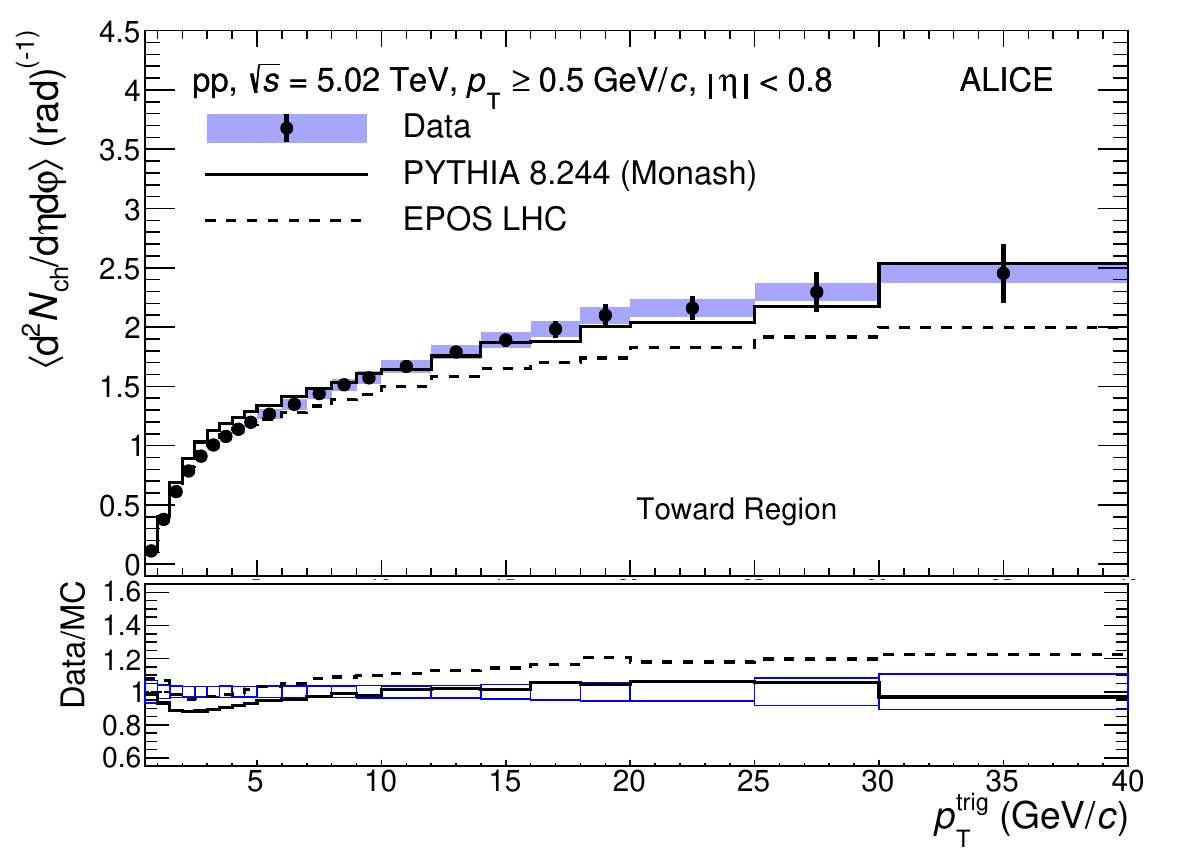}
  \includegraphics[width=7.90cm, height=6.80cm]{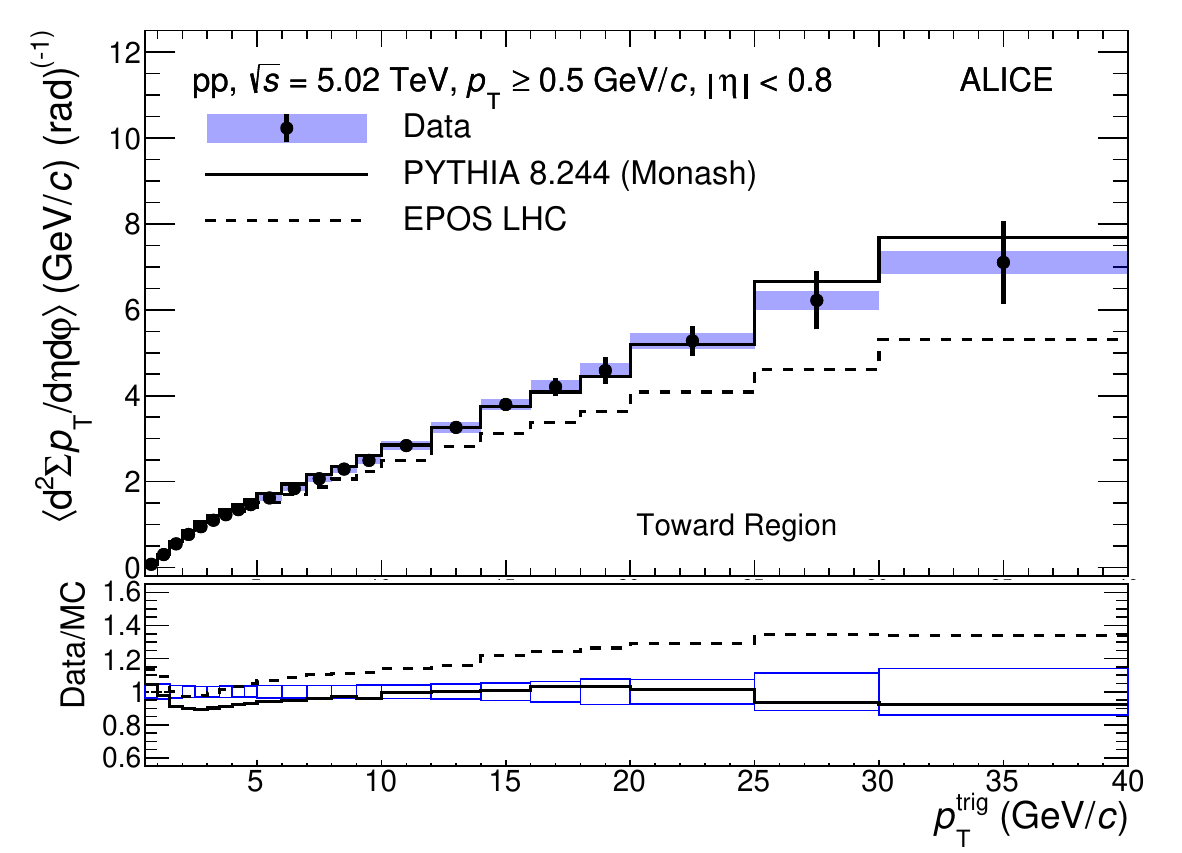}
  \end{center}
 \caption{The charged-particle number (left) and summed-$p_{\rm T}$ (right) densities as a function of \ptt in \pp collisions at $\sqrt{s}=5.02$\,\TeV are displayed. Results for the transverse (top), away (middle), and toward (bottom) regions were obtained for the transverse momentum threshold $\pt>0.5$\,\GeV/$c$. The shaded area and the error bars around the data points represent the systematic and statistical uncertainties, respectively. Data are compared with \py/Monash (solid line) and \ep (dashed line) predictions. The data-to-model ratios are displayed in the bottom panel of each plot. The boxes around unity represent the statistical and systematic uncertainties added in quadrature. }
\label{fig:f581016} 
 \end{figure}
\begin{figure}[!hp]
  \includegraphics[width=7.90cm, height=6.90cm]{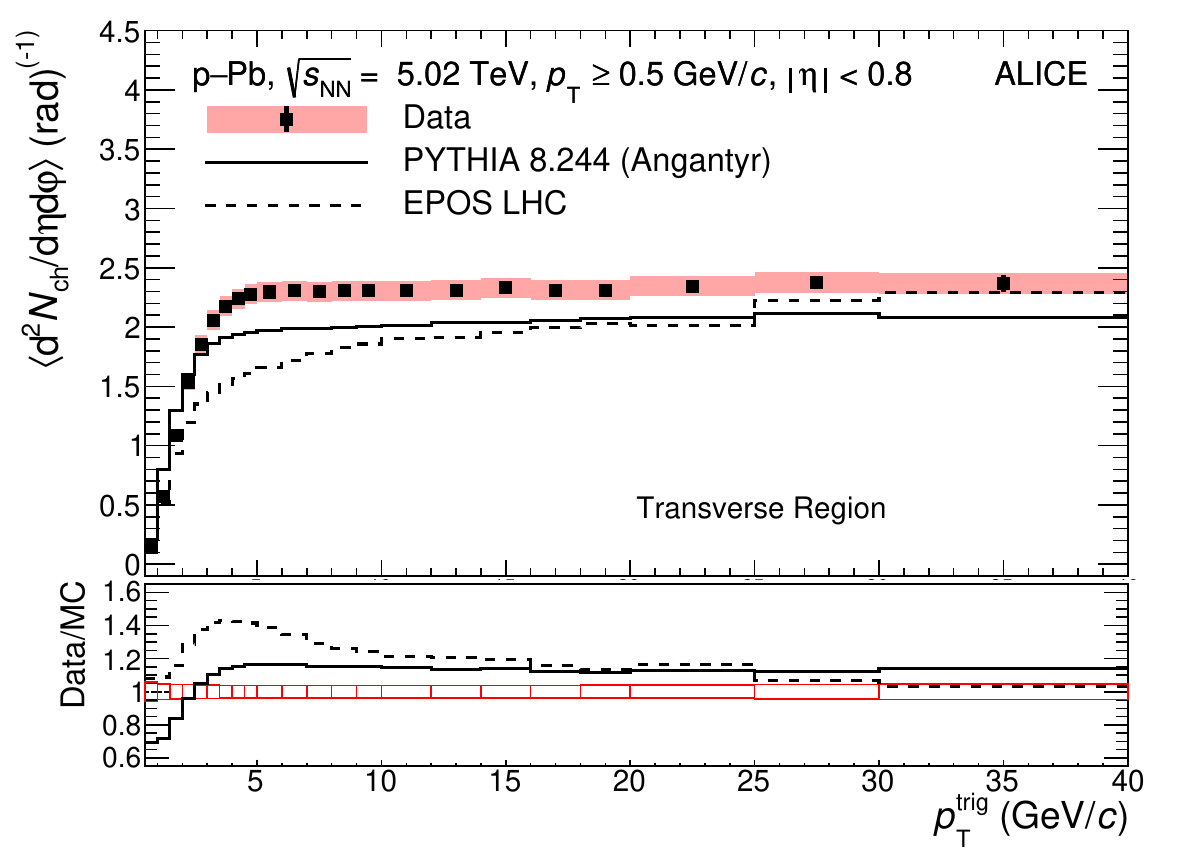}
  \includegraphics[width=7.90cm, height=6.90cm]{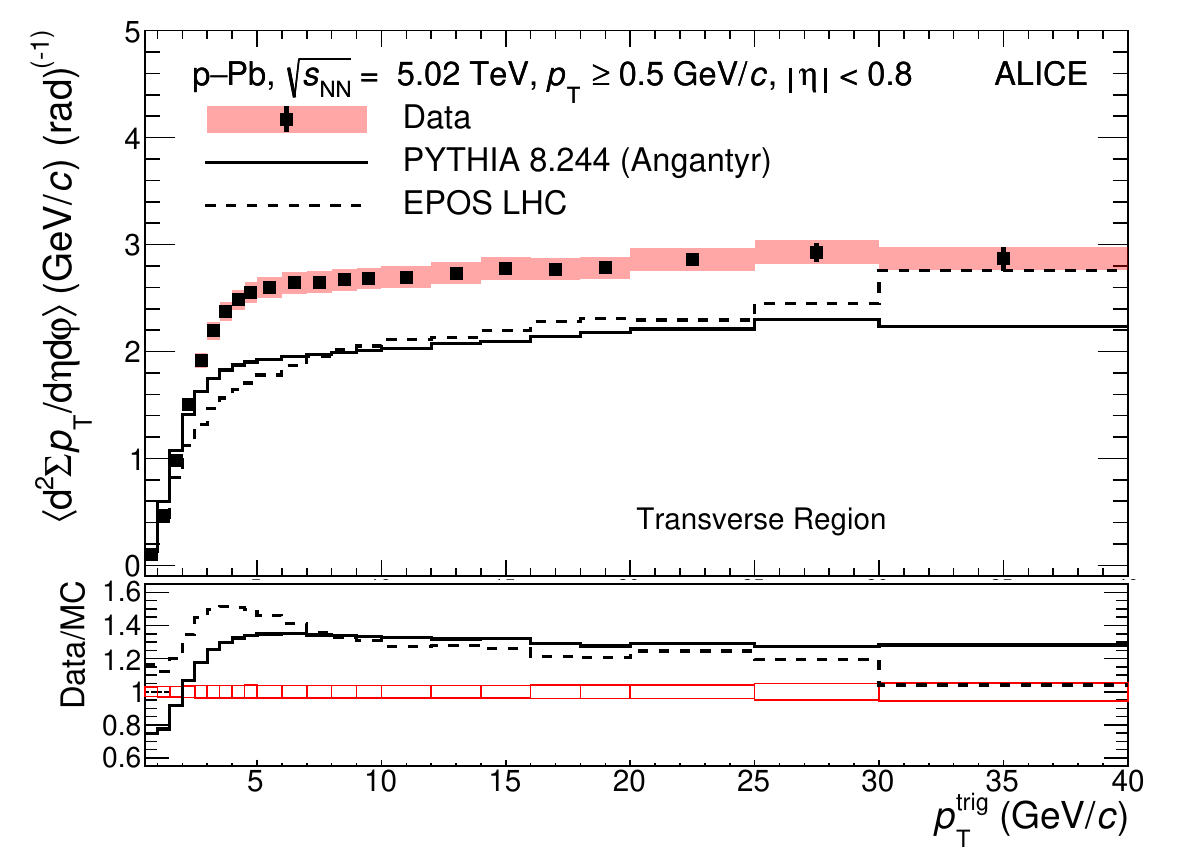}
  \includegraphics[width=7.90cm, height=6.90cm]{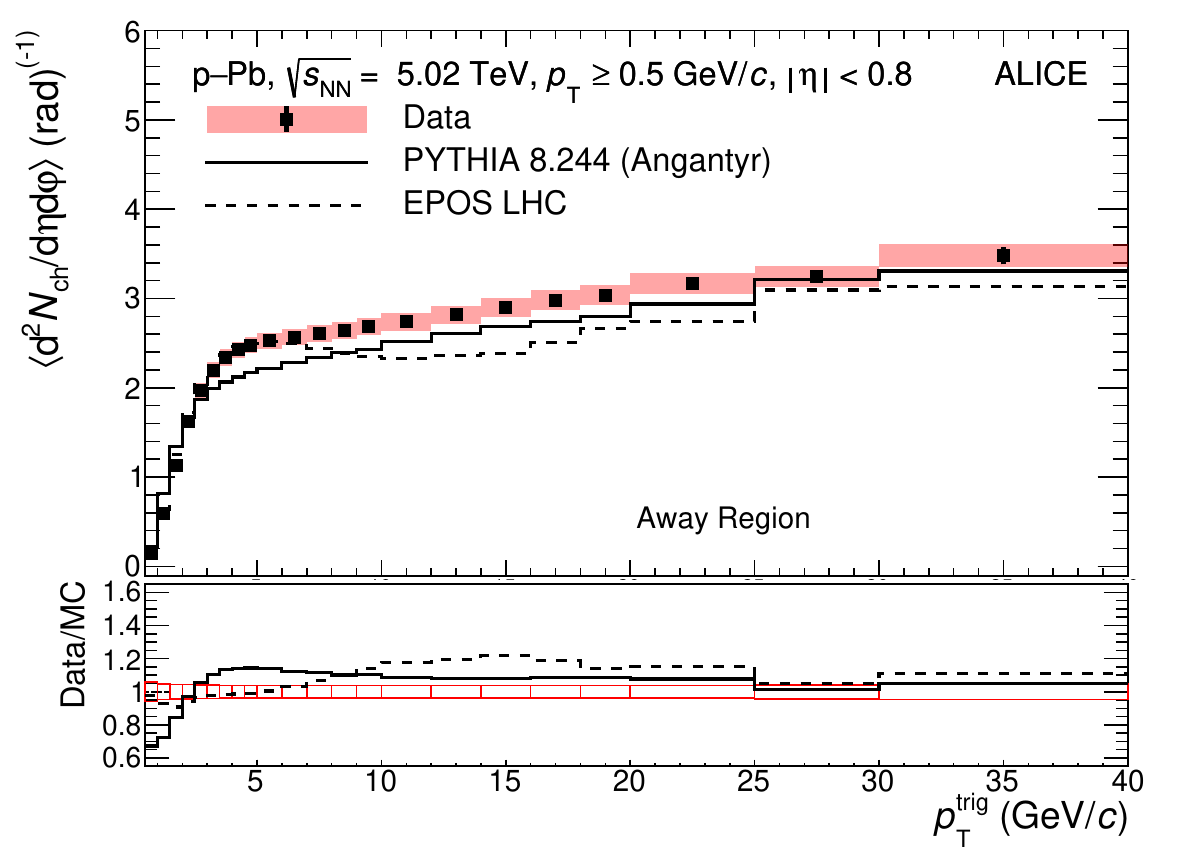}
  \includegraphics[width=7.90cm, height=6.90cm]{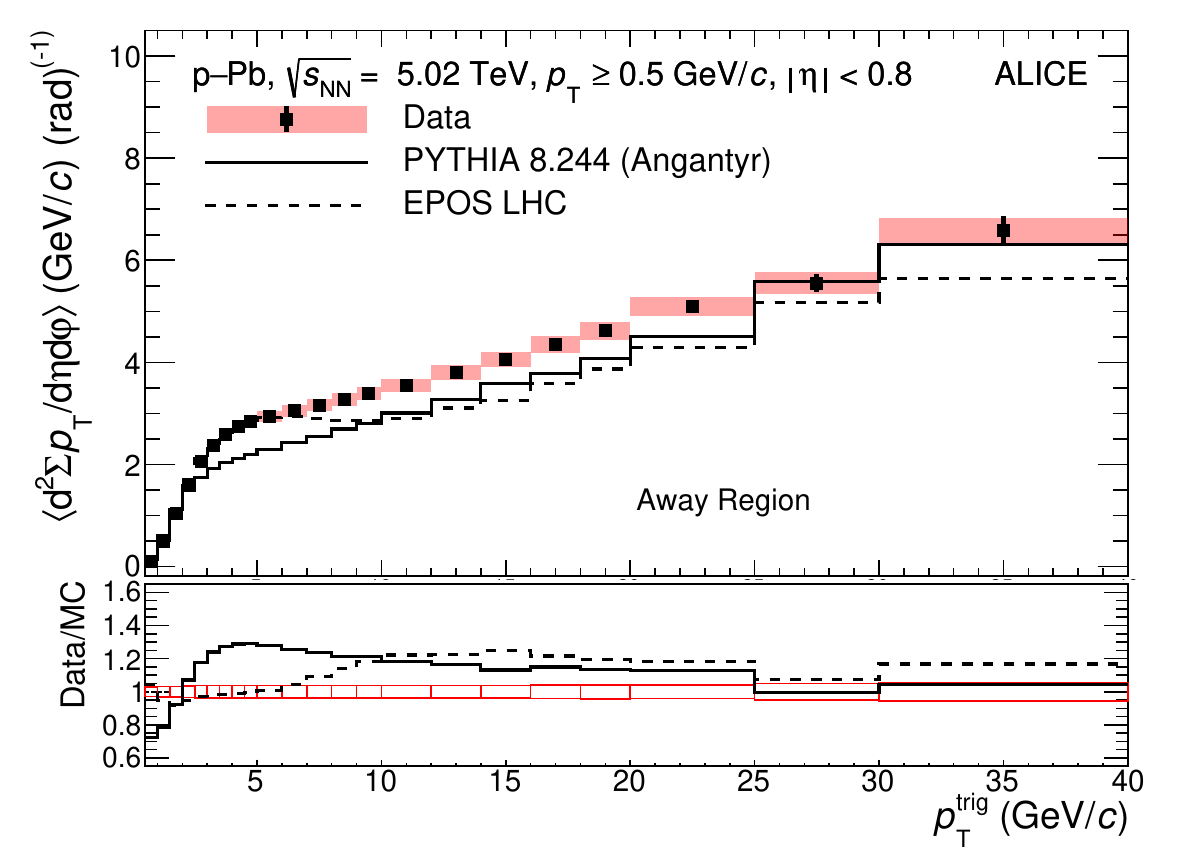}
  \includegraphics[width=7.90cm, height=6.90cm]{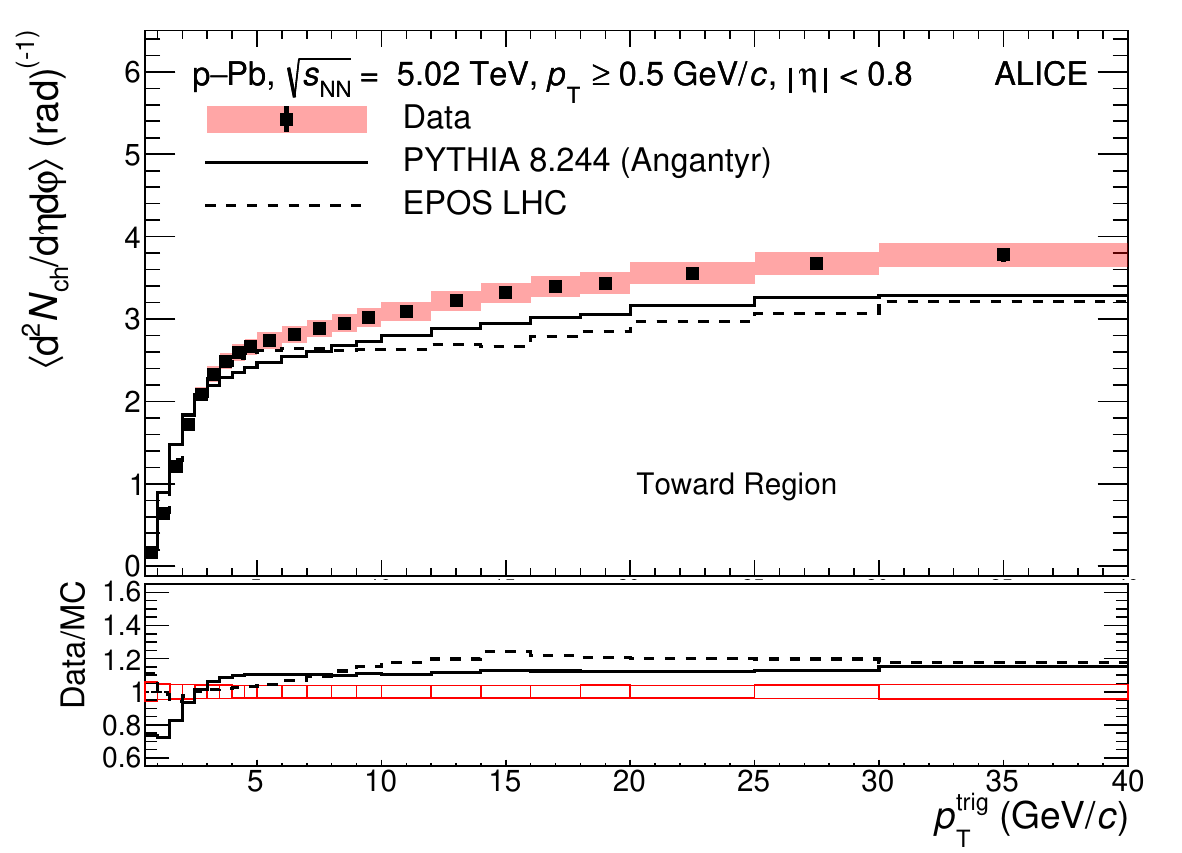}
  \includegraphics[width=7.90cm, height=6.90cm]{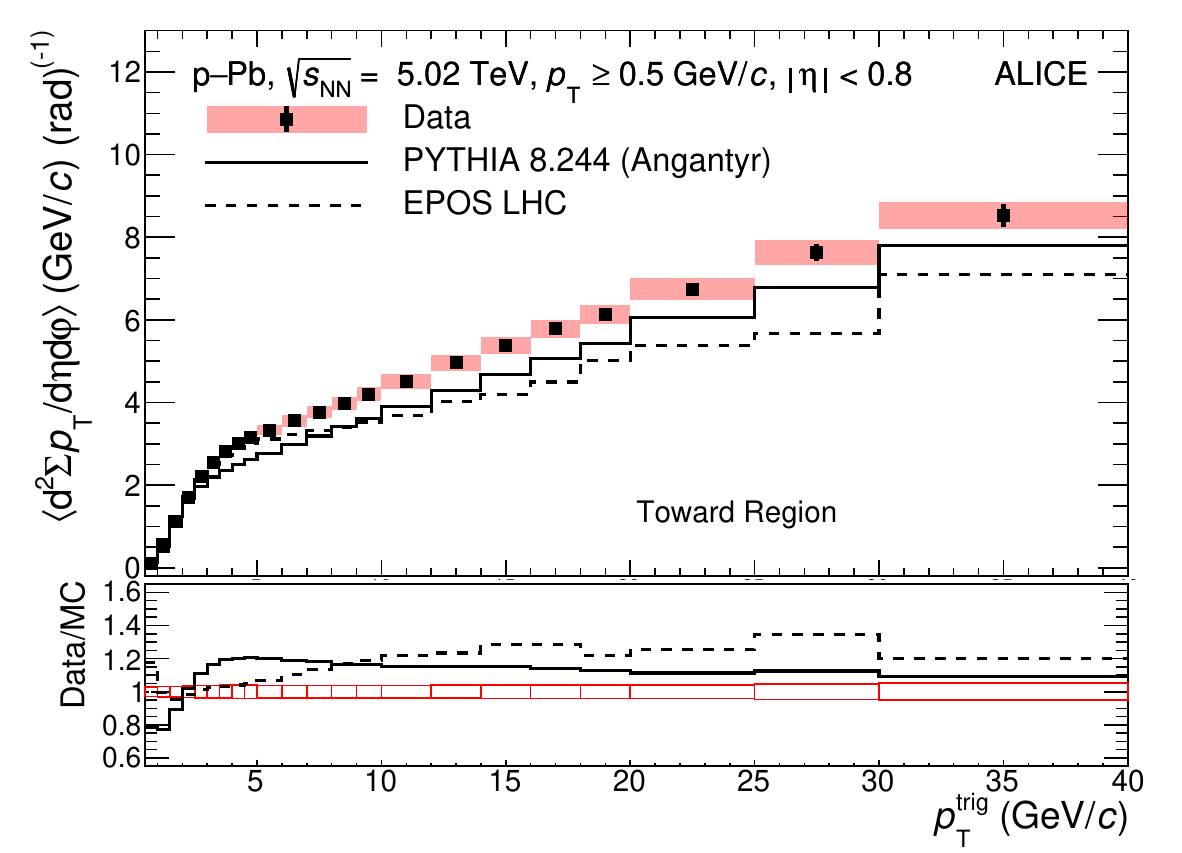}
 \caption{The charged-particle number (left) and summed-$p_{\rm T}$ (right) densities as a function of \ptt in \pPb collisions at $\sqrt{s_{\rm NN}}=5.02$\,\TeV are displayed. Results for the transverse (top), away (middle), and toward (bottom) regions were obtained for the transverse momentum threshold $\pt>0.5$\,\GeV/$c$. The shaded area and the error bars around the data points represent the systematic and statistical uncertainties, respectively. Data are compared with \py/Angantyr (solid line) and \ep (dashed line) predictions. The data-to-model ratios are displayed in the bottom panel of each plot. The boxes around unity represent the statistical and systematic uncertainties added in quadrature.}
\label{fig:f581017} 
 \end{figure}

Figure~\ref{fig:10011} shows the charged-particle number density as a function of \ptt measured in \pp and \pPb collisions at $\sqrt{s_{\rm NN}}=5.02$\,\TeV. Results are presented for the toward, transverse, and away regions. The $p_{\rm T}^{\rm trig}$ dependence in all regions is similar for both collision systems. For $\ptt\gtrsim5$\,GeV/$c$ the charged-particle number density becomes almost independent of \ptt (plateau) in the transverse region, as already pointed out above, while in the toward and away regions it continues to rise with increasing \ptt. The continuous rise observed for the toward and away regions can be attributed to the fact that produced particles in these regions do not originate only from the UE, but have also a contribution due to fragments from hard scatterings, which are mostly collimated in azimuth. The contribution from fragments increases with increasing \ptt causing the rise of event activity. A qualitatively similar behaviour in \pp and \pPb collisions is observed.  However, the event activity in the toward and away regions in \pp collisions increases faster with \ptt than in \pPb collisions, namely, the increase of the particle density from $\ptt=5$\,\GeVc up to $\ptt=40$\,\GeVc amounts to a factor of $\approx2$ and $\approx 1.4$ in \pp and \pPb collisions, respectively. Moreover, at $\ptt=35$\,GeV/$c$ the relative level of the event activity in the transverse region with respect to that in the toward (away) region is $\approx0.4$ and $\approx0.60$ ($\approx0.5$ and $\approx0.65$) for \pp and \pPb collisions, respectively.  This indicates that the UE contribution to the toward and away regions is larger in \pPb than in \pp collisions, which is expected because of multiple nucleon--nucleon collisions in a single \pPb collision that give a large additional UE, with respect to MPI in the same \pp collision.

\begin{figure}[!tb]
 \begin{center}
  \includegraphics[width=7.40cm, height=6.40cm]{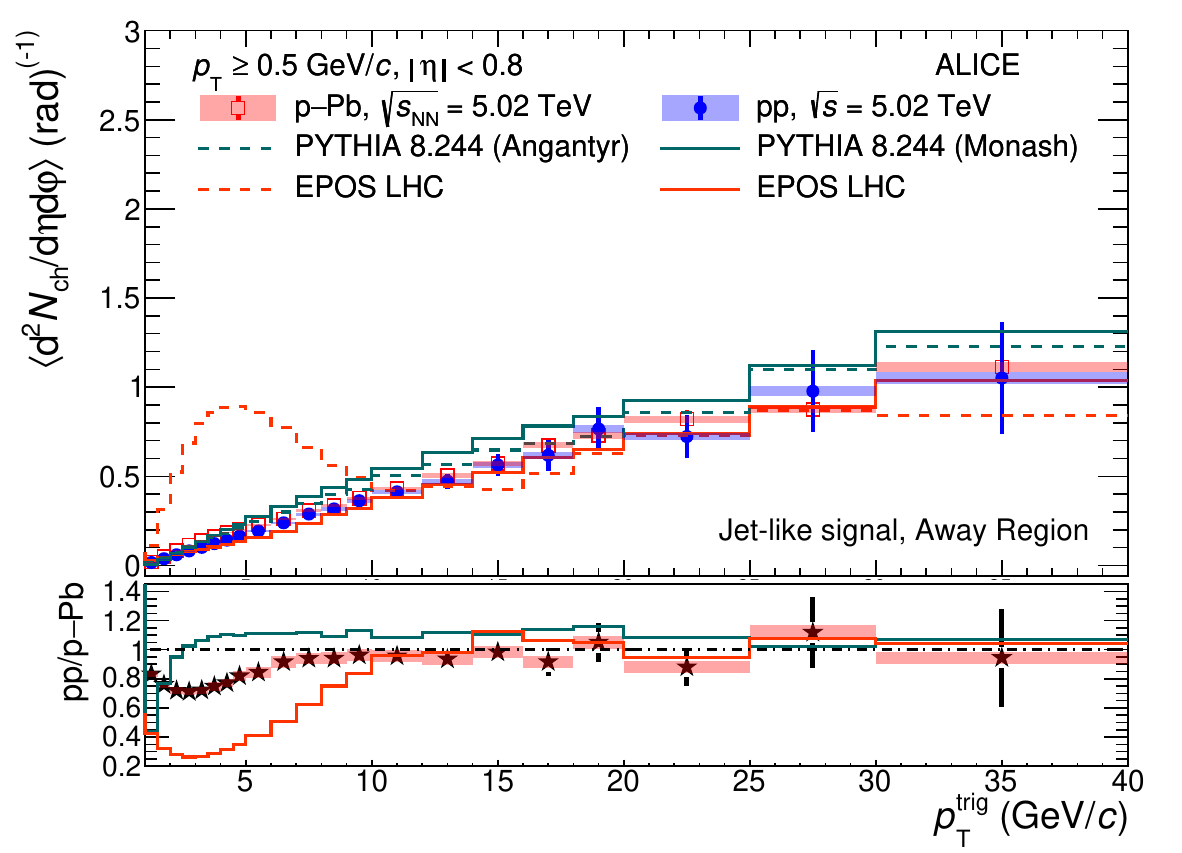}
  \includegraphics[width=7.40cm, height=6.40cm]{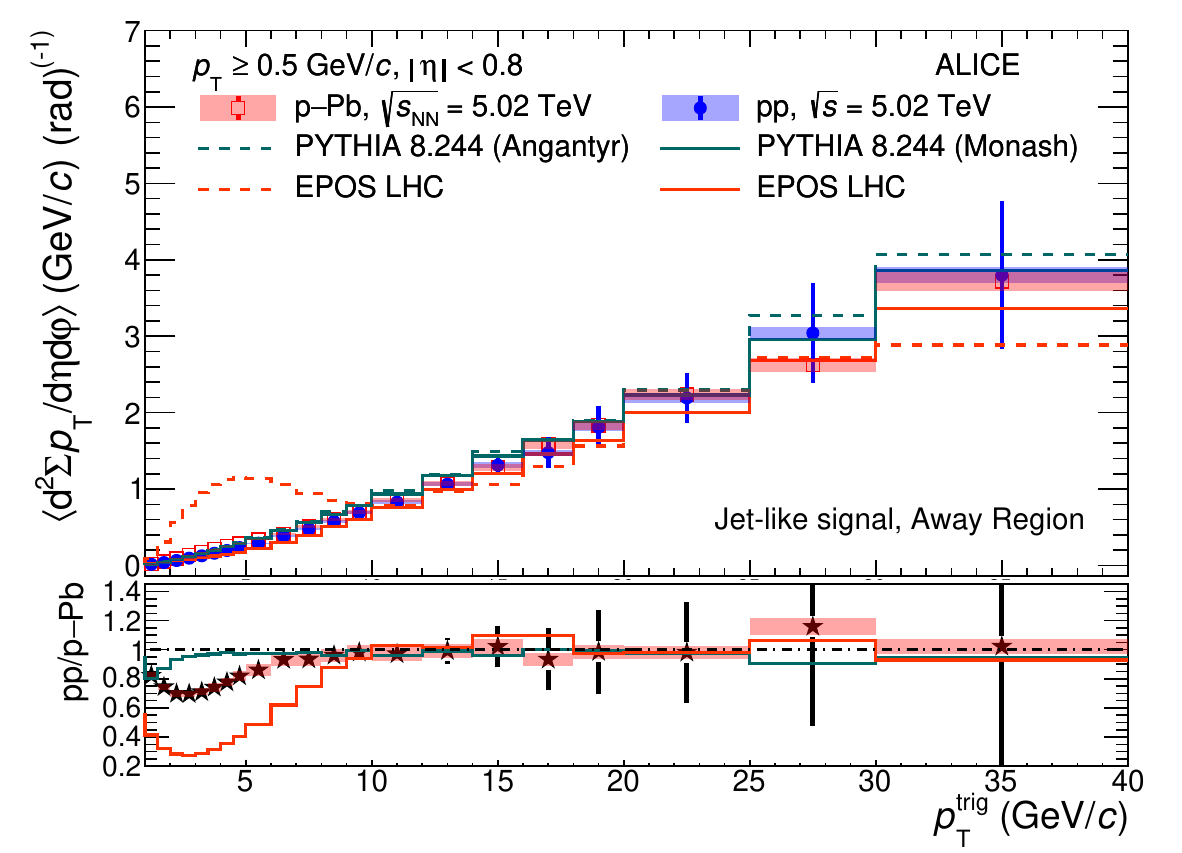}
  \includegraphics[width=7.40cm, height=6.40cm]{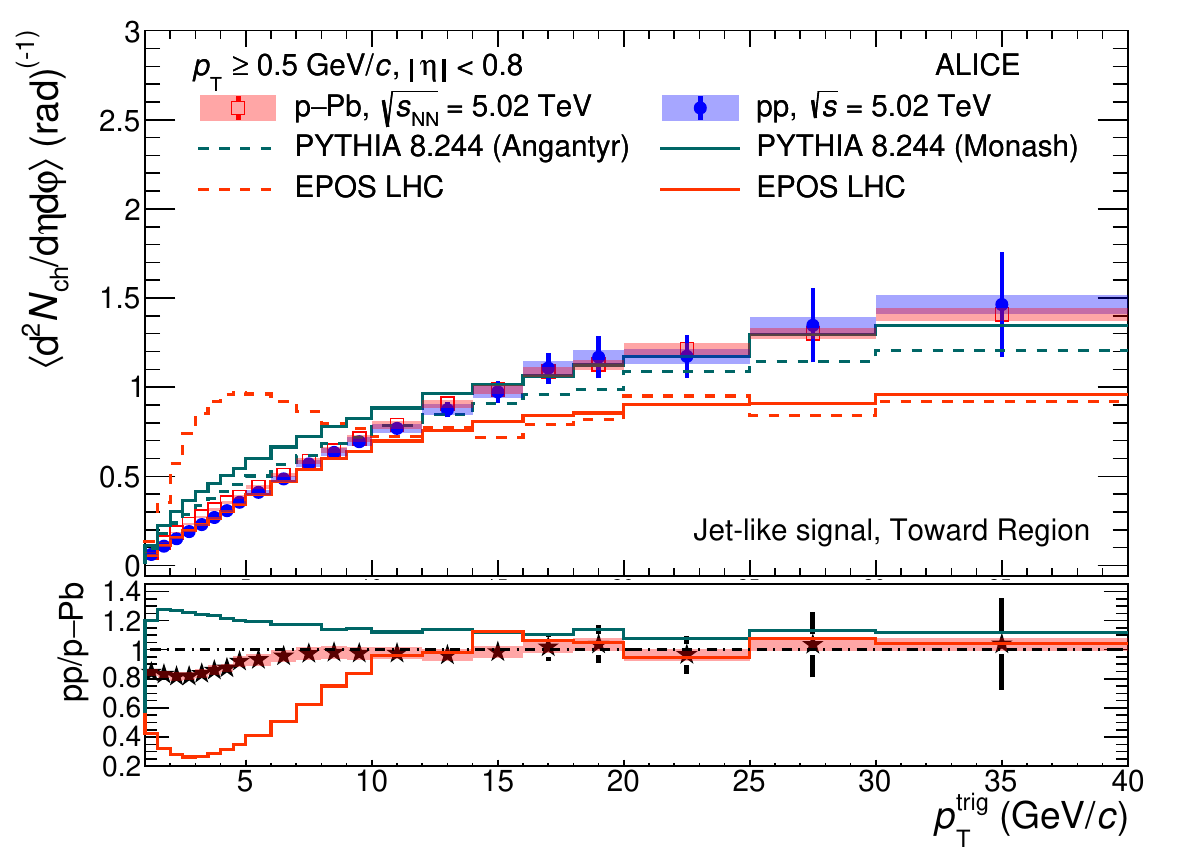}  
  \includegraphics[width=7.40cm, height=6.40cm]{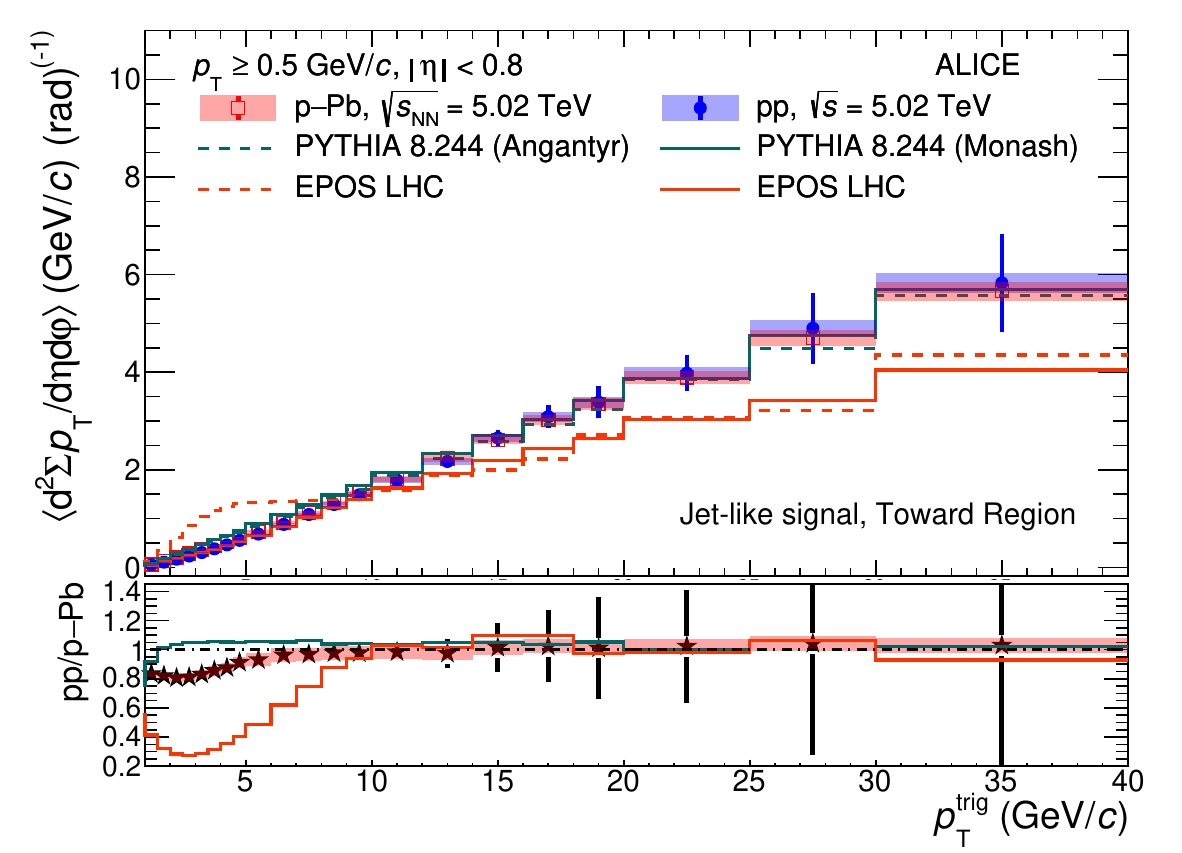}
  \caption{Upper panels: charged-particle number (left) and summed-$p_{\rm T}$ (right) densities as a function of \ptt in \pp (blue) and \pPb (red) collisions at $\sqrt{s_{\rm NN}}=5.02$\,\TeV. Results for data and comparison with models \py (green) and \ep (red) predictions for the away (upper) and toward (bottom) regions, after the subtraction of the charged-particle number (left) and summed-$p_{\rm T}$ (right) densities in the transverse region, are shown. Bottom panels: charged-particle number and summed-$p_{\rm T}$ densities measured in \pp collisions divided by those in \pPb collisions are displayed for both data and models.}
  \label{fig:f57212}
 \end{center}  
\end{figure}

Figure~\ref{fig:f581016} shows comparisons between the data from \pp collisions and the predictions of event generators for both the primary charged-particle number, and the summed-\pt densities in the three considered azimuthal regions (toward, away, and transverse). Although the modelling of the UE activity in \py/Monash is completely different with respect to that implemented in \ep, both models qualitatively describe the measured charged-particle densities in the three azimuthal regions. In the away region, within uncertainties, \py/Monash describes the data better than \ep in the full \ptt range of the measurement. The maximum deviation of \ep with respect to data is around 10\% and 20\% for the number density and summed-\pt density, respectively, in the \ptt interval 5--15\,\GeVc. Regarding the toward region, \py/Monash predictions overestimate the event activity by 10\% for $\ptt<5$\,\GeVc, whereas for higher \ptt \py/Monash describes the data quite well. The situation is the opposite for EPOS~LHC: at low \ptt \ep describes well the event activity, but it significantly underestimates the particle densities for higher \ptt ($>8$\,\GeVc) by $\approx30$\%.

Figure~\ref{fig:f581017} shows data-to-model comparisons for the case of \pPb collisions. For the transverse region, as already pointed out above, \py/Angantyr provides a better qualitative description of the measured trend of the charged-particle number densities as compared to \ep. However, both \py/Angantyr and \ep underestimate the charged-particle summed-\pt (number) density for $\ptt>5$\,\GeVc by more than 20\% (10\%). While for \py/Angantyr this discrepancy stays roughly constant up to $\pt \approx 3$\,\GeVc, for \ep the discrepancy increases up to 50\% at $\pt\approx3$\,\GeVc. For the toward and away regions, as visible from the ratio plots in the bottom panels of Fig.~\ref{fig:f581017}, \py/Angantyr does not describe the trend with \ptt of both the event-activity variables, in particular in the range $1<\ptt<5$\,GeV/$c$, where the event activity increases more steeply in the data than in the \py/Angantyr predictions. At higher \ptt ($5<\ptt<10$\,GeV/$c$) the ratio between the data and \py/Angantyr flattens, however a discrepancy by about 10\% (30\%) between the data and the model predictions is observed for the charged-particle number (summed-\pt) densities. The description of the data by \ep is slightly better that of \py/Angantyr for $\ptt<8$\,\GeVc. However, in that \ptt interval \ep predicts bump structures in the toward and away regions which are not seen in data. For higher \ptt \ep overestimates the event activity.  The inclusion of these data in future MC tunings would be relevant to improve the modelling of the UE in \pPb collisions.

\subsection{Event activity in the jet-like signals}

Figure~\ref{fig:f57212} shows the jet-like contribution to the charged-particle number and summed-\pt densities in the toward and away regions as a function of \ptt for \pp and \pPb collisions at $\sqrt{s_{\rm NN}}=5.02$\,\TeV. As discussed earlier, the event activity for the jet-like signals is obtained from the event activity in the toward and away regions after subtracting the event activity in the transverse region (see Eq.~\ref{eq3} and Eq.~\ref{eq4}). In contrast to the behaviour observed for the toward and away regions, where at $\ptt\approx5$\,\GeVc the event activity tends to flatten out, the densities in the jet-like signals rise  with increasing \ptt in the entire range of the measurement.  

At high \ptt ($\ptt>10$\,\GeVc), the event activity in the jet-like signals exhibits a remarkable similarity between measurements in \pp and \pPb collisions for both charged-particle multiplicity and summed-\pt densities. Within 10\%, both \py/Angantyr and \ep reproduce this feature. At low \ptt ($\ptt<10$\,\GeVc), the models overestimate the event activity in the jet-like signals measured in \pPb collisions. The disagreement is more remarkable for \ep than for \py/Angantyr. For \pp collisions, \py slightly overestimates the event activity, while \ep underestimates the particle densities. For $\ptt< 10$\,\GeVc, the event activity in \pp collisions scaled to that in \pPb collisions is smaller than unity, reaching a minimum of $\approx 0.8$ at $\ptt \approx 3$\,\GeVc. This behaviour is not reproduced by \py/Angantyr,  which gives a ratio above unity for $\ptt>1$\,\GeVc. In contrast, \ep exhibits a similar pattern, but the size of the effect is much larger than in data. The main difference between \py/Angantyr and \ep is that \ep incorporates collective flow, which is expected to be significant in the \ptt interval (3-4\,GeV/$c$~\cite{ALICE:2012vgf}) where we observe the differences between measurements in \pp and \pPb collisions. Given that (radial and elliptic) flow is larger in \pPb than in \pp collisions~\cite{ALICE:2013wgn,ALICE:2019zfl}, its contribution to the toward and away regions is expected to be higher in \pPb than in \pp collisions. In particular, the elliptic azimuthal correlations modulate the background according to: $B(\Delta\varphi)=B_{0}\Big(  1 + 2V_{2}\cos{(2\Delta\varphi)} \Big)$, where $V_{2}\approx v_{2}^{\rm trig}v_{2}^{\rm assoc}$ is approximately given by the product of anisotropic flow coefficients for trigger and associated particles at their respective momenta~\cite{ALICE:2012vgf}. From \PbPb results we expect the effect to be the largest at intermediate transverse momenta and to decrease for high transverse momentum particles~\cite{Aamodt:2011vg}.    

\begin{figure}[!tb]
 \begin{center}
   \includegraphics[width=7.40cm, height=6.40cm]{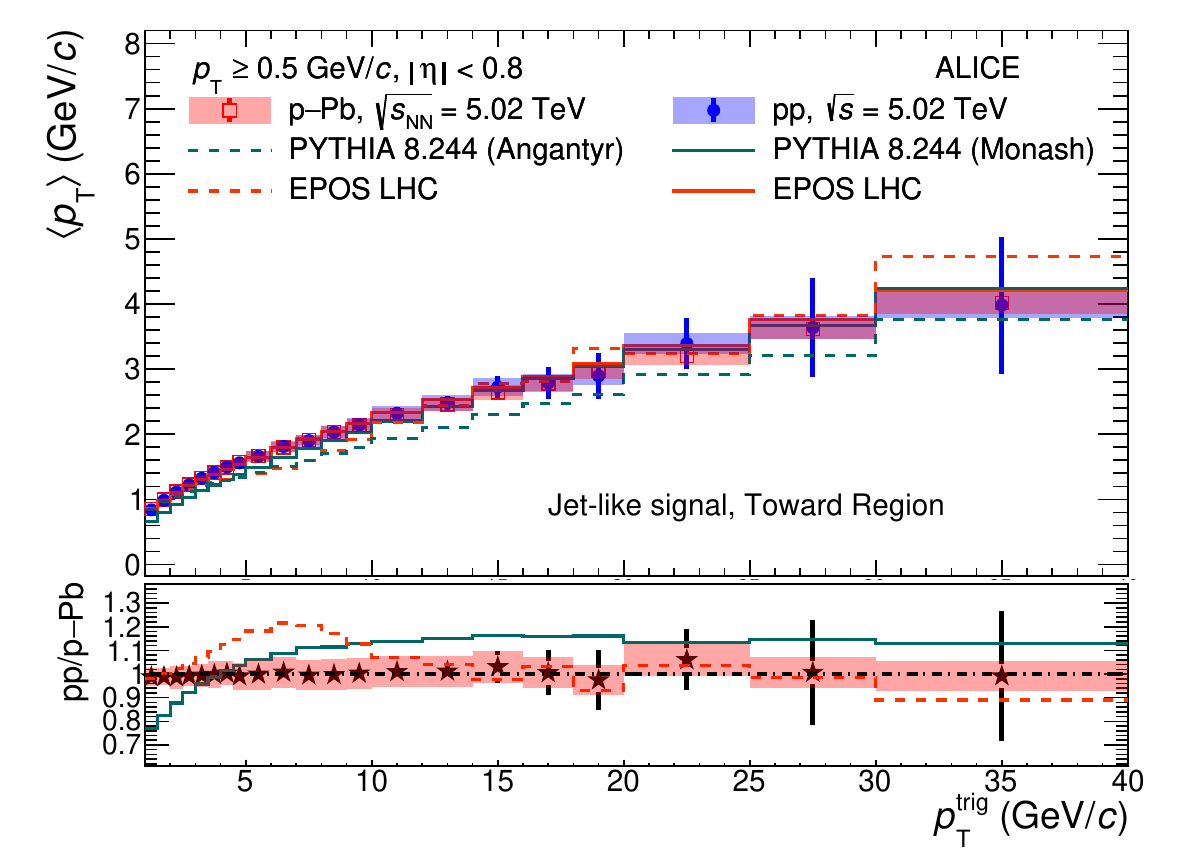}
  \includegraphics[width=7.40cm, height=6.40cm]{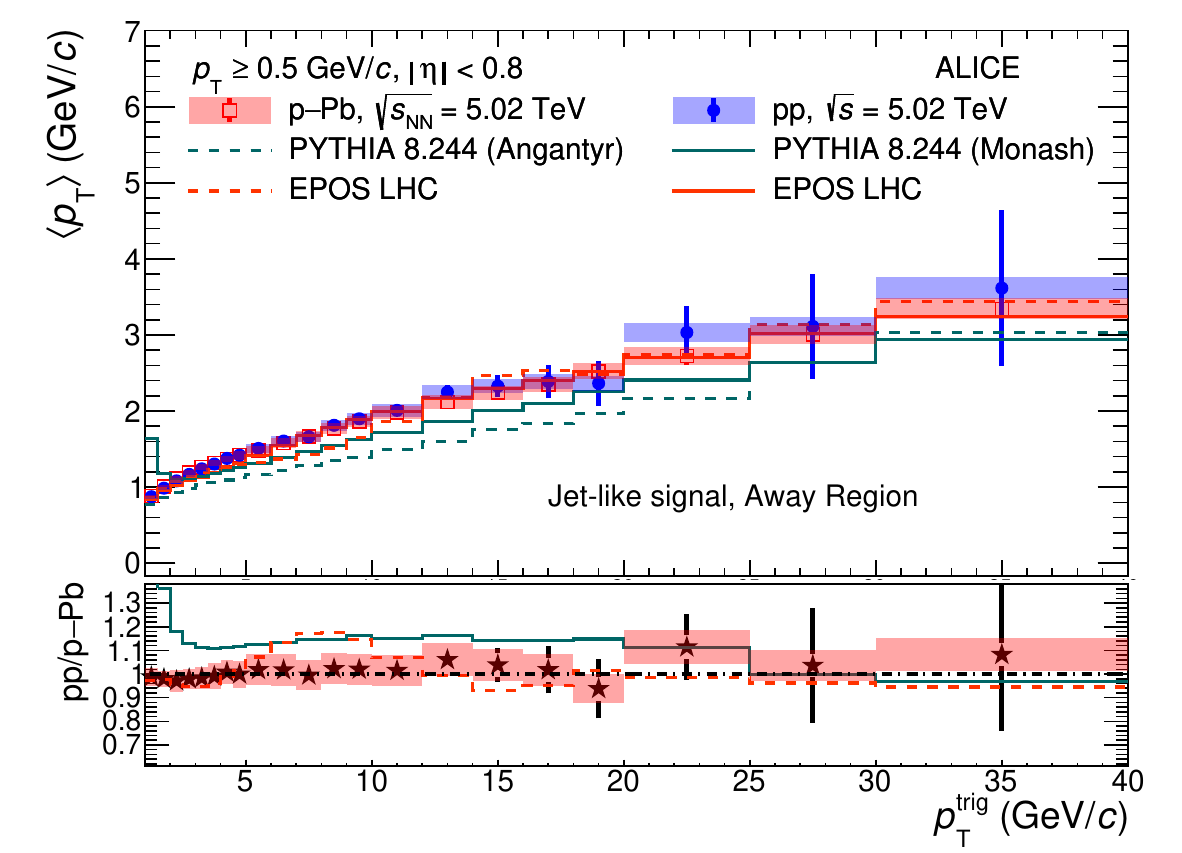}
  \caption{Upper: average transverse momentum as a function of \ptt in the toward (left) and away (right) regions measured in \pp and \pPb collisions at $\sqrt{s_{\rm NN}}=5.02$\,\TeV. Results for data and comparison with models \py (green) and \ep (red) predictions are shown. Bottom: average transverse momentum measured in \pp collisions divided by that measured in \pPb collisions. A similar ratio is shown for model predictions.}

\label{fig:f57216}  
 \end{center}  
\end{figure}

\begin{figure}[tb]
    \begin{center}
    \includegraphics[width=7.80cm, height=6.85cm]{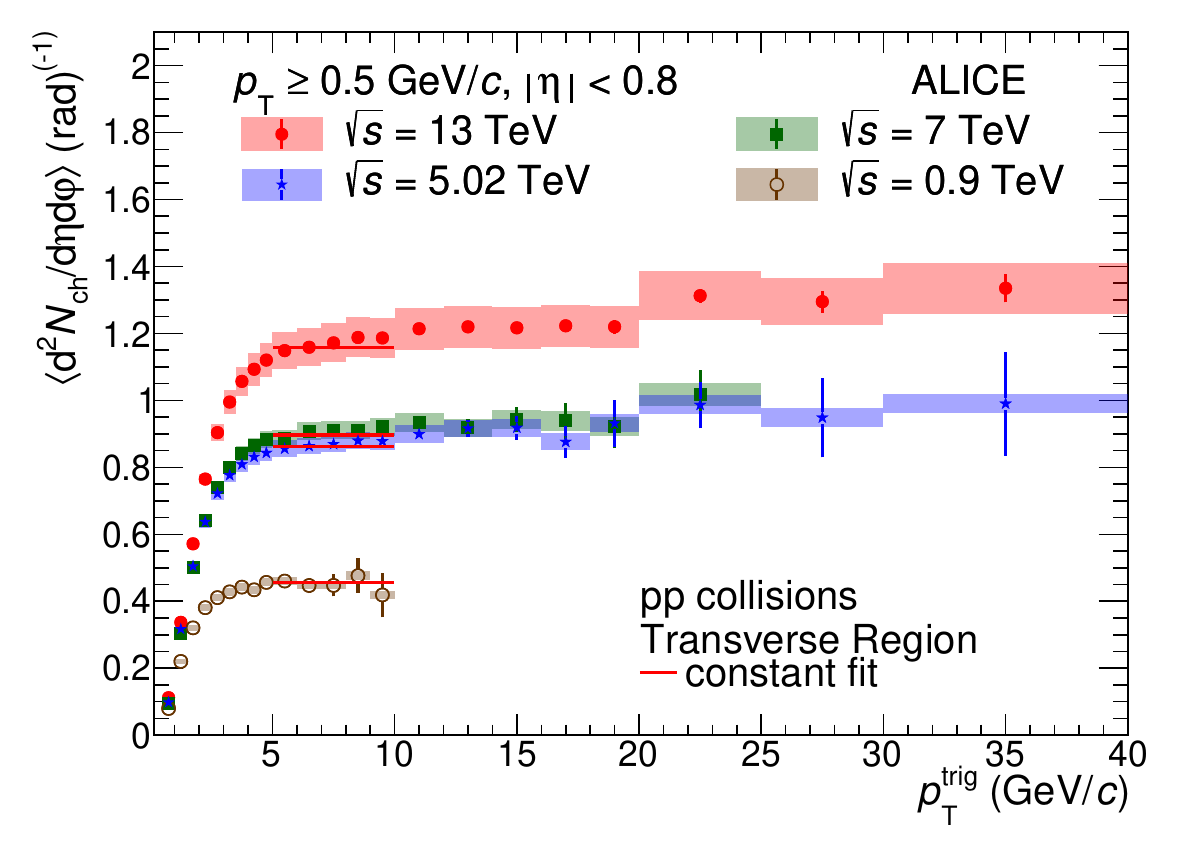}
    \includegraphics[width=7.80cm, height=6.85cm]{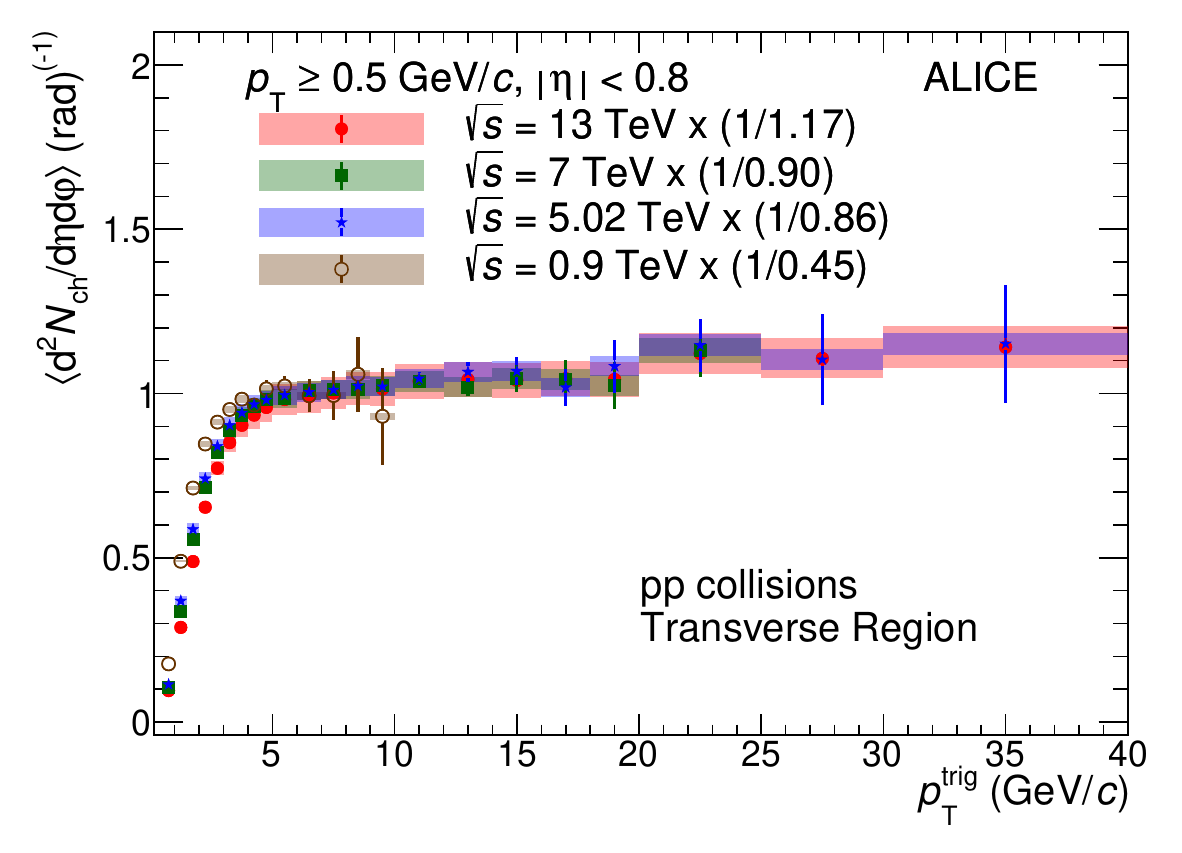}
    \end{center}
    \caption{Left: charged-particle number density in the transverse region as a function of \ptt for pp collisions at $\sqrt{s}= 0.9$, 5.02, 7, and 13\,\TeV~\cite{ALICE:2011ac,ALICE:2019mmy}. A constant function (solid lines) is used to fit the data in the range $5 < \ptt < 10$\,\GeV/$c$. Right: number densities scaled by the plateau values obtained from the fit to compare the shapes. The coloured boxes represent the systematic uncertainties, and vertical error bars indicate statistical uncertainties.}
    \label{fig:10013}
\end{figure}

Finally, the average transverse momentum \meanpt of particles in the toward and away regions after subtracting the UE contribution estimated from the transverse region is shown in Fig.~\ref{fig:f57216} as a function of \ptt for \pp and \pPb collisions at $\sqrt{s_{\rm NN}} = 5.02$\,\TeV. Within uncertainties, the \meanpt values are consistent in \pp and \pPb collisions in the measured \ptt interval. The \py tunes considered in this paper do not reproduce this behaviour. They predict that in the away region the average \pt for the jet-like signal in \pp collisions is about 20\% (10\%) larger than in \pPb collisions for $\ptt<2$\,\GeVc ($\ptt>5$\,\GeVc). For the toward region, the situation is similar at high \ptt; however, for $\ptt<2$\,\GeVc, the \meanpt in \pPb collisions is predicted to be about 20\% larger than in \pp collisions. Although the main source of this discrepancy is the underestimation of the measured \meanpt in \pPb collisions by \py/Angantyr, the prediction for \pp collisions is lower than the measured \meanpt in the away region. The agreement of \py/Monash with \pp data is better for the toward region. On the other hand, \ep reproduces the average \pt for the two collision systems better than \py/Angantyr. Although, for  $\ptt < 10$\,GeV/c, \ep predicts the \meanpt in the toward region to be larger in \pp as compared to \pPb collisions, leading to a bump-like structure in the ratio of pp over p--Pb results, which is not observed in the data. 

\subsection{Energy dependence of the underlying event in pp collisions}\label{sec:4.3}

This subsection discusses the collision-energy dependence of the charged-particle number density in the transverse region. Given that data from experiments at RHIC, the Tevatron and the LHC, are available for the \pt threshold, $\pt>0.5$\,GeV/$c$, our results for this \pt threshold are compared with existing measurements at other centre-of-mass energies. Figure~\ref{fig:10013} (left) compares the UE activity obtained in \pp collisions at $\sqrt{s} = 5.02$\,\TeV to those obtained at other LHC energies, namely $\sqrt{s}=0.9$, 7, and 13\,\TeV~\cite{ALICE:2011ac,ALICE:2019mmy}. Between the two higher energies, $\sqrt{s}=7$ and 13\,\TeV, the number density in the plateau increases by about 30\%. A similar increase was reported considering associated particles with $\pt>0.15$\,\GeVc~\cite{Acharya:2019nqn}. More information about the $\sqrt{s}$-dependence in the transverse region can be obtained by comparing the shapes of the number density as a function of \ptt. One attempt using the data provided by the ATLAS collaboration has been reported in Ref.~\cite{Ortiz:2017jaz}; a similar comparison was performed by the ALICE collaboration in Ref.~\cite{Acharya:2019nqn}. Following the approach presented in Ref.~\cite{Acharya:2019nqn}, the height of the plateau for different collision energies is quantified by fitting a constant function in the range $5 <\ptt <10$\,\GeVc (the fit functions are also shown in the left panel of Fig.~\ref{fig:10013}). The fitting range was
restricted to that common range in order to be consistent with the procedure used for the measurements at other centre-of-mass energies. Larger fitting ranges were also considered, and consistent results were obtained. The shapes of
the particle densities as a function of \ptt are then compared after dividing the densities by the level of the plateau, as estimated from the fit to a constant value. The results are shown in Fig.~\ref{fig:10013} (right). For the two higher energies, the \ptt coverage extends beyond the fitting range, i.e. to $\ptt >10$\,\GeVc. In this range, the densities agree within the statistical and systematic uncertainties. In the rise region ( $\ptt <5$\,\GeVc), one observes a clear ordering among
the four collision energies, the lowest energy having the highest density relative to the plateau. Moreover, at lower $\sqrt{s}$, the plateau values seem to be reached at a slightly lower \ptt. This feature is also observed in \pp collisions simulated with the \py event generator~\cite{Ortiz:2017jaz}.

\begin{figure}[t]
\centering
\includegraphics[width=0.50\textwidth]{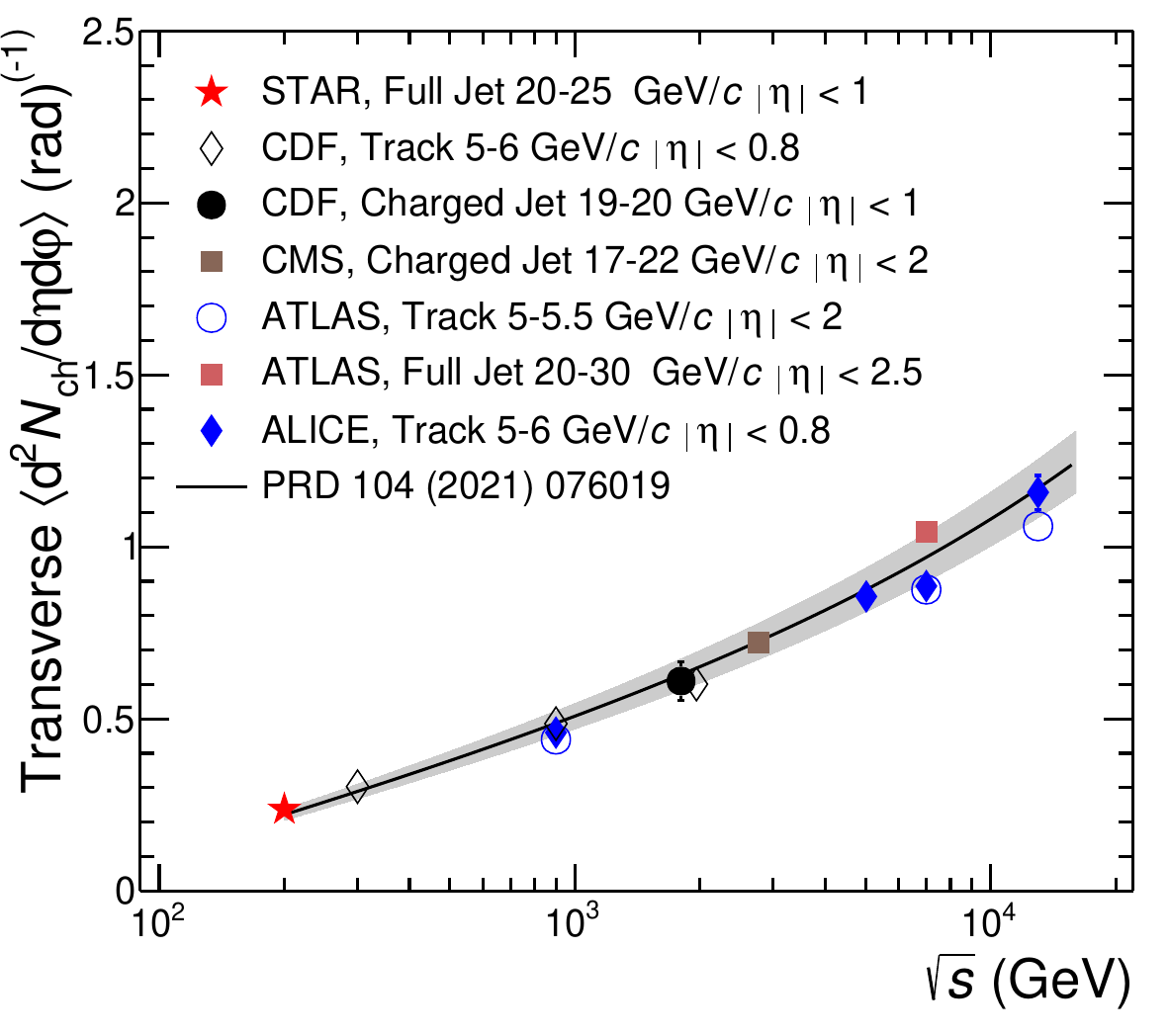}
\caption{Centre-of-mass energy dependence of the high \ptt plateau value of the charged-particle number density in the transverse region. The ATLAS~\cite{Aad:2010fh,Aad:2014hia,Aaboud:2017fwp}, CDF~\cite{Affolder:2001xt,Aaltonen:2015aoa}, CMS~\cite{Khachatryan:2015jza} and STAR~\cite{Adam:2019xpp} data points were taken from the compilation reported by the STAR collaboration~\cite{Adam:2019xpp}. Error bars represent statistical and systematic uncertainties summed in quadrature. The data are compared with a parametrisation of the form $s^{0.27}+0.14\log(s)$~\cite{Ortiz:2021gcr}. }
\label{fig:f51001f7}
\end{figure}

Figure~\ref{fig:f51001f7} shows the $\sqrt{s}$ dependence of the charged-particle number density measured in the transverse azimuthal interval and in the high \ptt interval of the plateau region. Results from various experiments at RHIC~\cite{Adam:2019xpp}, the Tevatron~\cite{Affolder:2001xt,Aaltonen:2015aoa}, and the LHC~\cite{Aad:2010fh,ALICE:2011ac,Aad:2014hia,Khachatryan:2015jza,Aaboud:2017fwp,Acharya:2019nqn} are displayed. The ATLAS, CDF,  CMS, and STAR data points are taken from the compilation reported by the STAR collaboration~\cite{Adam:2019xpp}. The event activity shows a modest increase from $\sqrt{s}=0.2$ up to 0.9\,\TeV, while for higher energies it exhibits a steeper rise. 
This behaviour is described by a function of the form $\propto s^{0.27}+0.14\log(s)$, in which the power-law term describes the UE contribution, whereas the logarithmic term describes the contribution from ISR and FSR. The parametrisation was taken from Ref.~\cite{Ortiz:2021gcr}. A comparison to the charged particle multiplicity at mid-pseudorapidity in minimum-bias pp data, where $\langle {\rm d}N_{\rm ch}/{\rm d}\eta \rangle$ can be parameterised as $\langle {\rm d}N_{\rm ch}/{\rm d}\eta \rangle \propto s^{0.114}$~\cite{ALICE:2015olq}, suggests that the UE contribution increases faster with centre-of-mass energy than the charged particle multiplicity in minimum-bias pp collisions.   

\section{Conclusions}\label{sec:7}

In this paper, the measurements of underlying-event observables performed in \pp and \pPb collisions at a centre-of-mass energy per nucleon--nucleon collisions of 5.02\,TeV were reported. The analysis was carried out following the strategy introduced by the CDF collaboration consisting of the definition of three azimuthal regions relative to the highest transverse momentum particle in the collision (\ptt). The charged-particle production is measured within the pseudorapidity interval $|\eta|<0.8$; and it is quantified with the number and summed-\pt densities considering particles above a given \pt threshold. Three \pt thresholds are considered: 0.15, 0.5, and 1\,\GeVc. These quantities are reported as a function of \ptt, and for the toward, away, and transverse azimuthal regions. The transverse region is the most sensitive to the underlying event; while the toward and away regions include both the underlying-event and jet fragments from the main partonic scattering. For the isolation of the jet-like signal, the event activity in the transverse region is subtracted from those measured in the toward and away regions. Results for \pp collisions are compared with data at other centre-of-mass energies and with MC predictions. In addition, the event activities measured in \pp and \pPb collisions are compared with each other at the same \ptt value. The main conclusions of the present work are listed below. 
 
 \begin{itemize}
     \item The underlying-event observables in \pp collisions follow the same behaviour as observed at lower centre-of-mass energies. In the transverse region the charged-particle densities measured in the three azimuthal regions exhibit a fast rise for $\ptt<5$\,\GeVc followed by a flattening at higher \ptt (plateau). Data for the three azimuthal regions relative to the leading particle are reproduced by the \py event generator with the Monash tune. \ep predicts a slightly different behaviour in particular at \ptt around 3\,\GeVc where a bump structure is present in the three azimuthal regions which is not observed in the data.
     \item The underlying-event observables in \pPb collisions qualitatively behave like in \pp interactions. The particle densities in the transverse region exhibit a saturation at $\ptt\approx5$\,\GeVc. \py/Angantyr qualitatively reproduces this saturation but underestimates the particle densities. The EPOS LHC model does not describe the saturation and underestimates the event activity within the measured \ptt interval. For the toward and away regions, above the onset of the plateau, data exhibit a slower increase of the particle densities with increasing \ptt than that observed in \pp collisions. \ep and \py/Angantyr underestimate the particle densities at high \ptt ($>8$\,\GeVc). At lower \ptt, \ep predicts a bump structure at $\ptt\approx4$\,\GeVc which is not seen in data. For \ptt $< 3$\,\GeVc, \ep describes the particle densities, whereas \py/Angantyr overestimates those in data by up to 30\%.
     \item The particle densities in the toward and away regions after subtraction of the UE contribution as a function of \ptt in \pp collisions are consistent with those measured in \pPb collisions for $\ptt>10$\,\GeVc, i.e., no modification of the jet-like yield in \pPb collisions relative to \pp collisions is found. At lower \ptt, the charged-particle densities are larger in \pPb collisions relative to \pp collisions. This behaviour is expected given the larger collective flow effects in \pPb collisions relative to \pp collisions. This feature is qualitatively captured by the \ep generator, which incorporates collective flow effects in the modelling of the system created in the collision. However, the size of the effect is significantly larger in \ep than in data. An opposite trend is instead predicted by simulations with \py/Angantyr, which do not include collective effects. The average \pt as a function of \ptt was also measured. The average \pt values measured in \pp and \pPb collision are found to be consistent between each other in the entire \ptt interval of the measurement. Simulations with \ep predict instead a slightly lower average \pt for jet-like signal in \pPb collisions as compared to \pp collisions for \ptt $< 10$\,\GeVc, while \py with the Monash and Angantyr tunes do not provide a good description for this observable.
 \end{itemize}

The measurements reported in this article represent important input for the tuning of some of the parameters of the event generators in order to improve the modelling of soft particle production in \pp and \pPb collisions. Moreover, they can contribute to the understanding of the origin of signals resembling a collective behaviour in \pp and \pPb collisions.  



\newenvironment{acknowledgement}{\relax}{\relax}
\begin{acknowledgement}
\section*{Acknowledgements}

The ALICE Collaboration would like to thank all its engineers and technicians for their invaluable contributions to the construction of the experiment and the CERN accelerator teams for the outstanding performance of the LHC complex.
The ALICE Collaboration gratefully acknowledges the resources and support provided by all Grid centres and the Worldwide LHC Computing Grid (WLCG) collaboration.
The ALICE Collaboration acknowledges the following funding agencies for their support in building and running the ALICE detector:
A. I. Alikhanyan National Science Laboratory (Yerevan Physics Institute) Foundation (ANSL), State Committee of Science and World Federation of Scientists (WFS), Armenia;
Austrian Academy of Sciences, Austrian Science Fund (FWF): [M 2467-N36] and Nationalstiftung f\"{u}r Forschung, Technologie und Entwicklung, Austria;
Ministry of Communications and High Technologies, National Nuclear Research Center, Azerbaijan;
Conselho Nacional de Desenvolvimento Cient\'{\i}fico e Tecnol\'{o}gico (CNPq), Financiadora de Estudos e Projetos (Finep), Funda\c{c}\~{a}o de Amparo \`{a} Pesquisa do Estado de S\~{a}o Paulo (FAPESP) and Universidade Federal do Rio Grande do Sul (UFRGS), Brazil;
Ministry of Education of China (MOEC) , Ministry of Science \& Technology of China (MSTC) and National Natural Science Foundation of China (NSFC), China;
Ministry of Science and Education and Croatian Science Foundation, Croatia;
Centro de Aplicaciones Tecnol\'{o}gicas y Desarrollo Nuclear (CEADEN), Cubaenerg\'{\i}a, Cuba;
Ministry of Education, Youth and Sports of the Czech Republic, Czech Republic;
The Danish Council for Independent Research | Natural Sciences, the VILLUM FONDEN and Danish National Research Foundation (DNRF), Denmark;
Helsinki Institute of Physics (HIP), Finland;
Commissariat \`{a} l'Energie Atomique (CEA) and Institut National de Physique Nucl\'{e}aire et de Physique des Particules (IN2P3) and Centre National de la Recherche Scientifique (CNRS), France;
Bundesministerium f\"{u}r Bildung und Forschung (BMBF) and GSI Helmholtzzentrum f\"{u}r Schwerionenforschung GmbH, Germany;
General Secretariat for Research and Technology, Ministry of Education, Research and Religions, Greece;
National Research, Development and Innovation Office, Hungary;
Department of Atomic Energy Government of India (DAE), Department of Science and Technology, Government of India (DST), University Grants Commission, Government of India (UGC) and Council of Scientific and Industrial Research (CSIR), India;
National Research and Innovation Agency - BRIN, Indonesia;
Istituto Nazionale di Fisica Nucleare (INFN), Italy;
Japanese Ministry of Education, Culture, Sports, Science and Technology (MEXT) and Japan Society for the Promotion of Science (JSPS) KAKENHI, Japan;
Consejo Nacional de Ciencia (CONACYT) y Tecnolog\'{i}a, through Fondo de Cooperaci\'{o}n Internacional en Ciencia y Tecnolog\'{i}a (FONCICYT) and Direcci\'{o}n General de Asuntos del Personal Academico (DGAPA), Mexico;
Nederlandse Organisatie voor Wetenschappelijk Onderzoek (NWO), Netherlands;
The Research Council of Norway, Norway;
Commission on Science and Technology for Sustainable Development in the South (COMSATS), Pakistan;
Pontificia Universidad Cat\'{o}lica del Per\'{u}, Peru;
Ministry of Education and Science, National Science Centre and WUT ID-UB, Poland;
Korea Institute of Science and Technology Information and National Research Foundation of Korea (NRF), Republic of Korea;
Ministry of Education and Scientific Research, Institute of Atomic Physics, Ministry of Research and Innovation and Institute of Atomic Physics and University Politehnica of Bucharest, Romania;
Ministry of Education, Science, Research and Sport of the Slovak Republic, Slovakia;
National Research Foundation of South Africa, South Africa;
Swedish Research Council (VR) and Knut \& Alice Wallenberg Foundation (KAW), Sweden;
European Organization for Nuclear Research, Switzerland;
Suranaree University of Technology (SUT), National Science and Technology Development Agency (NSTDA), Thailand Science Research and Innovation (TSRI) and National Science, Research and Innovation Fund (NSRF), Thailand;
Turkish Energy, Nuclear and Mineral Research Agency (TENMAK), Turkey;
National Academy of  Sciences of Ukraine, Ukraine;
Science and Technology Facilities Council (STFC), United Kingdom;
National Science Foundation of the United States of America (NSF) and United States Department of Energy, Office of Nuclear Physics (DOE NP), United States of America.
In addition, individual groups or members have received support from:
Marie Sk\l{}odowska Curie, Strong 2020 - Horizon 2020 (grant nos. 824093, 896850), European Union;
Academy of Finland (Center of Excellence in Quark Matter) (grant nos. 346327, 346328), Finland;
Programa de Apoyos para la Superaci\'{o}n del Personal Acad\'{e}mico, UNAM, Mexico.

\end{acknowledgement}

\bibliographystyle{utphys}   
\bibliography{bibliography}

\newpage
\appendix
\section{Appendix}\label{app:101}
\subsection{Charged-particle densities as a function of \ptt for others \pt thresholds}

\begin{figure}[!h]
\centering
  \includegraphics[width=7.90cm, height=6.90cm]{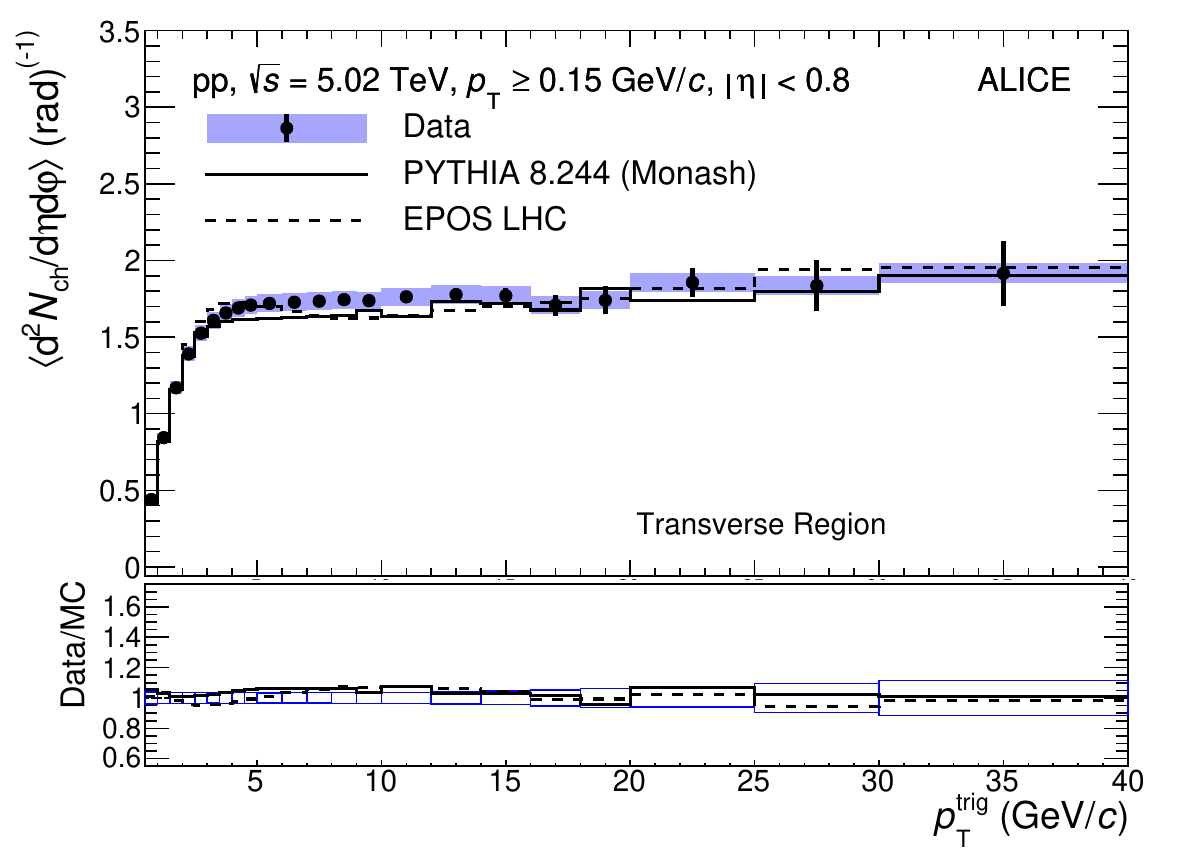}
  \includegraphics[width=7.90cm, height=6.90cm]{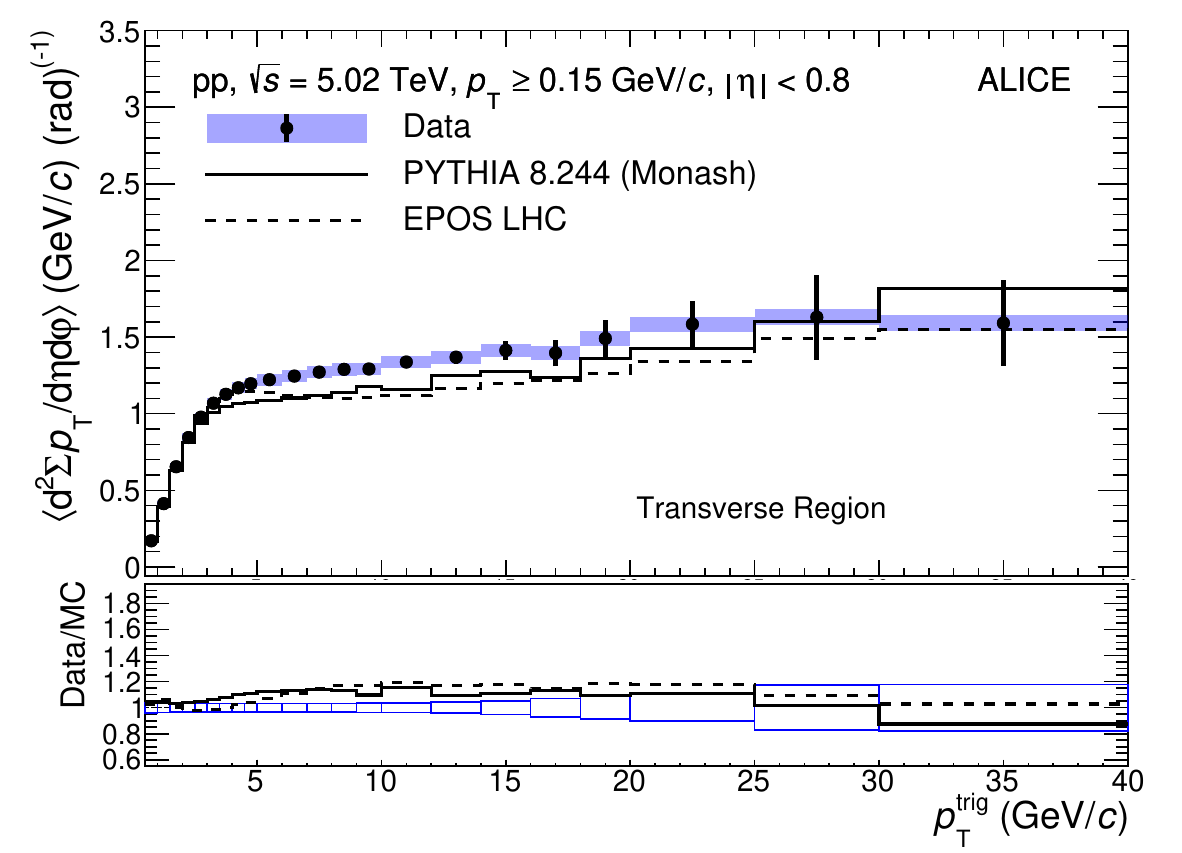}
  \includegraphics[width=7.90cm, height=6.90cm]{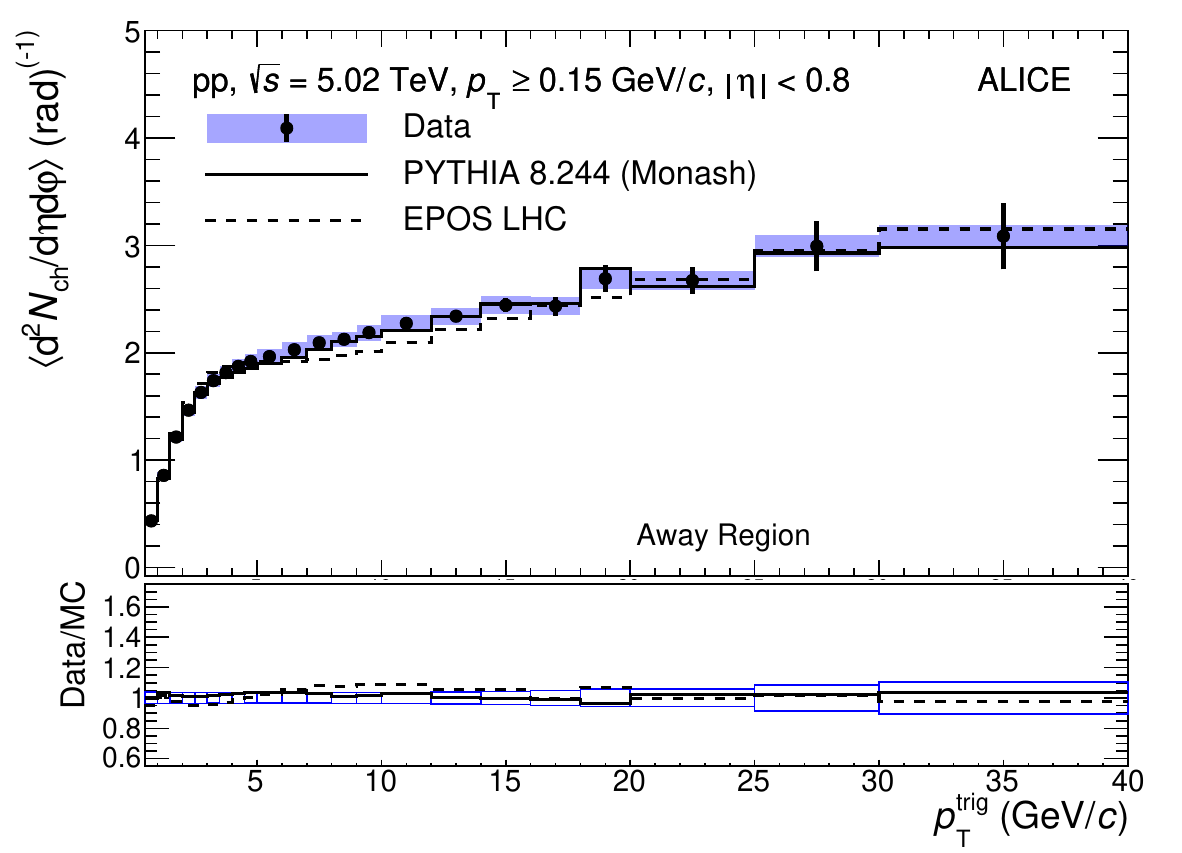}
  \includegraphics[width=7.90cm, height=6.90cm]{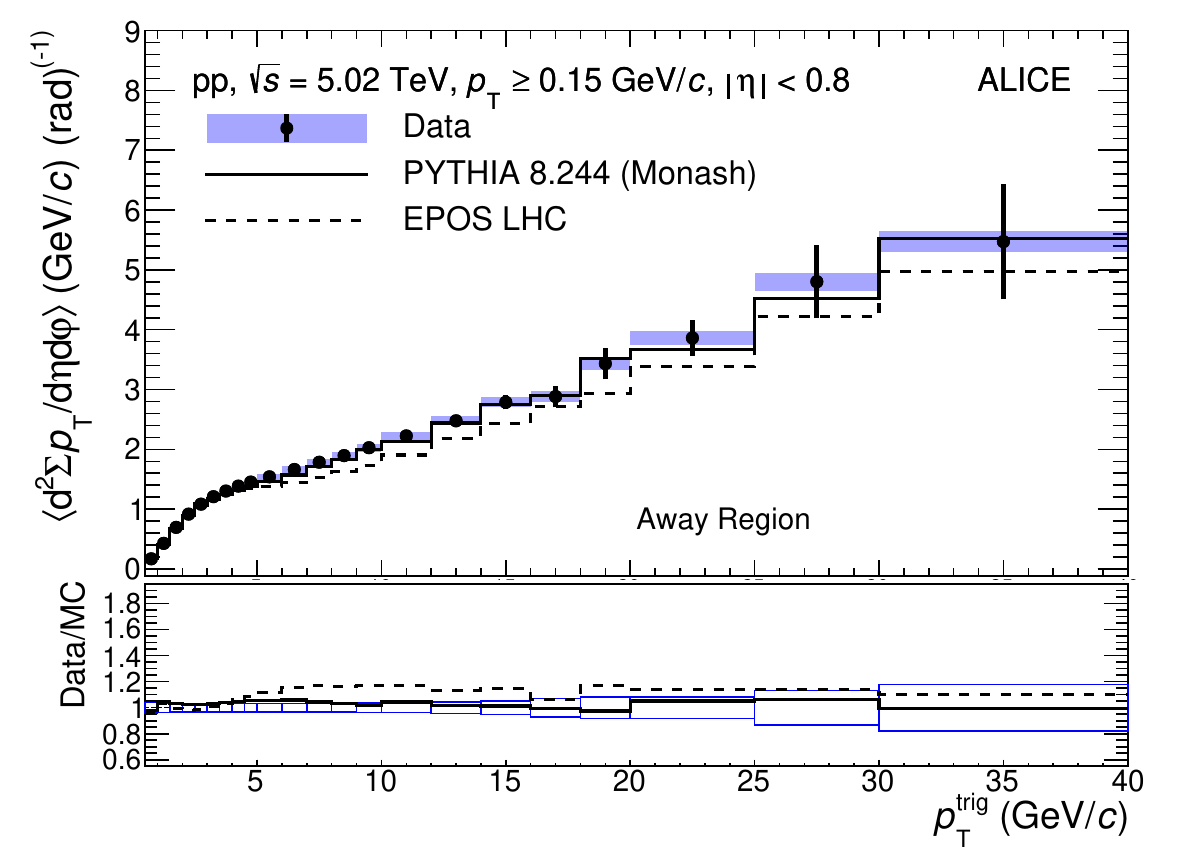}
  \includegraphics[width=7.90cm, height=6.90cm]{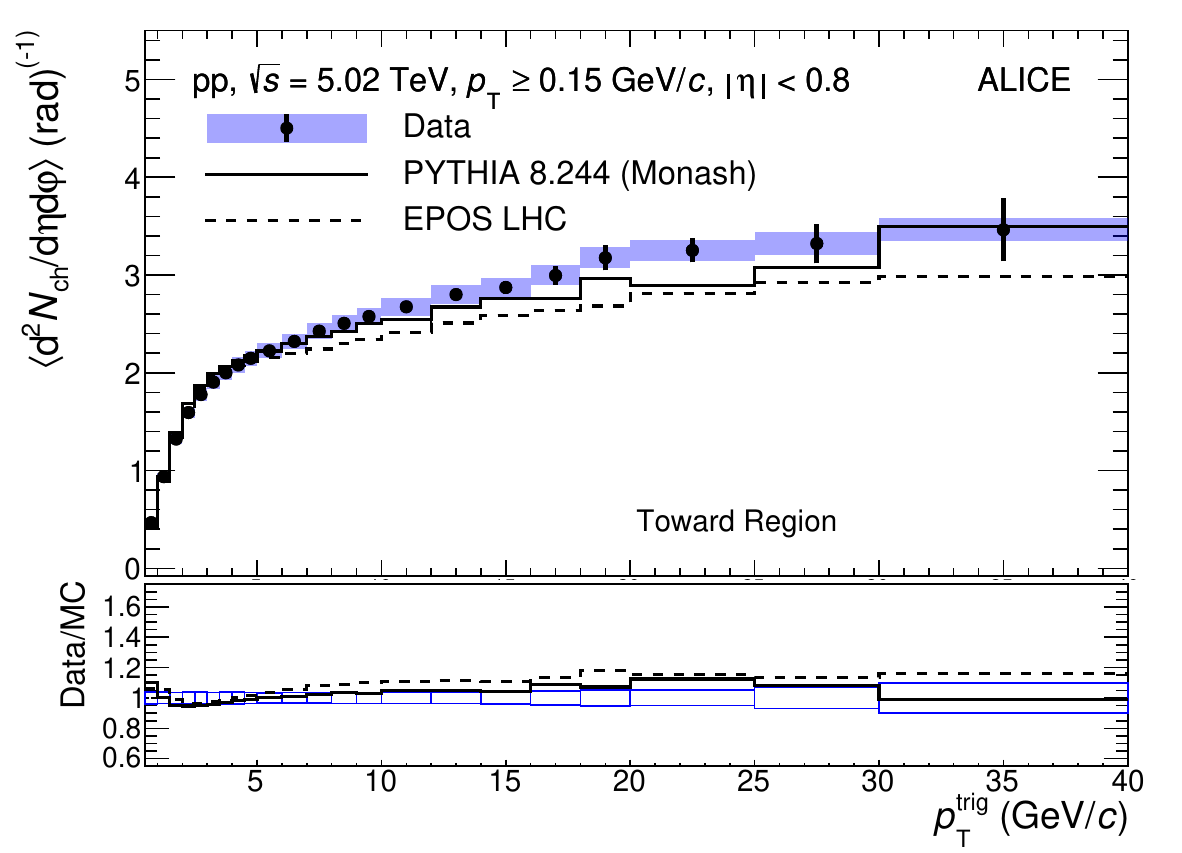}
 \includegraphics[width=7.90cm, height=6.90cm]{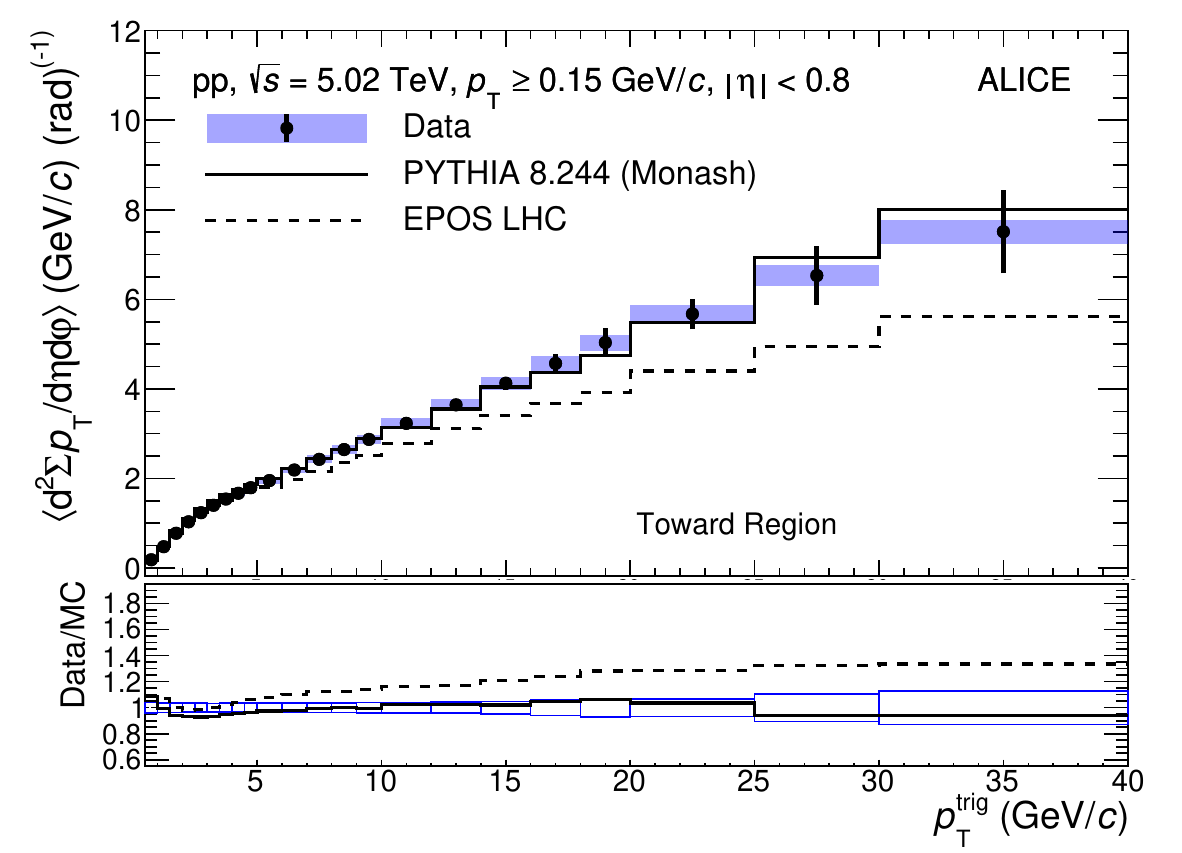}
 \caption{The charged-particle number (left) and summed-$p_{\rm T}$ (right) densities as a function of \ptt in \pp collision at $\sqrt{s}=5.02$\,\TeV are displayed. Results for the transverse (top), away (middle), and toward (bottom) regions were obtained for the transverse momentum threshold $\pt>0.15$\,GeV/$c$. The shaded area and the error bars around the data points represent the systematic and statistical uncertainties, respectively. Data are compared with \py/Monash (solid line) and \ep (dashed line) predictions. The data-to-model ratios are displayed in the bottom panel of each plot. The boxes around unity represent the statistical and systematic uncertainties added in quadrature.}
\label{fig:f581014} 
\end{figure}

\begin{figure}[!ht]
  \includegraphics[width=7.90cm, height=6.90cm]{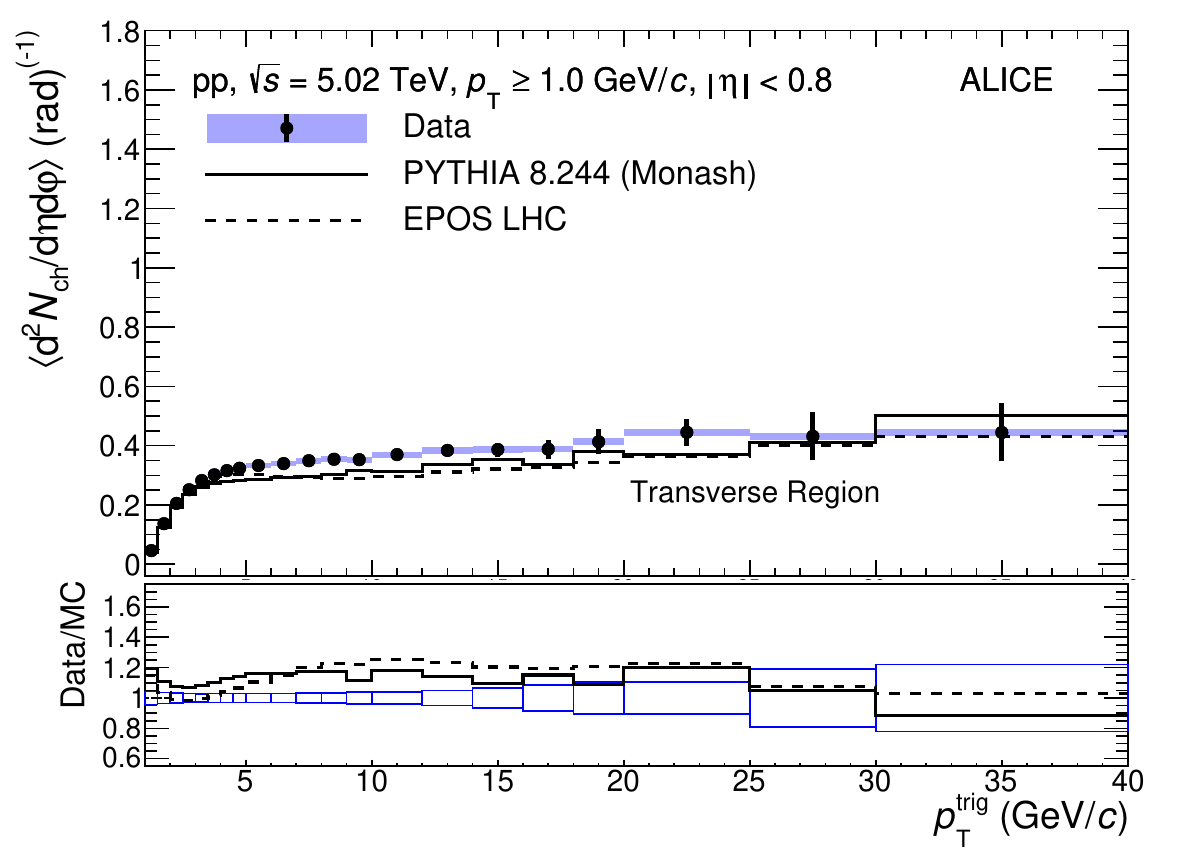}
  \includegraphics[width=7.90cm, height=6.90cm]{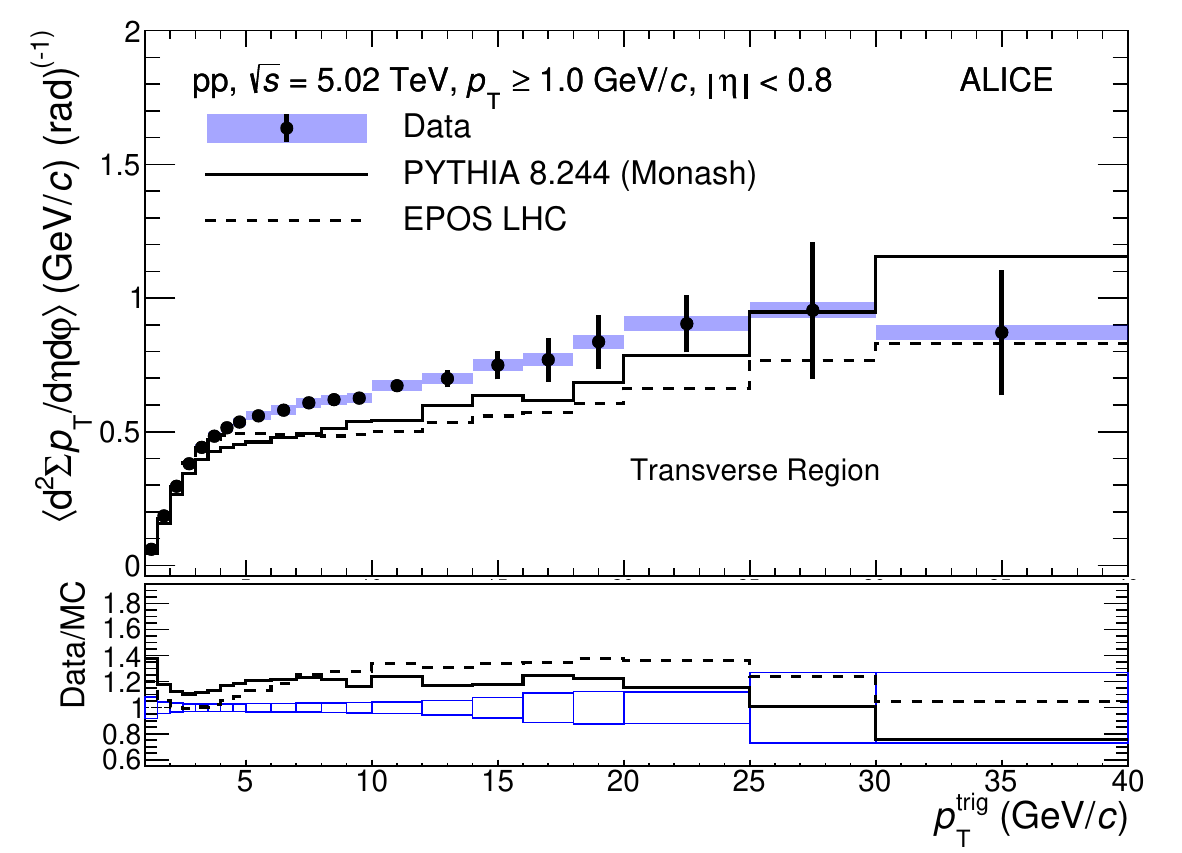}
  \includegraphics[width=7.90cm, height=6.90cm]{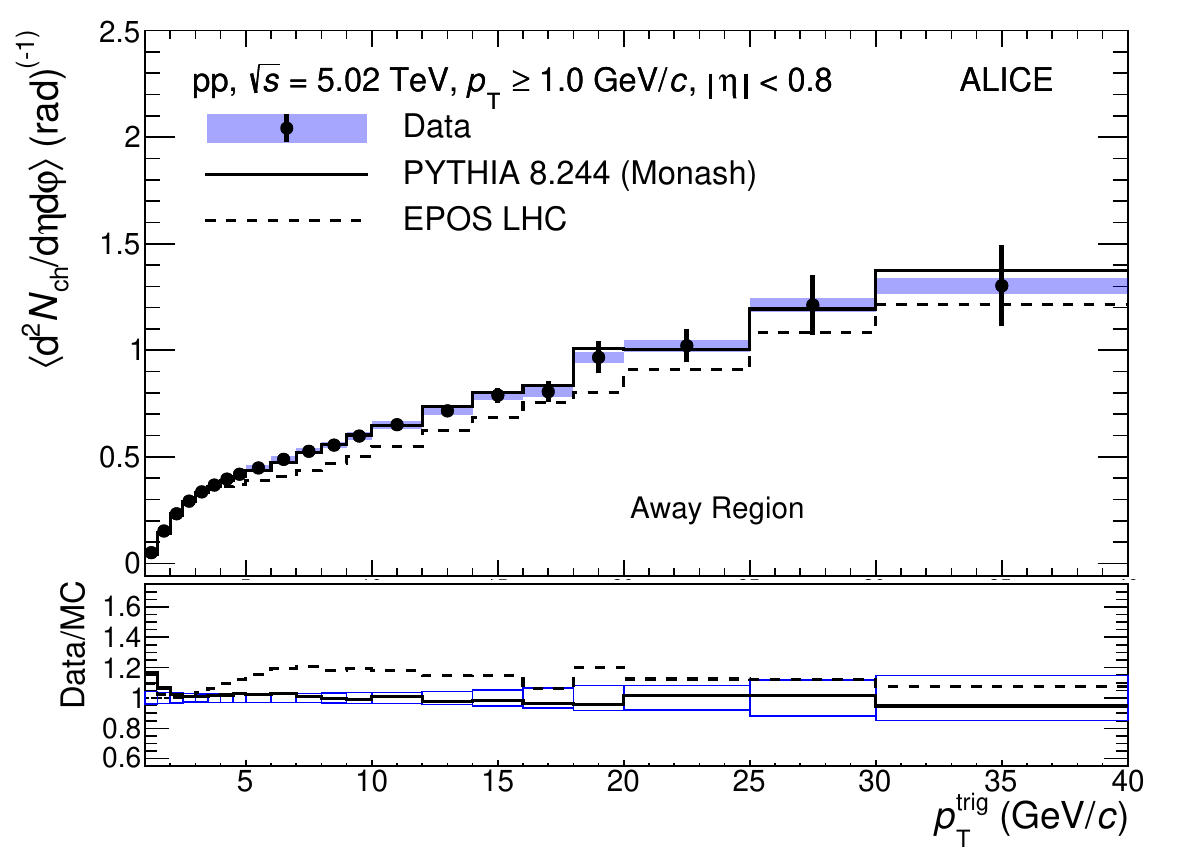}
  \includegraphics[width=7.90cm, height=6.90cm]{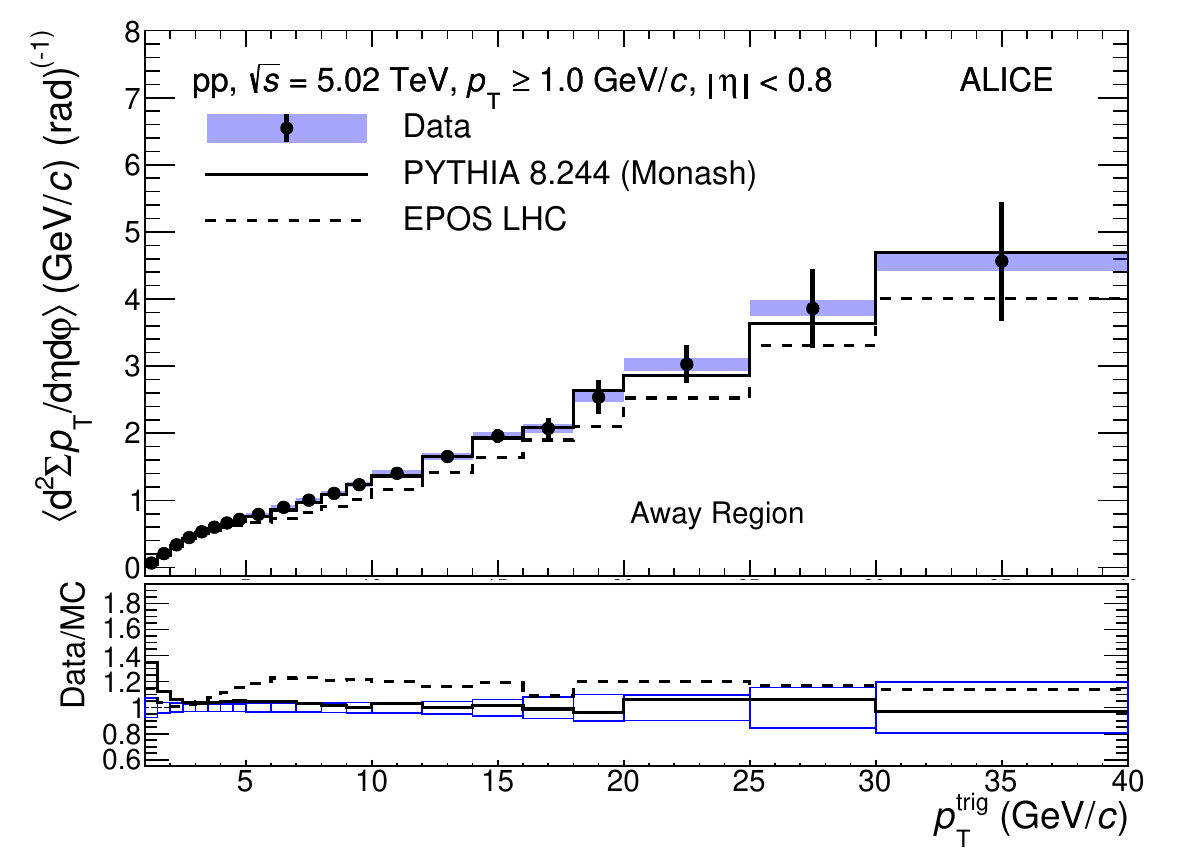}
  \includegraphics[width=7.90cm, height=6.90cm]{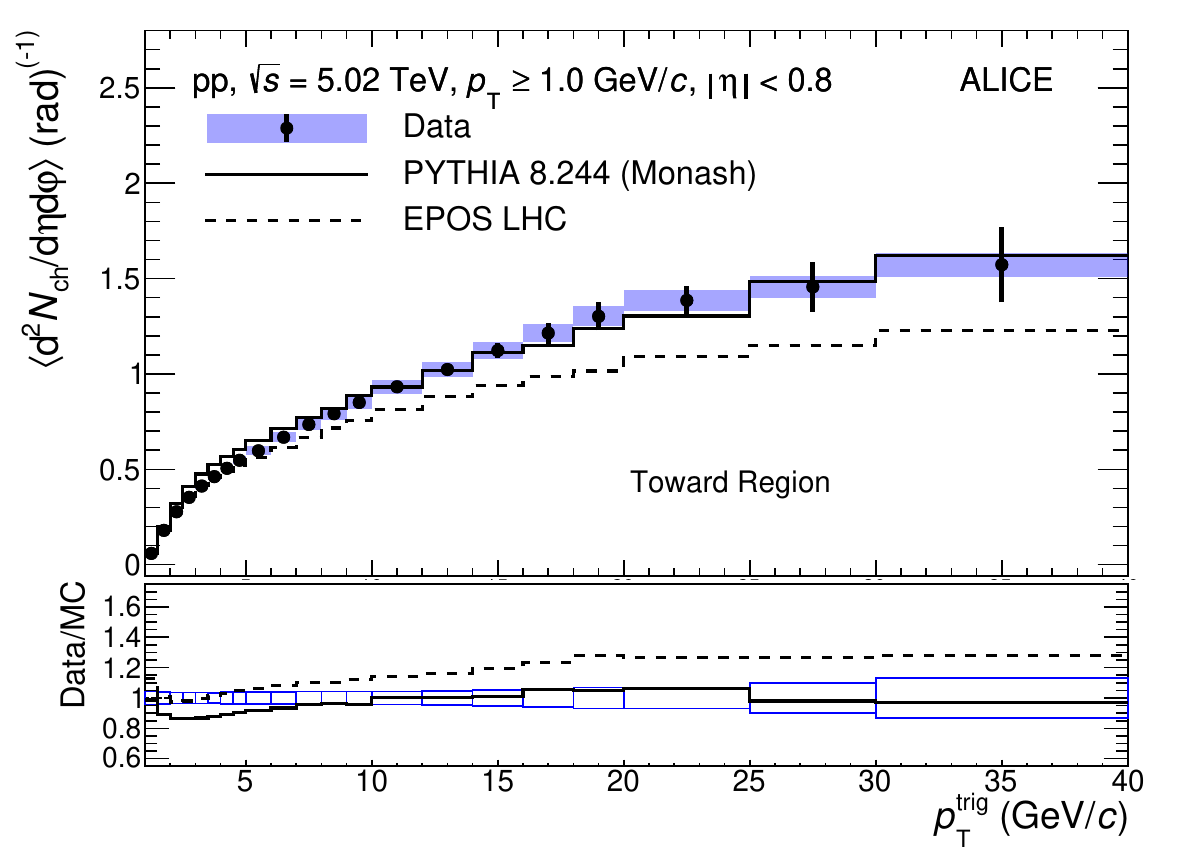}
  \includegraphics[width=7.90cm, height=6.90cm]{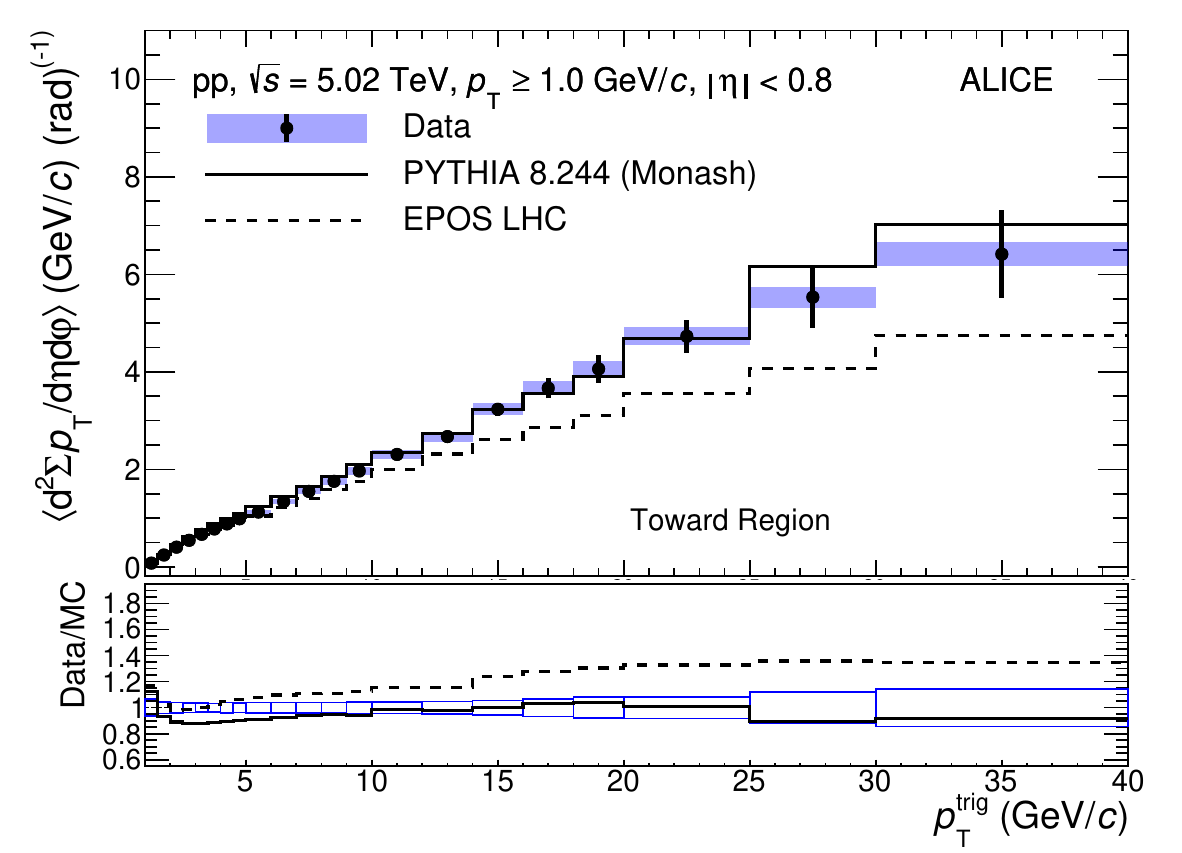}
 \caption{The charged-particle number (left) and summed-$p_{\rm T}$ (right) densities as a function of \ptt in \pp collision at $\sqrt{s}=5.02$\,\TeV are displayed. Results for the transverse (top), away (middle), and toward (bottom) regions were obtained for the transverse momentum threshold $\pt>1$\,GeV/$c$. The shaded area and the error bars around the data points represent the systematic and statistical uncertainties, respectively. Data are compared with \py/Monash (solid line) and \ep (dashed line) predictions. The data-to-model ratios are displayed in the bottom panel of each plot. The boxes around unity represent the statistical and systematic uncertainties added in quadrature.}
\label{fig:f581018} 
 \end{figure}

\begin{figure}[!ht]
\centering
  \includegraphics[width=7.90cm, height=6.90cm]{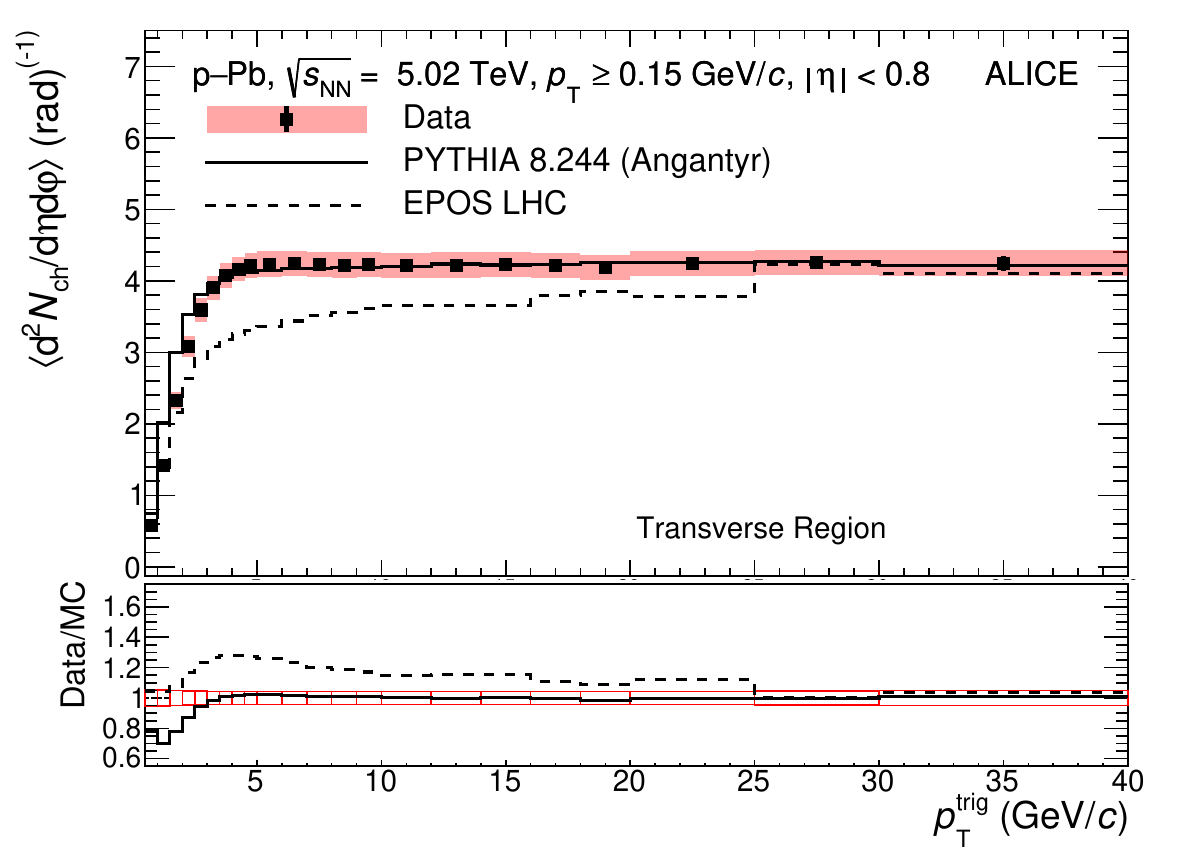}
  \includegraphics[width=7.90cm, height=6.90cm]{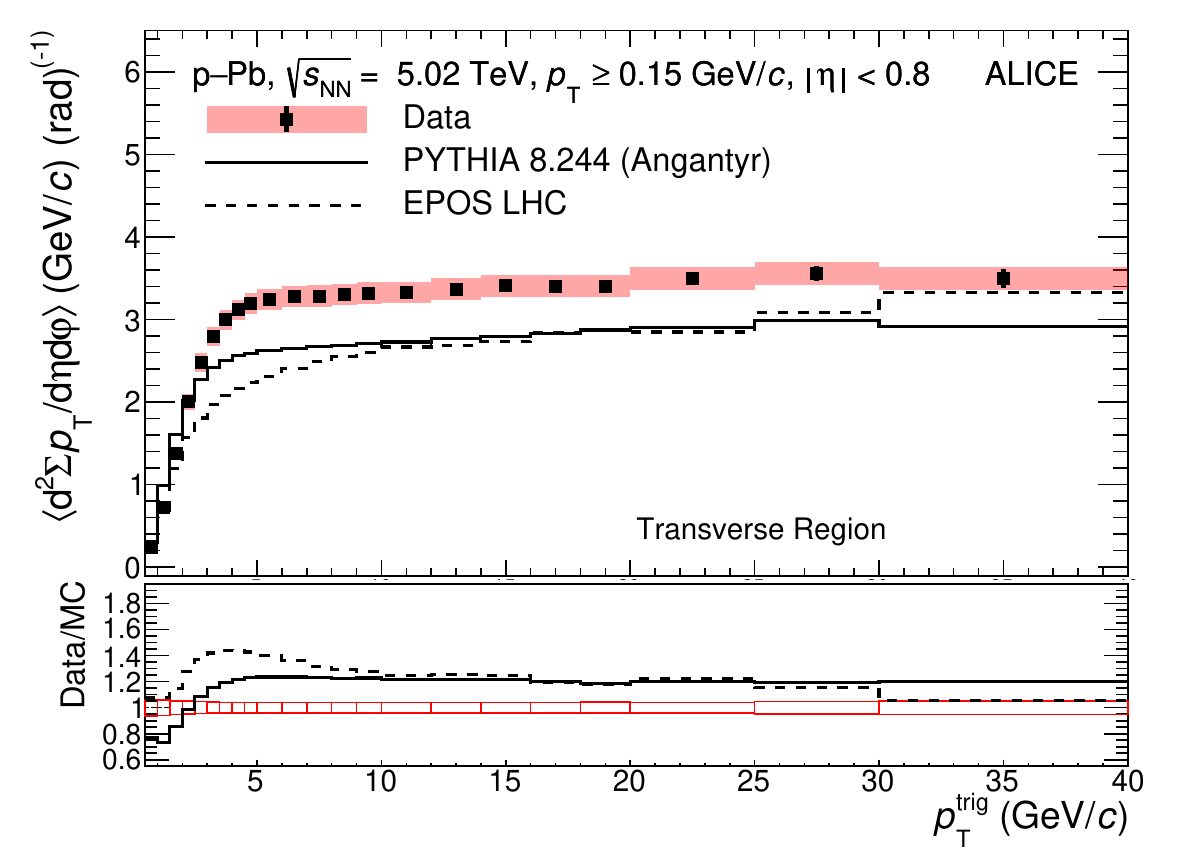}
  \includegraphics[width=7.90cm, height=6.90cm]{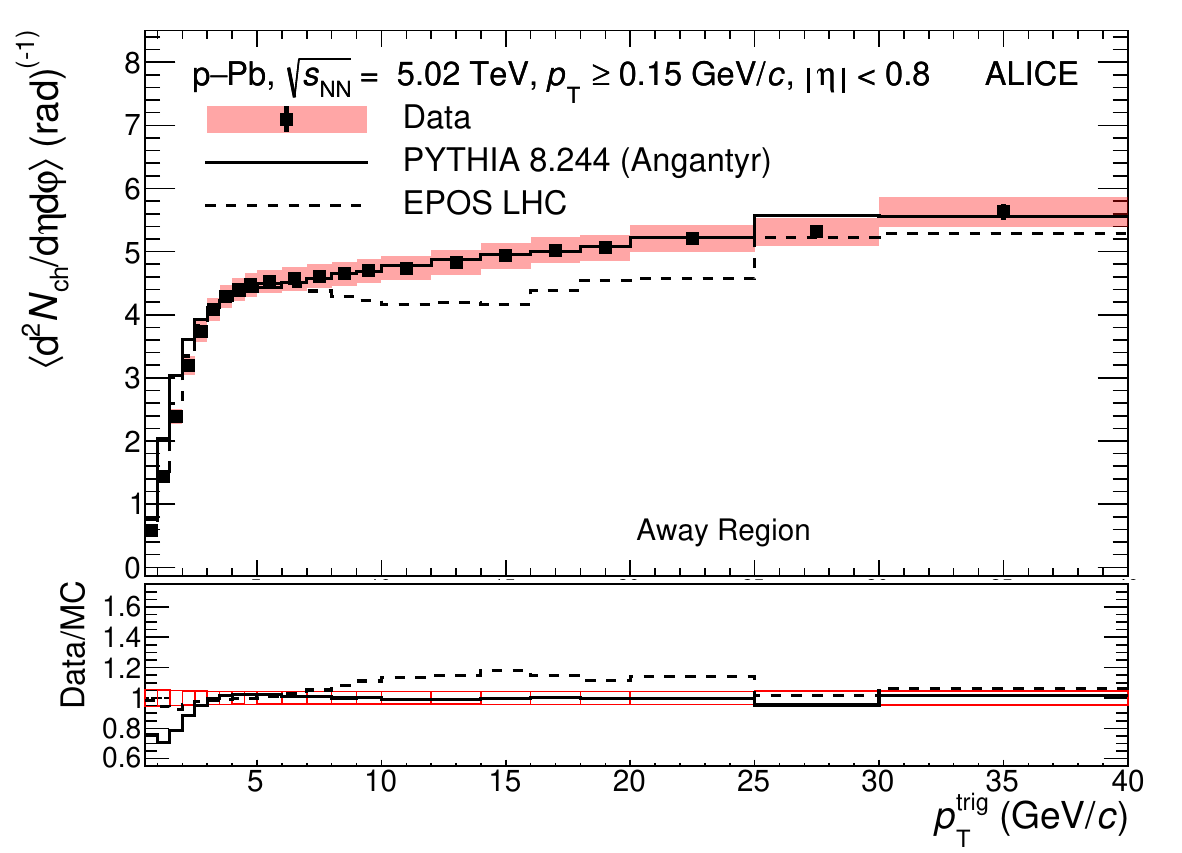}
  \includegraphics[width=7.90cm, height=6.90cm]{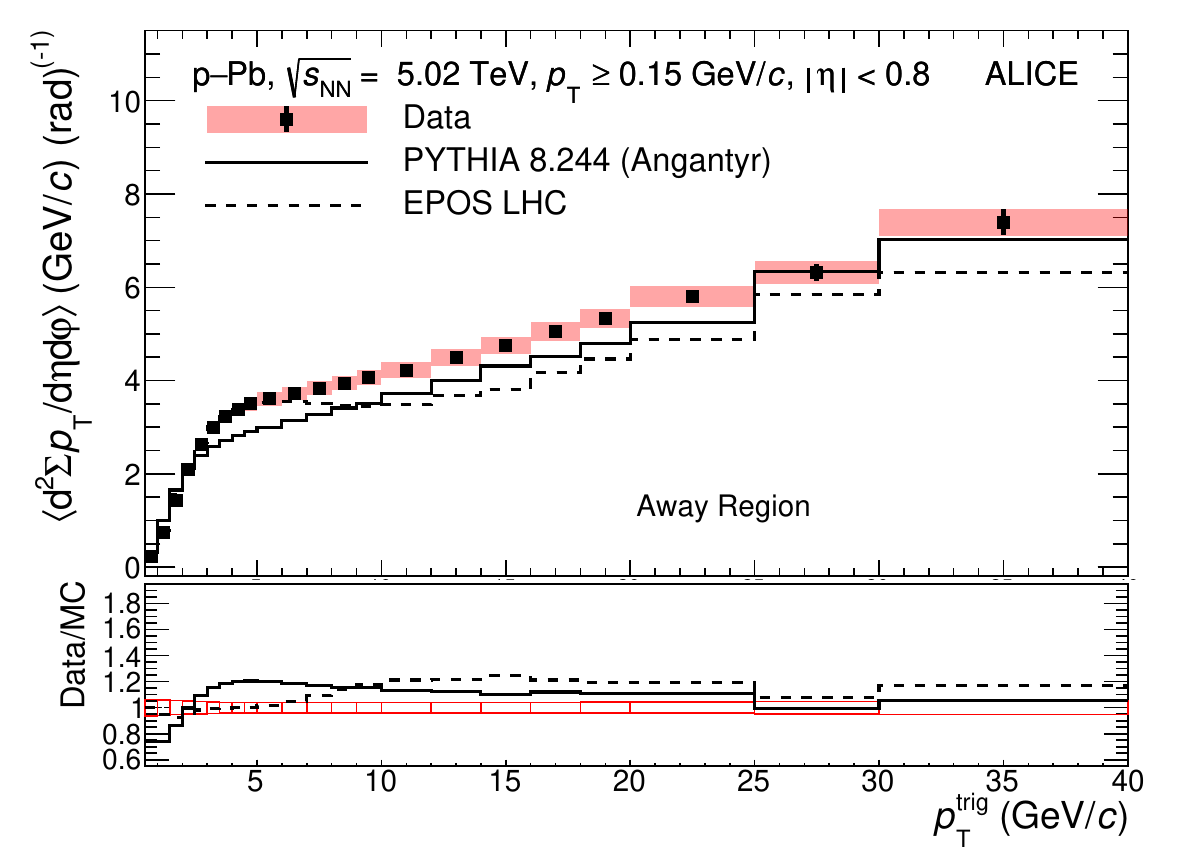}
  \includegraphics[width=7.90cm, height=6.90cm]{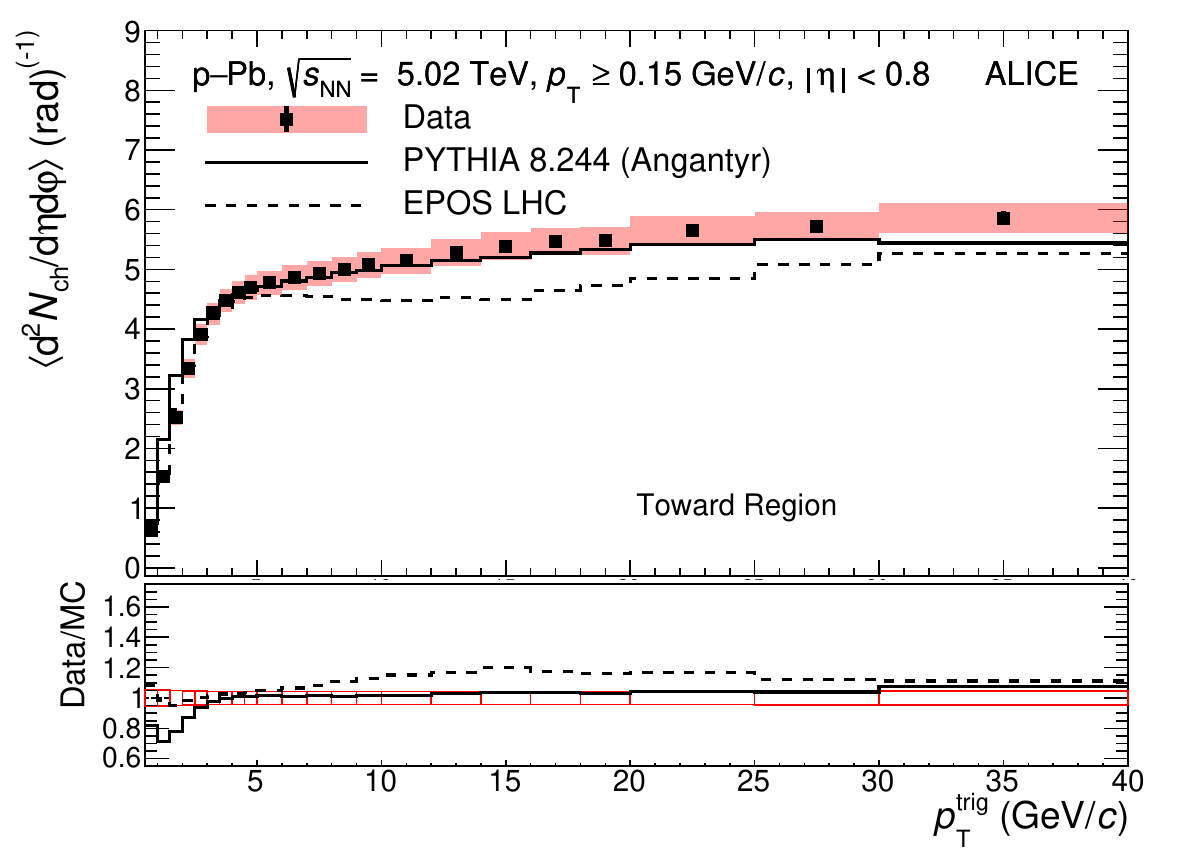}
  \includegraphics[width=7.90cm, height=6.90cm]{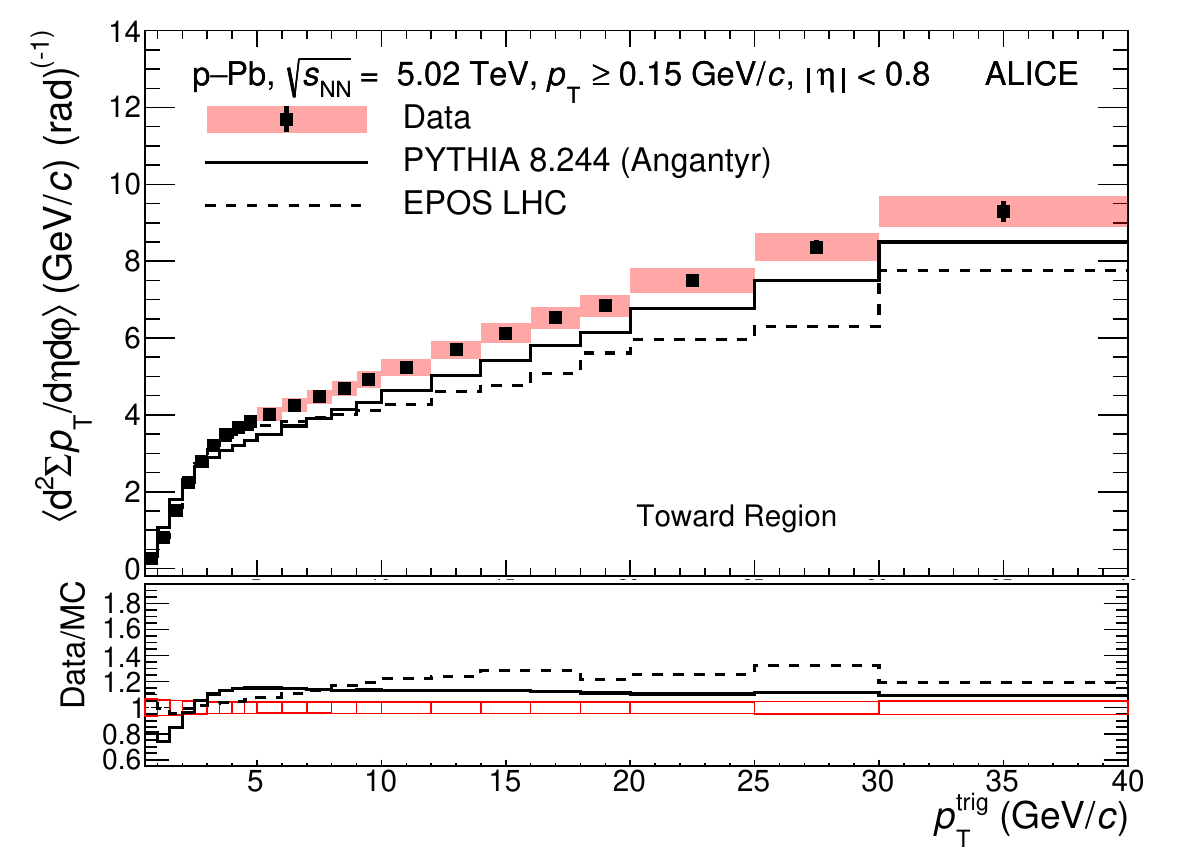}
 \caption{The charged-particle number (left) and summed-$p_{\rm T}$ (right) densities as a function of \ptt in \pPb collision at $\sqrt{s_{\rm NN}}=5.02$\,\TeV are displayed. Results for the transverse (top), away (middle), and toward (bottom) regions were obtained for the transverse momentum threshold $\pt>0.15$\,GeV/$c$. The shaded area and the error bars around the data points represent the systematic and statistical uncertainties, respectively. Data are compared with \py/Angantyr (solid line) and \ep (dashed line) predictions. The data-to-model ratios are displayed in the bottom panel of each plot. The boxes around unity represent the statistical and systematic uncertainties added in quadrature.}
\label{fig:f581015} 
 \end{figure}

\begin{figure}[!ht]
  \includegraphics[width=7.90cm, height=6.90cm]{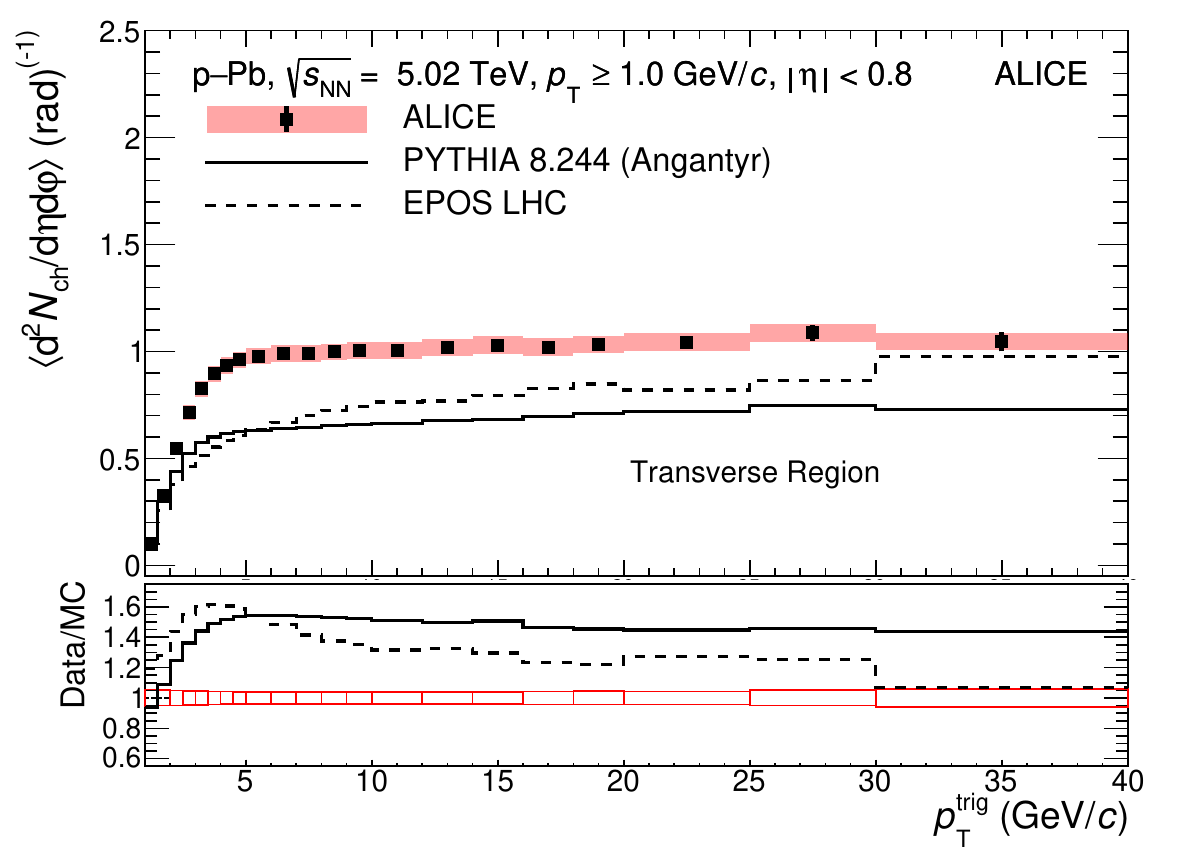}
  \includegraphics[width=7.90cm, height=6.90cm]{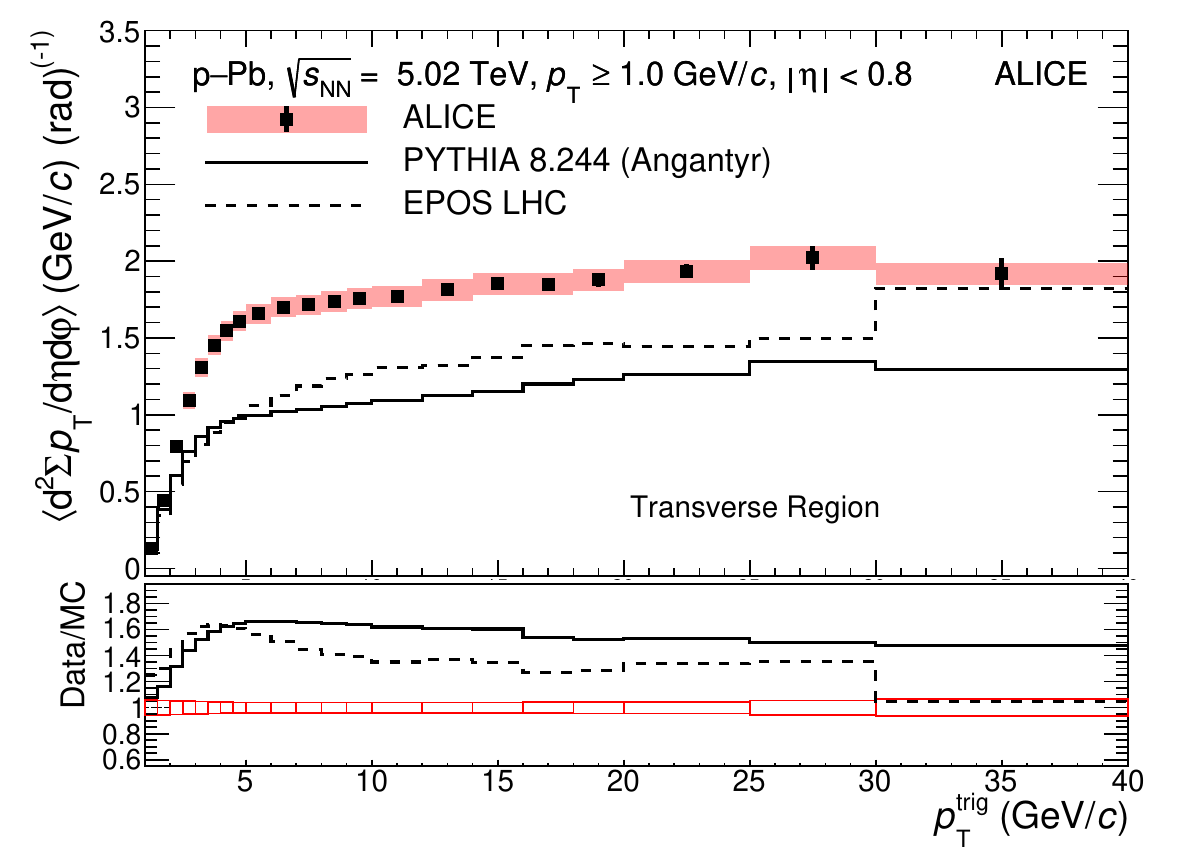}
  \includegraphics[width=7.90cm, height=6.90cm]{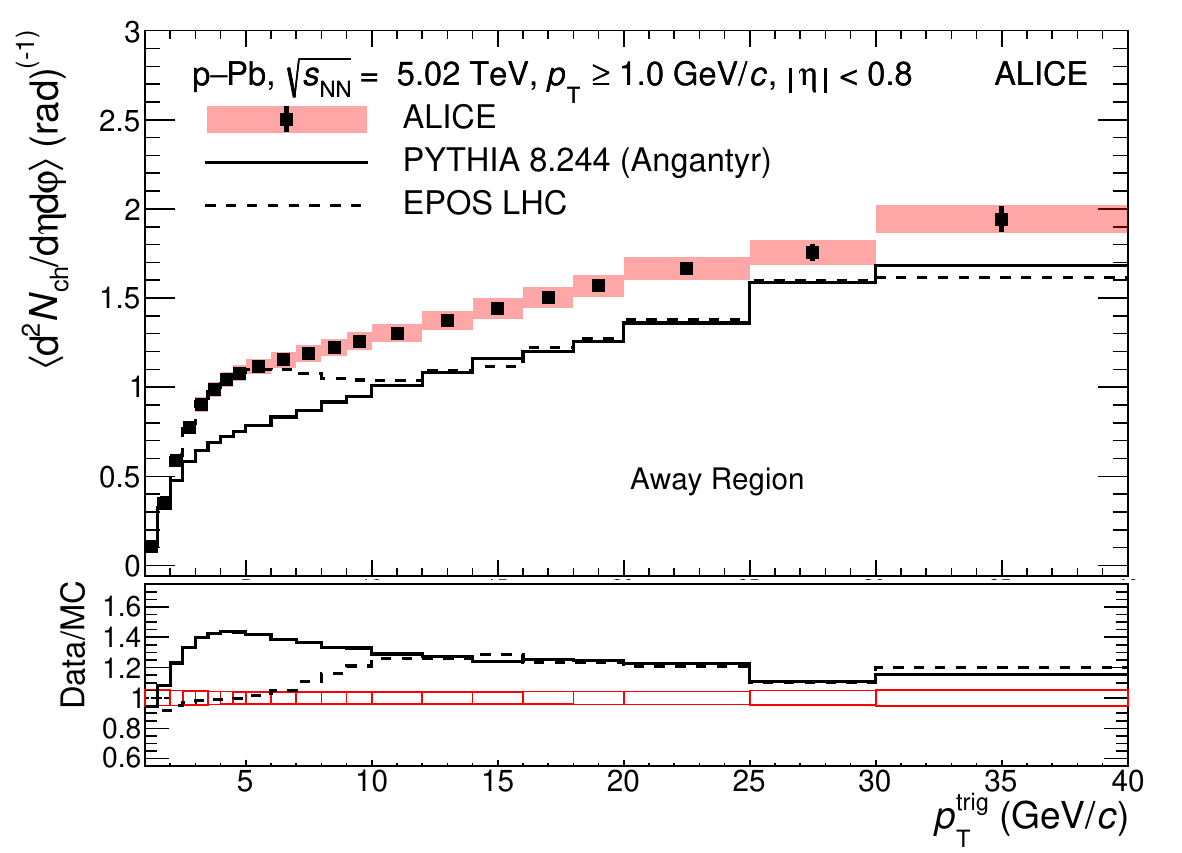}
  \includegraphics[width=7.90cm, height=6.90cm]{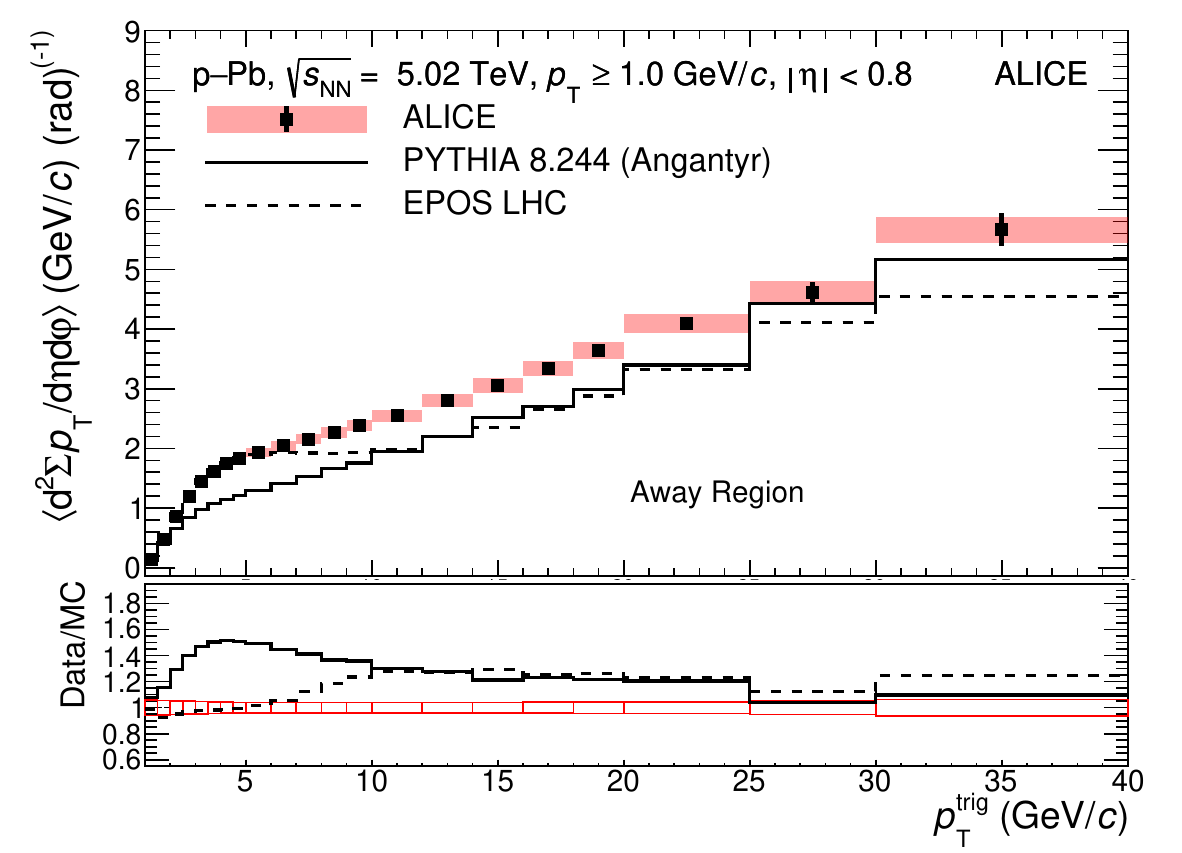}
  \includegraphics[width=7.90cm, height=6.90cm]{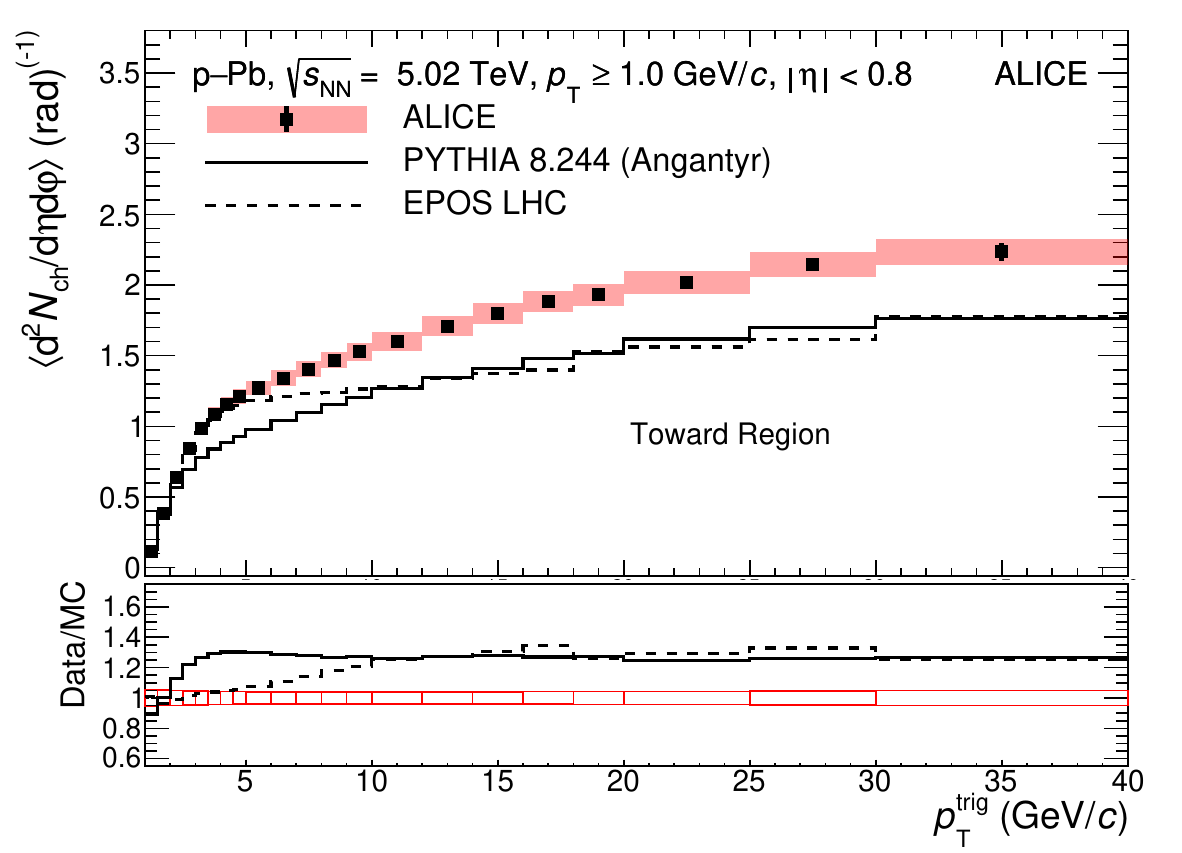}
  \includegraphics[width=7.90cm, height=6.90cm]{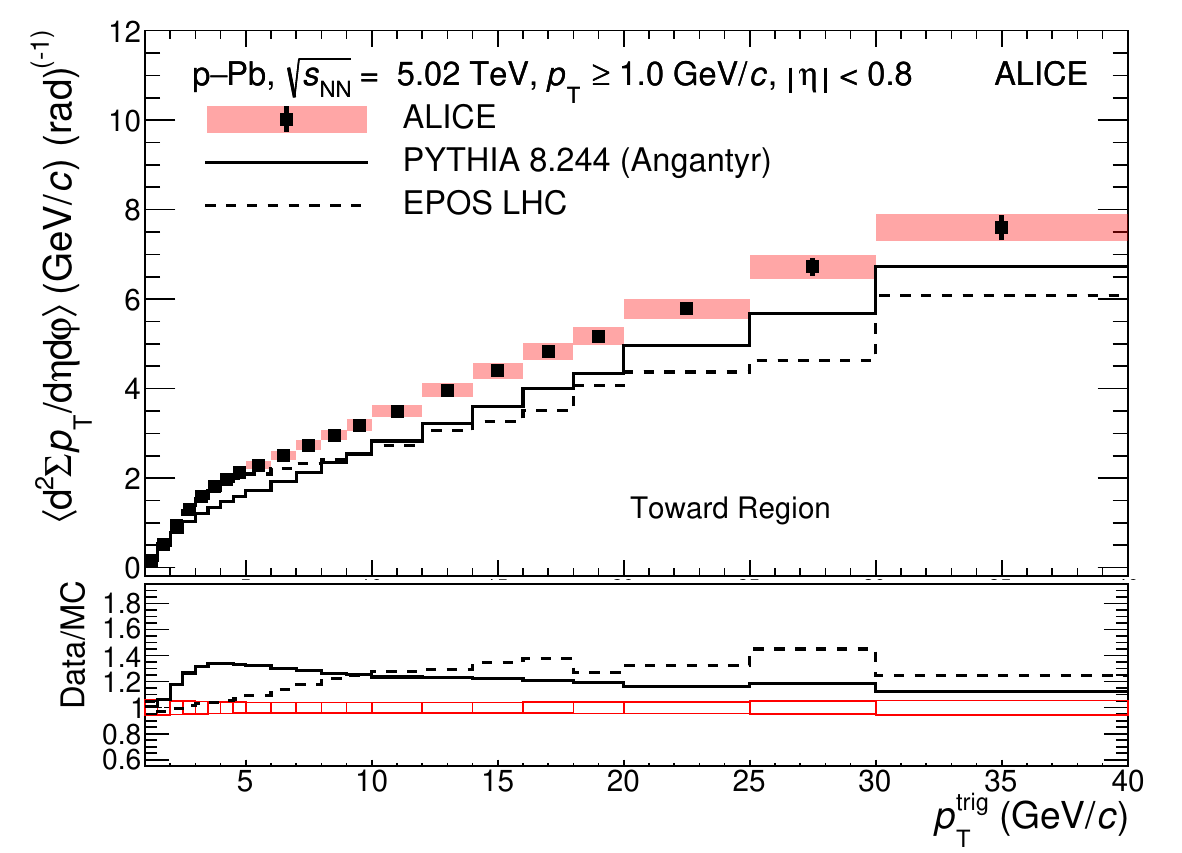}
 \caption{The charged-particle number (left) and summed-$p_{\rm T}$ (right) densities as a function of \ptt in \pPb collision at $\sqrt{s_{\rm NN}}=5.02$\,\TeV are displayed. Results for the transverse (top), away (middle), and toward (bottom) regions were obtained for the transverse momentum threshold $\pt>1$\,GeV/$c$. The shaded area and the error bars around the data points represent the systematic and statistical uncertainties, respectively. Data are compared with \py/Angantyr (solid line) and \ep (dashed line) predictions. The data-to-model ratios are displayed in the bottom panel of each plot. The boxes around unity represent the statistical and systematic uncertainties added in quadrature.}
\label{fig:f581019} 
 \end{figure}

The charged-particle number and summed-\pt densities as a function of \ptt measured in \pp collisions at $\sqrt{s}=5.02$\,\TeV, in the transverse, away, and toward regions for the transverse momentum thresholds $\pt>0.15$\,\GeV/$c$ and $\pt>1$\,\GeV/$c$ are shown in figures~\ref{fig:f581014} and ~\ref{fig:f581018}, respectively.
The charged-particle number and summed-\pt densities as a function of \ptt measured in \pPb collisions at $\sqrt{s_{\rm NN}}=5.02$\,\TeV, in the transverse, away, and toward regions for the transverse momentum thresholds $\pt>0.15$\,GeV/$c$ and $\pt>1$\,GeV/$c$ are shown in figures~\ref{fig:f581015} and ~\ref{fig:f581019}, respectively.

\newpage

 \newpage
 
\clearpage
\newpage

\section{The ALICE collaboration}
\label{app:collab}
\begin{flushleft} 
\small

S.~Acharya\,\orcidlink{0000-0002-9213-5329}\,$^{\rm 123,130}$, 
D.~Adamov\'{a}\,\orcidlink{0000-0002-0504-7428}\,$^{\rm 85}$, 
A.~Adler$^{\rm 68}$, 
G.~Aglieri Rinella\,\orcidlink{0000-0002-9611-3696}\,$^{\rm 32}$, 
M.~Agnello\,\orcidlink{0000-0002-0760-5075}\,$^{\rm 29}$, 
N.~Agrawal\,\orcidlink{0000-0003-0348-9836}\,$^{\rm 49}$, 
Z.~Ahammed\,\orcidlink{0000-0001-5241-7412}\,$^{\rm 130}$, 
S.~Ahmad\,\orcidlink{0000-0003-0497-5705}\,$^{\rm 15}$, 
S.U.~Ahn\,\orcidlink{0000-0001-8847-489X}\,$^{\rm 69}$, 
I.~Ahuja\,\orcidlink{0000-0002-4417-1392}\,$^{\rm 36}$
% ...

\section*{Affiliation Notes}

$^{\rm I}$ Deceased\\
$^{\rm II}$ Also at: Max-Planck-Institut f\"{u}r Physik, Munich, Germany\\
$^{\rm III}$ Also at: Italian National Agency for New Technologies, Energy and Sustainable Economic Development (ENEA), Bologna, Italy\\
$^{\rm IV}$ Also at: Dipartimento DET del Politecnico di Torino, Turin, Italy\\
$^{\rm V}$ Also at: Department of Applied Physics, Aligarh Muslim University, Aligarh, India\\
$^{\rm VI}$ Also at: Institute of Theoretical Physics, University of Wroclaw, Poland\\
$^{\rm VII}$ Also at: An institution covered by a cooperation agreement with CERN\\

\section*{Collaboration Institutes}

$^{1}$ A.I. Alikhanyan National Science Laboratory (Yerevan Physics Institute) Foundation, Yerevan, Armenia\\
$^{2}$ AGH University of Science and Technology, Cracow, Poland\\
$^{3}$ Bogolyubov Institute for Theoretical Physics, National Academy of Sciences of Ukraine, Kiev, Ukraine\\
$^{4}$ Bose Institute, Department of Physics  and Centre for Astroparticle Physics and Space Science (CAPSS), Kolkata, India\\
$^{5}$ California Polytechnic State University, San Luis Obispo, California, United States\\
$^{6}$ Central China Normal University, Wuhan, China\\
$^{7}$ Centro de Aplicaciones Tecnol\'{o}gicas y Desarrollo Nuclear (CEADEN), Havana, Cuba\\
$^{8}$ Centro de Investigaci\'{o}n y de Estudios Avanzados (CINVESTAV), Mexico City and M\'{e}rida, Mexico\\
$^{9}$ Chicago State University, Chicago, Illinois, United States\\
$^{10}$ China Institute of Atomic Energy, Beijing, China\\
$^{11}$ Chungbuk National University, Cheongju, Republic of Korea\\
$^{12}$ Comenius University Bratislava, Faculty of Mathematics, Physics and Informatics, Bratislava, Slovak Republic\\
$^{13}$ COMSATS University Islamabad, Islamabad, Pakistan\\
$^{14}$ Creighton University, Omaha, Nebraska, United States\\
$^{15}$ Department of Physics, Aligarh Muslim University, Aligarh, India\\
$^{16}$ Department of Physics, Pusan National University, Pusan, Republic of Korea\\
$^{17}$ Department of Physics, Sejong University, Seoul, Republic of Korea\\
$^{18}$ Department of Physics, University of California, Berkeley, California, United States\\
$^{19}$ Department of Physics, University of Oslo, Oslo, Norway\\
$^{20}$ Department of Physics and Technology, University of Bergen, Bergen, Norway\\
$^{21}$ Dipartimento di Fisica, Universit\`{a} di Pavia, Pavia, Italy\\
$^{22}$ Dipartimento di Fisica dell'Universit\`{a} and Sezione INFN, Cagliari, Italy\\
$^{23}$ Dipartimento di Fisica dell'Universit\`{a} and Sezione INFN, Trieste, Italy\\
$^{24}$ Dipartimento di Fisica dell'Universit\`{a} and Sezione INFN, Turin, Italy\\
$^{25}$ Dipartimento di Fisica e Astronomia dell'Universit\`{a} and Sezione INFN, Bologna, Italy\\
$^{26}$ Dipartimento di Fisica e Astronomia dell'Universit\`{a} and Sezione INFN, Catania, Italy\\
$^{27}$ Dipartimento di Fisica e Astronomia dell'Universit\`{a} and Sezione INFN, Padova, Italy\\
$^{28}$ Dipartimento di Fisica `E.R.~Caianiello' dell'Universit\`{a} and Gruppo Collegato INFN, Salerno, Italy\\
$^{29}$ Dipartimento DISAT del Politecnico and Sezione INFN, Turin, Italy\\
$^{30}$ Dipartimento di Scienze MIFT, Universit\`{a} di Messina, Messina, Italy\\
$^{31}$ Dipartimento Interateneo di Fisica `M.~Merlin' and Sezione INFN, Bari, Italy\\
$^{32}$ European Organization for Nuclear Research (CERN), Geneva, Switzerland\\
$^{33}$ Faculty of Electrical Engineering, Mechanical Engineering and Naval Architecture, University of Split, Split, Croatia\\
$^{34}$ Faculty of Engineering and Science, Western Norway University of Applied Sciences, Bergen, Norway\\
$^{35}$ Faculty of Nuclear Sciences and Physical Engineering, Czech Technical University in Prague, Prague, Czech Republic\\
$^{36}$ Faculty of Science, P.J.~\v{S}af\'{a}rik University, Ko\v{s}ice, Slovak Republic\\
$^{37}$ Frankfurt Institute for Advanced Studies, Johann Wolfgang Goethe-Universit\"{a}t Frankfurt, Frankfurt, Germany\\
$^{38}$ Fudan University, Shanghai, China\\
$^{39}$ Gangneung-Wonju National University, Gangneung, Republic of Korea\\
$^{40}$ Gauhati University, Department of Physics, Guwahati, India\\
$^{41}$ Helmholtz-Institut f\"{u}r Strahlen- und Kernphysik, Rheinische Friedrich-Wilhelms-Universit\"{a}t Bonn, Bonn, Germany\\
$^{42}$ Helsinki Institute of Physics (HIP), Helsinki, Finland\\
$^{43}$ High Energy Physics Group,  Universidad Aut\'{o}noma de Puebla, Puebla, Mexico\\
$^{44}$ Horia Hulubei National Institute of Physics and Nuclear Engineering, Bucharest, Romania\\
$^{45}$ Indian Institute of Technology Bombay (IIT), Mumbai, India\\
$^{46}$ Indian Institute of Technology Indore, Indore, India\\
$^{47}$ INFN, Laboratori Nazionali di Frascati, Frascati, Italy\\
$^{48}$ INFN, Sezione di Bari, Bari, Italy\\
$^{49}$ INFN, Sezione di Bologna, Bologna, Italy\\
$^{50}$ INFN, Sezione di Cagliari, Cagliari, Italy\\
$^{51}$ INFN, Sezione di Catania, Catania, Italy\\
$^{52}$ INFN, Sezione di Padova, Padova, Italy\\
$^{53}$ INFN, Sezione di Pavia, Pavia, Italy\\
$^{54}$ INFN, Sezione di Torino, Turin, Italy\\
$^{55}$ INFN, Sezione di Trieste, Trieste, Italy\\
$^{56}$ Inha University, Incheon, Republic of Korea\\
$^{57}$ Institute for Gravitational and Subatomic Physics (GRASP), Utrecht University/Nikhef, Utrecht, Netherlands\\
$^{58}$ Institute of Experimental Physics, Slovak Academy of Sciences, Ko\v{s}ice, Slovak Republic\\
$^{59}$ Institute of Physics, Homi Bhabha National Institute, Bhubaneswar, India\\
$^{60}$ Institute of Physics of the Czech Academy of Sciences, Prague, Czech Republic\\
$^{61}$ Institute of Space Science (ISS), Bucharest, Romania\\
$^{62}$ Institut f\"{u}r Kernphysik, Johann Wolfgang Goethe-Universit\"{a}t Frankfurt, Frankfurt, Germany\\
$^{63}$ Instituto de Ciencias Nucleares, Universidad Nacional Aut\'{o}noma de M\'{e}xico, Mexico City, Mexico\\
$^{64}$ Instituto de F\'{i}sica, Universidade Federal do Rio Grande do Sul (UFRGS), Porto Alegre, Brazil\\
$^{65}$ Instituto de F\'{\i}sica, Universidad Nacional Aut\'{o}noma de M\'{e}xico, Mexico City, Mexico\\
$^{66}$ iThemba LABS, National Research Foundation, Somerset West, South Africa\\
$^{67}$ Jeonbuk National University, Jeonju, Republic of Korea\\
$^{68}$ Johann-Wolfgang-Goethe Universit\"{a}t Frankfurt Institut f\"{u}r Informatik, Fachbereich Informatik und Mathematik, Frankfurt, Germany\\
$^{69}$ Korea Institute of Science and Technology Information, Daejeon, Republic of Korea\\
$^{70}$ KTO Karatay University, Konya, Turkey\\
$^{71}$ Laboratoire de Physique des 2 Infinis, Ir\`{e}ne Joliot-Curie, Orsay, France\\
$^{72}$ Laboratoire de Physique Subatomique et de Cosmologie, Universit\'{e} Grenoble-Alpes, CNRS-IN2P3, Grenoble, France\\
$^{73}$ Lawrence Berkeley National Laboratory, Berkeley, California, United States\\
$^{74}$ Lund University Department of Physics, Division of Particle Physics, Lund, Sweden\\
$^{75}$ Nagasaki Institute of Applied Science, Nagasaki, Japan\\
$^{76}$ Nara Women{'}s University (NWU), Nara, Japan\\
$^{77}$ National and Kapodistrian University of Athens, School of Science, Department of Physics , Athens, Greece\\
$^{78}$ National Centre for Nuclear Research, Warsaw, Poland\\
$^{79}$ National Institute of Science Education and Research, Homi Bhabha National Institute, Jatni, India\\
$^{80}$ National Nuclear Research Center, Baku, Azerbaijan\\
$^{81}$ National Research and Innovation Agency - BRIN, Jakarta, Indonesia\\
$^{82}$ Niels Bohr Institute, University of Copenhagen, Copenhagen, Denmark\\
$^{83}$ Nikhef, National institute for subatomic physics, Amsterdam, Netherlands\\
$^{84}$ Nuclear Physics Group, STFC Daresbury Laboratory, Daresbury, United Kingdom\\
$^{85}$ Nuclear Physics Institute of the Czech Academy of Sciences, Husinec-\v{R}e\v{z}, Czech Republic\\
$^{86}$ Oak Ridge National Laboratory, Oak Ridge, Tennessee, United States\\
$^{87}$ Ohio State University, Columbus, Ohio, United States\\
$^{88}$ Physics department, Faculty of science, University of Zagreb, Zagreb, Croatia\\
$^{89}$ Physics Department, Panjab University, Chandigarh, India\\
$^{90}$ Physics Department, University of Jammu, Jammu, India\\
$^{91}$ Physics Department, University of Rajasthan, Jaipur, India\\
$^{92}$ Physics Program and International Institute for Sustainability with Knotted Chiral Meta Matter (SKCM2), Hiroshima University, Hiroshima, Japan\\
$^{93}$ Physikalisches Institut, Eberhard-Karls-Universit\"{a}t T\"{u}bingen, T\"{u}bingen, Germany\\
$^{94}$ Physikalisches Institut, Ruprecht-Karls-Universit\"{a}t Heidelberg, Heidelberg, Germany\\
$^{95}$ Physik Department, Technische Universit\"{a}t M\"{u}nchen, Munich, Germany\\
$^{96}$ Politecnico di Bari and Sezione INFN, Bari, Italy\\
$^{97}$ Research Division and ExtreMe Matter Institute EMMI, GSI Helmholtzzentrum f\"ur Schwerionenforschung GmbH, Darmstadt, Germany\\
$^{98}$ Saha Institute of Nuclear Physics, Homi Bhabha National Institute, Kolkata, India\\
$^{99}$ School of Physics and Astronomy, University of Birmingham, Birmingham, United Kingdom\\
$^{100}$ Secci\'{o}n F\'{\i}sica, Departamento de Ciencias, Pontificia Universidad Cat\'{o}lica del Per\'{u}, Lima, Peru\\
$^{101}$ Stefan Meyer Institut f\"{u}r Subatomare Physik (SMI), Vienna, Austria\\
$^{102}$ SUBATECH, IMT Atlantique, Nantes Universit\'{e}, CNRS-IN2P3, Nantes, France\\
$^{103}$ Suranaree University of Technology, Nakhon Ratchasima, Thailand\\
$^{104}$ Technical University of Ko\v{s}ice, Ko\v{s}ice, Slovak Republic\\
$^{105}$ The Henryk Niewodniczanski Institute of Nuclear Physics, Polish Academy of Sciences, Cracow, Poland\\
$^{106}$ The University of Texas at Austin, Austin, Texas, United States\\
$^{107}$ Universidad Aut\'{o}noma de Sinaloa, Culiac\'{a}n, Mexico\\
$^{108}$ Universidade de S\~{a}o Paulo (USP), S\~{a}o Paulo, Brazil\\
$^{109}$ Universidade Estadual de Campinas (UNICAMP), Campinas, Brazil\\
$^{110}$ Universidade Federal do ABC, Santo Andre, Brazil\\
$^{111}$ University of Cape Town, Cape Town, South Africa\\
$^{112}$ University of Houston, Houston, Texas, United States\\
$^{113}$ University of Jyv\"{a}skyl\"{a}, Jyv\"{a}skyl\"{a}, Finland\\
$^{114}$ University of Kansas, Lawrence, Kansas, United States\\
$^{115}$ University of Liverpool, Liverpool, United Kingdom\\
$^{116}$ University of Science and Technology of China, Hefei, China\\
$^{117}$ University of South-Eastern Norway, Kongsberg, Norway\\
$^{118}$ University of Tennessee, Knoxville, Tennessee, United States\\
$^{119}$ University of the Witwatersrand, Johannesburg, South Africa\\
$^{120}$ University of Tokyo, Tokyo, Japan\\
$^{121}$ University of Tsukuba, Tsukuba, Japan\\
$^{122}$ University Politehnica of Bucharest, Bucharest, Romania\\
$^{123}$ Universit\'{e} Clermont Auvergne, CNRS/IN2P3, LPC, Clermont-Ferrand, France\\
$^{124}$ Universit\'{e} de Lyon, CNRS/IN2P3, Institut de Physique des 2 Infinis de Lyon, Lyon, France\\
$^{125}$ Universit\'{e} de Strasbourg, CNRS, IPHC UMR 7178, F-67000 Strasbourg, France, Strasbourg, France\\
$^{126}$ Universit\'{e} Paris-Saclay Centre d'Etudes de Saclay (CEA), IRFU, D\'{e}partment de Physique Nucl\'{e}aire (DPhN), Saclay, France\\
$^{127}$ Universit\`{a} degli Studi di Foggia, Foggia, Italy\\
$^{128}$ Universit\`{a} del Piemonte Orientale, Vercelli, Italy\\
$^{129}$ Universit\`{a} di Brescia, Brescia, Italy\\
$^{130}$ Variable Energy Cyclotron Centre, Homi Bhabha National Institute, Kolkata, India\\
$^{131}$ Warsaw University of Technology, Warsaw, Poland\\
$^{132}$ Wayne State University, Detroit, Michigan, United States\\
$^{133}$ Westf\"{a}lische Wilhelms-Universit\"{a}t M\"{u}nster, Institut f\"{u}r Kernphysik, M\"{u}nster, Germany\\
$^{134}$ Wigner Research Centre for Physics, Budapest, Hungary\\
$^{135}$ Yale University, New Haven, Connecticut, United States\\
$^{136}$ Yonsei University, Seoul, Republic of Korea\\
$^{137}$  Zentrum  f\"{u}r Technologie und Transfer (ZTT), Worms, Germany\\
$^{138}$ Affiliated with an institute covered by a cooperation agreement with CERN\\
$^{139}$ Affiliated with an international laboratory covered by a cooperation agreement with CERN.\\

\end{flushleft} 
  
\end{document}